\documentclass[12pt]{article}
\usepackage{amsmath, latexsym, amssymb, amscd, float, authblk, tensor, mathtools, tikz-cd}
\numberwithin{equation}{section}

\newcommand{\R}{\mathbb R}
\newcommand{\C}{\mathbb C}

\newcommand{\N}{\mathbb N}
\newcommand{\Ti}{\mathbb T}

\newcommand{\ZZ}{\mathbb Z}
\newcommand{\ZZZ}{\mathcal Z}
\newcommand{\p}{\partial}

\newcommand{\Ta}{T_a}

\newcommand{\T}{T_a'}

\newcommand{\Bold}[1]{{\boldsymbol{\mathit{#1}}}}
\newcommand{\dual}{\langle\tau',\tau\rangle}

\newcommand{\ti}{\mathrm{t}}
\newcommand{\kk}{\mathrm{k}}

\newcommand{\x}{\mathrm{\mathsf{x}}}
\newcommand{\s}{\mathrm{s}}

\newcommand{\zz}{\bar{z}}

\newcommand{\m}{\mathrm{m}}

\newcommand{\M}{\mathbb{M}}
\newcommand{\U}{\mathbb{U}}

\newcommand{\D}{\mathbb{D}}

\newcommand{\z}{\mathrm{z}}

\newcommand{\QED}{\hspace{.2in}\square\newline}
\newcommand{\qed}{\hspace{.2in}\boxminus\newline}

\newcommand{\OO}{\mathit{\Omega}}
\newtheorem{theorem}{Theorem}[section]
\newtheorem{corollary}[theorem]{Corollary}
\newtheorem{proposition}[theorem]{Proposition}
\newtheorem{definition}[theorem]{Definition}

\newtheorem{lemma}[theorem]{Lemma}
\newtheorem{application}[theorem]{Application}

\newtheorem{conjecture}[theorem]{Conjecture}
\newtheorem{remark}[theorem]{Remark}

\begin{document}

\title{A Framework for Non-Gaussian Functional Integrals \\ with Applications to Quantum Field Theory and Number Theory}
\author{J. LaChapelle}
\maketitle

\begin{abstract}
We define and develop a framework to understand functional integrals as countable families of Banach-valued Haar integrals on locally compact topological groups. The definition forgoes the goal of constructing a genuine measure on an infinite-dimensional space of functions, and instead provides for a topological realization of localization in the infinite-dimensional domain. This yields measurable subspaces that characterize meaningful functional integrals and a scheme that possesses significant potential for  representing non-commutative Banach algebras suitable for mathematical physics applications. The framework includes, within a broader structure, other successful approaches that define functional integrals in restricted cases, and it suggests new and potentially useful functional integrals that go beyond the standard Gaussian case. In particular, functional integrals based on skew-Hermitian and K\"{a}hler quadratic forms are defined and developed. Also defined are gamma-type and Poisson-type functional integrals based on linear forms suggested by the gamma probability distribution. These non-Gaussian functional integrals are expected to play an important role in generating $C^\ast$-algebras of quantum systems. To illustrate and test the framework, examples and applications are presented in the contexts of quantum field theory and number theory.
\end{abstract}

\vskip 1em



\section{Introduction}
\subsection{Background and motivation}
One of the most discordant objects in mathematical physics is the functional integral. On one hand, heuristic techniques employing functional integrals
 have had remarkable success both in physics and mathematics. The successes in the context of quantum mechanics are myriad and well-known. But for the most part they merely
  reproduce results that can be obtained through operator methods, albeit often more directly and intuitively. However, in the context of quantum field theory(QFT) their formulation has gone beyond reproducing operator results; spawning important mathematical developments (see e.g. \cite{AG}--\cite{T}). Such applications tend to be quite sophisticated and non-trivial. And yet, on the other hand, there is still no widely accepted mathematically rigorous formulation encompassing all types of functional integrals that can validate (and explain) the many evident, successful applications. It would be satisfying to have a definition of functional integrals that promises the possibility of mathematical rigor and broad applicability while maintaining the pragmatic heuristics that is their hallmark. Such a definition is often expected to be an integration theory on infinite-dimensional function spaces.

There are, of course, already some rigorous constructions of
functional integrals of limited type that have been developed
\cite{ITO}--\cite{FT}.  An excellent up-to-date synopsis of various approaches and a good source of authoritative references is \cite{AL}.
 But it is fair to say these constructions are not generally viewed as definitive; in part because they are restricted to subclasses of functions that have limited applicability \cite{ITO,IM,HPS,KPS,JL,KL,FT} or because they abandon the  notion of integration with respect to an orthodox measure \cite{D-W1,AL/KR,AL/BR,CA/D-W3}.

In searching for a definition, taking stock of shared characteristics among the various existing approaches  is a good place to start. Doing so, one can make two rather obvious observations: i) functional integrals are typically defined in terms of a limiting sequence of finite dimensional objects and/or by some Fourier-type duality; and ii) evaluating integrals invariably involves some kind of reduction/localization in the integration domain that eventually leads to finite-dimensional, or at least measurable, subspaces. Indeed, from a rigorous mathematical perspective, it has proven impossible (so far) to devise a consistent and general scheme any other way. So perhaps the aspiration for a generic infinite-dimensional integration theory \emph{in the context of quantum physics} is a misstep --- more than once quantum mechanics has taught us what we can and cannot measure.

In essence, the two observations suggest perhaps one should adopt the attitude that a single functional integral associated with an infinite-dimensional function space is materially realized by a whole family of bona fide integrals corresponding to  different `questions'\footnote{By `question' we mean some kind of restriction or constraint pertinent to one's problem that singles out a subclass of measurable functions in subspaces of the integration domain.} one may ask. Much like the case of general versus particular solutions in the theory of differential equations; the entire family represents a tool to probe the function space, and measure-theoretic aspects come into play only \emph{after} a specific `question' has been posed. This idea is not really new. It (seemingly) is always implicit in any functional integral evaluation, and it is often even explicitly stated in the form of localization principles (see e.g. \cite{BT,SZ,LOC}). Our aim  is to identify a mathematical structure that captures this essence.

With these observations in mind, we propose to define functional integrals in terms of data $(G, \mathfrak{B}, G_\Lambda)$ where $G$ is a topological group, $\mathfrak{B}$ is a Banach space, and $G_\Lambda:=\{G_\lambda,\lambda\in\Lambda\}$ is a countable family of locally compact topological groups indexed by surjective homomorphisms $\lambda:G\rightarrow G_\lambda$. Given this data we define integral operators $\mathrm{int}_\lambda$ on a suitable space of integrable functionals $\mathbf{F}(G)\ni\mathrm{F}:G\rightarrow\mathfrak{B}$ by
\begin{equation}\label{definition}
\mathrm{int}_\lambda(\mathrm{F})\equiv\int_G\mathrm{F}(g)\mathcal{D}_\lambda
g:=\int_{G_\lambda}f(g_\lambda)\;d\nu(g_\lambda)
\end{equation}
with $f$ denoting the restriction $\mathrm{F}|_{G_\lambda}$ such that $f\in L^1(G_\lambda,\mathfrak{B})$ with respect to Haar measure $\nu(G_\lambda)$ for all $\lambda\in\Lambda$. We call $\mathrm{int}_\Lambda$ a family of integral operators on $\mathbf{F}(G)$ and $\mathcal{D}_\Lambda g$ its associated family of integrators.

The right-hand side of (\ref{definition}) is clearly well defined once a choice of Haar measure $\nu(G_\lambda)$ is made. So the definition will be meaningful if the set of surjective homomorphisms $\Lambda$ and integrable $\mathrm{F}\in\mathbf{F}(g)$ can be quantified. In the sequel we will give some examples of well-known embodiments of $\Lambda$, and we will see that $\mathrm{int}_\Lambda$ is really nothing more than a shorthand notation for the two observations discussed above.

Nevertheless, the shorthand $\mathrm{int}_\Lambda$ has value. First, it serves as a vehicle to transfer algebraic structure between $\mathfrak{B}$ and
 $\mathbf{F}(G)$. Second, it simultaneously incorporates both measure-theoretic and Fourier-duality approaches. And third, it begs for the introduction of non-Gaussian integrators that have practical potential.

\subsection{Outline and main results}
The plan of this paper is to give evidence to these three attributes of $\mathrm{int}_\Lambda$. We first state the proposed definition of functional integral and investigate its algebraic properties when they exist. In fact, the definition is actually rather trivial --- almost. As already discussed, it does not attempt to formulate a rigorous measure theory in infinite dimensions. Instead, it incorporates the notion that measure theory should apply only after suitable localization is effected in some topological group. Again, this notion is not new and in a sense the definition is just bookkeeping, but it does provide a framework to realize $\mathbf{F}(G)$ as a Banach algebra through its relation to $\mathfrak{B}$; and this is important, particularly for physics.

Next, we develop the prototypic Gaussian integrator family in detail including integrators induced by symmetric \emph{and} skew-symmetric quadratic forms. It is significant that the latter (which we call skew-Gaussian) are characterized by Pfaffians; thus enabling the construction of functional integrals that model Grassmann-like variables as an alternative to Berezin functional integrals. To illustrate, we construct a skew-Gaussian functional integral representation of the Mathai-Quillen Thom class representative in the spirit of \cite{BT}.

The Gaussian and skew-Gaussian can be combined into what can be characterized as a Liouville integrator. It localizes via the Berline-Vergne and Duistermaat-Heckman theorems, and its heuristic counterpart has found application in topological and cohomological QFT (for a review see e.g. \cite{SZ}--\cite{W4}). Under suitable restrictions Liouville-type functional integrals mimic supersymmetric functional integrals but with a notably unorthodox interpretation of supersymmetric fields and their physical meaning: The $\mathbb{Z}_2$ grading comes  not from a supergroup but from an extended complex phase space comprised of degrees of freedom, dual with respect to a sesquilinear form, that represent two complementary descriptions of physical interactions; viz. correlations versus dynamics.  Beyond SUSY, Liouville exhibits BRST symmetry and instances of $S$-duality that we briefly explore in an appendix.

Finally, we develop an integrator family closely allied with gamma statistics and use it to define precursors of `distributionals'. We formulate the gamma-type integrator also for non-abelian topological groups; rendering matrix functional integrals accessible. For certain parameters, the gamma-type integrator family can be interpreted as Poisson-type. And, to present a novel application, we use it to give functional integral representations of some average counting functions involving both single primes and prime $\kk$-tuples. The accuracy of these counting functions suggests the toy physical model underlying these functional integral representations may provide a useful perspective in number theory. For physicists, the approach may be useful for counting  single and correlated prime cycles/geodesics in the context of spectral determinants and dynamical zeta functions (see e.g. \cite{BO1}--\cite{MS}).

We want to stress this point: The goal here is \emph{not} to rigorously define a measure on general infinite-dimensional spaces of functions. Instead, we claim that meaningful questions of measure (e.g. those associated with a specific quantum system) necessarily distinguish specific locally compact subspaces thereof. Our notion of functional integral is not ``a number-generating machine" but rather ``a generator of number-generating machines": There is no single rigorous functional integral per se but a collection of realizable Haar integrals.  As such, there is no new measure-theoretic value in our definition. The value lies in the realization that $\mathbf{F}(G)$ inherits algebraic structure from $\Lambda$ and $\mathfrak{B}$ (or vice versa), and a point of view that suggests new integral types that should profit mathematical physics.

\section{Functional integral definition}
Our definition of functional integral is based on well-developed mathematical constructs. The relevant topics and details are collected in appendix \ref{topological groups}.

Suppose we are given $\left(G,\mathfrak{B},G_\Lambda\right)$
where $G$ is a Hausdorff topological group, $\mathfrak{B}$ is a
Banach space that may have additional \emph{associative} algebraic structure, and
$G_\Lambda:=\{G_{\lambda},\lambda\in\Lambda\}$ is a \emph{countable} family of
locally compact topological groups indexed by surjective homomorphisms $\lambda:G\rightarrow G_{\lambda}$. Insofar as $\lambda$ is surjective and $G$ contains at least one point that does not have a compact neighborhood, we can think of $G_\lambda$ as a subset $G_\lambda\subseteq G$ and loosely refer to $\lambda$ as a `topological localization'.

The idea is to use
the rigorous $\mathfrak{B}$-valued integration theory associated
with $\{G_{\lambda},\lambda\in\Lambda\}$ to define and characterize
functional integration on $G$.

\begin{definition}\label{int-def}
Let $\overline{\mathbf{F}}(G)$ represent a space of functionals\footnote{We use the term `functional' throughout this paper to emphasize the functional integral context.}  $\mathrm{F}:G\rightarrow\mathfrak{B}$, and denote the restriction of $\,\mathrm{F}$ to $G_\lambda$ by $f:=\mathrm{F}|_{G_\lambda}$. Let $\nu$ be a left Haar measure on $G_\lambda$.

A family (indexed by $\Lambda$) of integral operators
$\mathrm{int}_\Lambda:{\overline{\mathbf{F}}}(G)\rightarrow \mathfrak{B}$ is
defined by
\begin{equation}\label{FI}
\mathrm{int}_\lambda(\mathrm{F})\equiv\int_G\mathrm{F}(g)\mathcal{D}_\lambda
g:=\int_{G_\lambda}f(g_\lambda)\;d\nu(g_\lambda)
\end{equation}
if $f\in L^1(G_\lambda,\mathfrak{B})$ for all
$\lambda\in\Lambda$. We say that $\mathrm{F}$ is integrable with respect to the integrator family
$\mathcal{D}_\Lambda g$, and $\mathbf{F}(G)\subseteq\overline{\mathbf{F}}(G)$ is the subspace of all such integrable functionals.

Further, if $\mathfrak{B}$ is an algebra, define the functional $\ast$-convolution on $\mathbf{F}(G)$ by
\begin{equation}
\left(\mathrm{F}_1\ast \mathrm{F}_2\right)_{{\lambda}}(g)
:=\int_G\mathrm{F}_1(\tilde{g})\mathrm{F}_2(\tilde{g}^{-1}g)
\mathcal{D}_{{\lambda}}\tilde{g}\;\;\;\;\forall\;\lambda\in\Lambda.
\end{equation}
\end{definition}

According to the definition, for any given
$\lambda$, the integral operator is linear and bounded according to
\begin{equation}
\|\mathrm{int}_\lambda(\mathrm{F})\|\leq\int_{G_\lambda}\|f(g_\lambda)\|
\;d\nu(g_\lambda)=:\|f\|_{1}<\infty
\end{equation}
which follows from the Cauchy-Schwarz inequality  and Proposition \ref{banach integration}. This suggests to define the norm $\|\mathrm{F}\|_{\mathbf{F}}:=\sup_\lambda\|\mathrm{int}_\lambda(\mathrm{F})\|$.
It is not hard to see $\|\cdot\|_{\mathbf{F}}$ is a norm using the fact that $\|\cdot\|_{1}$ is a norm on $\mathfrak{B}$. But there may be more suitable norms depending on the context, and we don't insist on this particular choice.

Since $\mathbf{F}(G)$ is a normed space, its completion (which will be denoted by the same symbol) is a Banach space. Suppose $\mathfrak{B}$ is an algebra. The $\ast$-convolution then implies
\begin{eqnarray}\label{product}
\mathrm{int}_\lambda(\mathrm{F}_1\ast \mathrm{F}_2)
&=&\int_G(\mathrm{F}_1\ast \mathrm{F}_2)(g)\mathcal{D}_\lambda g\notag\\
&=&\int_{G_\lambda\times G_\lambda}f_1(\tilde{g}_\lambda)
f_2(\tilde{g}_\lambda^{-1}g_\lambda)
\;d\nu(\tilde{g}_\lambda,g_\lambda)\notag\\
&=&\int_{G_\lambda}\int_{G_\lambda}f_1(\tilde{g}_\lambda)
f_2(g_\lambda)
\;d\nu(\tilde{g}_\lambda)d\nu(g_\lambda)\notag\\
&=&\mathrm{int}_\lambda(\mathrm{F}_1)\,\mathrm{int}_\lambda(\mathrm{F}_2)
\end{eqnarray}
where we used  left-invariance of the Haar measure and Fubini. Hence, integral operators are algebra homomorphisms.

A similar computation (again using left-invariance and Fubini) establishes associativity $(\mathrm{F}_1\ast \mathrm{F}_2)\ast \mathrm{F}_3=\mathrm{F}_1\ast (\mathrm{F}_2\ast \mathrm{F}_3)$. Moreover, Banach $\mathfrak{B}$, eq. (\ref{product}), and the multiplicative property of $\mathrm{sup}$ imply $\|\mathrm{F}_1\ast \mathrm{F}_2\|_{\mathbf{F}}\leq \|\mathrm{F}_1\|_{\mathbf{F}}\,\|\mathrm{F}_2\|_{\mathbf{F}}$. Consequently, $\mathbf{F}(G)$ inherits the algebraic structure of $\mathfrak{B}$ and we have shown:

\begin{proposition}
$\mathbf{F}(G)$  equipped with the $\ast$-convolution is a Banach algebra when completed with respect to the norm
$\|\mathrm{F}\|_{\mathbf{F}}:=\sup_\lambda\|\mathrm{int}_\lambda(\mathrm{F})\|$ (or other suitable norm).
\end{proposition}

With an eye toward applications to quantum systems, we want to add the additional structure of $\ast$-algebras. To that end, suppose now $\mathfrak{B}$ is a Banach $\ast$-algebra, and define an involution on $\mathbf{F}(G)$ such that
\begin{equation}\label{involution}
\left.\mathrm{F}^\ast (g)\right|_{G_\lambda}:=\left.{\mathrm{F}(g^{-1})}^\ast\Delta(g^{-1})\right|_{G_\lambda}
:={f(g_\lambda^{-1})}^\ast\Delta(g_\lambda^{-1})\;\;\;\;\;\;\;\;\forall\lambda\in\Lambda\;;
\end{equation}
which defines a modular functional on $G$ through its restriction to $G_\Lambda$. Equipped with this involution we get
\begin{proposition}
 The integral operator $\mathrm{int}_\lambda$ is a $\ast$-homomorphism, and involutive $\mathbf{F}(G)$ --- endowed with a suitable topology and completed with respect to
the norm $\|\cdot\|_{\mathbf{F}}$ ---  is a Banach
$\ast$-algebra.
\end{proposition}
For quantum physics applications, restrict to $C^\ast$-algebras:
\begin{corollary}
If $\mathfrak{B}$ is a $C^\ast$-algebra, then involutive $\mathbf{F}(G)$ is a $C^\ast$-algebra.
\end{corollary}
\emph{Proof}:
First,
\begin{eqnarray}\label{star}
\mathrm{int}_\lambda(\mathrm{F}^\ast)
&=&\int_G\mathrm{F}^\ast(g)\mathcal{D}_\lambda g\notag\\
&=&\int_{G_\lambda}f^\ast(g_\lambda)\;d\nu(g_\lambda)\notag\\
&=&\int_{G_\lambda}f(g_\lambda^{-1})^\ast\Delta(g_\lambda^{-1})
\;d\nu(g_\lambda)\notag\\
&=&\int_{G_\lambda}f(g_\lambda)^\ast \;d\nu(g_\lambda)\notag\\
&=&\left(\int_{G_\lambda}f(g_\lambda)\;d\nu(g_\lambda)\right)^\ast\notag\\
&=&\mathrm{int}_\lambda(\mathrm{F})^\ast
\end{eqnarray}
where the fourth line follows by virtue of the Haar measure. Together with (\ref{product}), this shows the integral operators are $\ast$-homomorphisms.

From the previous proposition, we know that $\mathbf{F}(G)$ is a Banach algebra so it remains to verify the $\ast$-algebra axioms. First, the $\ast$-operation is continuous for a suitable choice of topology, and linearity is obvious. Next,
\begin{equation}
(\mathrm{F}^\ast)^\ast(g):={\mathrm{F}^\ast(g^{-1})}^\ast\Delta(g^{-1})=(\mathrm{F}(g)^\ast)^\ast\Delta(g)\Delta(g^{-1})=\mathrm{F}(g)
\end{equation}
and
\begin{eqnarray}
\left(\mathrm{F}_1^\ast\ast \mathrm{F}_2^\ast\right)_{{\lambda}}(g)
&:=&\int_{G_\lambda}f^\ast_1(\tilde{g}_\lambda)
f^\ast_2(\tilde{g}_\lambda^{-1}g_\lambda)
\;d\nu(\tilde{g}_\lambda)\notag\\
&=&\int_{G_\lambda}\left(f_2(g_\lambda^{-1}\tilde{g}_\lambda)
\Delta(g_\lambda^{-1}\tilde{g}_\lambda)f_1(\tilde{g}^{-1}_\lambda)
\Delta(\tilde{g}_\lambda^{-1})\right)^\ast
\;d\nu(\tilde{g}_\lambda)\notag\\
&=&\left(\int_{G_\lambda}f_2(g_\lambda^{-1}\tilde{g}_\lambda)
 f_1(\tilde{g}^{-1}_\lambda)
\Delta(g_\lambda^{-1})\;d\nu(\tilde{g}_\lambda)
\right)^\ast\notag\\
&=&\left((\mathrm{F}_2\ast \mathrm{F}_1)_{{\lambda}}(g^{-1})\right)^\ast\Delta(g^{-1})\notag\\
&=&\left(\mathrm{F}_2\ast \mathrm{F}_1\right)^\ast_{{\lambda}}(g)
\end{eqnarray}
where we used (\ref{involution}), left-invariance of the Haar measure, and the modular function $\Delta$ is a homomorphism. For the norm, Banach $\mathfrak{B}$ and eq. (\ref{star}) imply $\|\mathrm{int}_\lambda(\mathrm{F}^\ast)\|=\|\mathrm{int}_\lambda(\mathrm{F})^\ast\|
=\|\mathrm{int}_\lambda(\mathrm{F})\|$ which implies $\|\mathrm{F}^\ast\|_{\mathbf{F}}=\|\mathrm{F}\|_{\mathbf{F}}$. Conclude that $\mathbf{F}(G)$ is a $\ast$-algebra. Finally, if $\mathfrak{B}$ is a $C^\ast$-algebra, the corollary follows from (\ref{product}) and (\ref{star}) since now
\begin{equation}
\|\mathrm{int}_\lambda(\mathrm{F}\ast\mathrm{F}^\ast)\|
=\|\mathrm{int}_\lambda(\mathrm{F})\,\mathrm{int}_\lambda(\mathrm{F})^\ast\|
=\|\mathrm{int}_\lambda(\mathrm{F})\|\|\mathrm{int}_\lambda(\mathrm{F})^\ast\|
=\|\mathrm{int}_\lambda(\mathrm{F})\|^2
\end{equation}
implies $\|\mathrm{F}\ast\mathrm{F}^\ast\|_{\mathbf{F}}=\|\mathrm{F}\|^2_{\mathbf{F}}$.
$\QED$

The corollary is important, and note that the implication can be reversed. It means the as-defined functional integral provides a two-way bridge between two $C^\ast$-algebras. In particular, if $\mathfrak{B}$ happens to be the space of bounded, linear operators on a Hilbert space $L_B(\mathcal{H})$, this bridge implements a fusion of operator and functional integral methods in quantum physics.

\begin{remark}\label{measurement} Although the products in $\mathbf{F}(G)$ and $L^1(G_\lambda,\mathfrak{B})$ are trivially equivalent
 by definition, their respective norms are not. Our particular choice of norm on $\mathbf{F}(G)$ renders the restriction  of $\mathbf{F}(G)$ to the countable family $G_\Lambda$ a direct sum if the cardinality of $\Lambda$ is finite; namely
  $\mathbf{F}(G)|_{G_\Lambda}=\bigoplus_{\lambda\in\Lambda} L^1(G_\lambda,\mathfrak{B})$ \emph{(\cite[pg. 16--17]{BL})}. In this sense, a `question' --- which
   corresponds to a measurable element(observable) $\mathrm{O}\in\mathbf{F}(G)$ \textbf{and} a particular choice of $\lambda$ --- induces a projection.

 In the context of quantum mechanics, the above property and the topological framework provide a Heisenberg-picture interpretation of the measurement
 process. Suppose that $\mathbf{F}(G)$ is the $C^\ast$-algebra characterizing some quantum
 system, $G$ governs the system dynamics, and $\mathrm{O}\in\mathbf{F}(G)$ is an observable. The very notion of observation/measurement requires a restriction to locally compact $G_\Lambda$. Further, restricting to closed quantum systems, insist that the set
 $\Lambda_{meas}\subseteq\Lambda$ yields a finite family $G_{\Lambda_{meas}}$ that can be represented by \textbf{unitaries} in $\mathfrak{B}$.

 We have seen that, as far as measure is concerned, the functional $\mathrm{O}$  (represented as a functional integral) is realized as an
    entire family of functions. It is easy to imagine that the physical quantum states (more precisely, density matrices) of macroscopic observers correspond to this family of functions and, hence, are indexed by the set $\Lambda_{meas}$. Which member $\lambda\in\Lambda_{meas}$ is realized in a measurement of course is not predetermined reflecting quantum indeterminacy. Alternatively and equally valid, one can imagine that a particular `topological localization' $G\rightarrow G_\lambda$ represents a subjective ``choice" made by some observer.

Either way we get a topological interpretation: performing a measurement\footnote{By the phrase ``performing a measurement'' we mean not the system/measurement-device interaction, but the preparation of the system and subsequent query of the device's readout which takes place at some finite space-time interval away from the system/measurement-device interaction.} --- and thereby actualizing an observable ---
corresponds to a projection of $\mathbf{F}(G)|_{G_{\Lambda_{meas}}}$ onto
$L^1(G_\lambda,\mathfrak{B})$.\footnote{We do not mean to imply that this projection has any causal effect on
physical reality:  We are in the Heisenberg picture so the system's wave function can be ontological (if one insists; but we don't) while the observable representing a measurement is both deterministic and epistemic.} Precisely which projection is effected by the measurement cannot be known unless $\lambda$ is already known. Subsequent
 measurement will of course be referred to $L^1(G_\lambda,\mathfrak{B})\subset\mathbf{F}(G)$ unless interaction dynamics takes
  the system out of this subspace: The dynamics are generically governed by $G$ so, if there are external interactions, the initial subspace may be no longer relevant and measurement may settle on a new $G_{\lambda'}$ and consequently a new fiducial subspace $L^1(G_{\lambda'},\mathfrak{B})$. Evidently, evolution of a prepared \textbf{closed} quantum system would be modeled by
  $L^1(G_\lambda,\mathfrak{B})$ where $\lambda\in\Lambda_{meas}$ (corresponding to a known preparation) encodes its initial state.

In effect, the topological model supplies a family of isomorphic (under measurement) Hilbert spaces. The family represents indeterminacy; not of the system (observable $\mathrm{O}$) but of the system relative to some measuring ``ruler" (observable $\mathrm{O}_\lambda$). Once a measurement has been made, $\mathrm{O}$ is given only comparative meaning
   through a specific representation of the associated observable $\mathrm{O}_\lambda$ carried by
   some Hilbert space based on a specific, locally compact $G_\lambda$. Alternatively, we can imagine a single encompassing Hilbert space with unitarily-equivalent undetermined subspaces. An initial preparation/measurement allows identification of a fiducial subspace $L^1(G_\lambda,\mathfrak{B})$  by which comparisons with subsequent $G_\lambda$-preserving measurements can be made. This picture meshes well with the perspective of \emph{\cite{APR}} if we identify $\Lambda$ with their ``frames of reference and preparation devices''.
\end{remark}

\section{Quadratic-type integrators}\label{Quadratic-type integrators}
This section develops integrators based on sesquilinear forms on \emph{abelian} topological groups.\footnote{Strictly, sesquilinear forms restricted to the underlying complex abelian group of a complex topological vector space.} This is a substantial simplification of the general non-abelian case. However, such forms give rise to the ubiquitous Gaussian integrator family as well as the skew-Gaussian integrator family to be introduced below. In this section we restrict to `path integrals' because this simpler context allows attention to be focused on the properties of the integrators without the distractions and issues of fields. Nevertheless, extending these quadratic-type integrator families to include fields should be straightforward; regularization and local symmetry notwithstanding. The development can be extended to non-abelian linear Lie groups $G$ (see Definition \ref{linear Lie}) by considering sesquilinear forms on the cotangent bundle $T^\ast G$.

Our starting point is a topological space $\mathcal{P}_{ab}\C^{m}$ of $L^{2,1}$ piece-wise continuous paths $ z:([\ti_a,\ti_b]=:\mathbb{T}\subseteq\R)\rightarrow \C^m$ satisfying boundary conditions $c_B(z(\ti_a),\dot{z}(\ti_a))=c_a$ and $c_B(z(\ti_b),\dot{z}(\ti_b))=c_b$.\footnote{One often starts with the (contractible and hence Banach) space $\mathcal{P}_{a}\C^{m}$ of piece-wise continuous, pointed maps $ z:(\mathbb{T},\ti_a)\rightarrow (\C^m,c_a)$.(see e.g. \cite{CA/D-W3,LA2}) Being Banach, $\mathcal{P}_{a}\C^{m}$ is an abelian topological group. Subsequently, one often fixes the other endpoint $z(\ti_b)$ through functional integral manipulation. Here we want to add some variety and provide an alternative approach, but note that $\mathcal{P}_{ab}\C^{m}$ is not a Banach space.} The topological dual of $\mathcal{P}_{ab}\C^{m}$ is the space of continuous linear forms $z':\mathcal{P}_{ab}\C^{m}\rightarrow\C$ by $z\mapsto\langle z',z\rangle\in\C$. The involution and complex structure on $\C^m$ induce an involution and complex structure on $\mathcal{P}_{ab}\C^{m}$ according to the prescriptions
\begin{eqnarray}
(z^\ast)(\ti):=z(\ti)^\ast\;\;\;\;\;\;
(\mathrm{J}z)(\ti):=iz(\ti)\;\;\;\;\;\;
(\mathrm{J}z^\ast)(\ti):=-iz(\ti)^\ast\;.
\end{eqnarray}
Use these structures to  complexify $(\mathcal{P}_{ab}\C^{m})^\C=:Z_{ab}$. By duality these structures can be transferred to the topological dual space $Z'_{ab}$. For example ${\mathrm{J}'}z'=iz'$ where the transpose ${\mathrm{J}'}$ is determined by $\langle{\mathrm{J}'}z',z\rangle=\langle z',\mathrm{J}z\rangle$.

Let $ Z'_{ab}$ be endowed with a continuous sesquilinear form
\begin{eqnarray}
\mathrm{F}': Z'_{ab}\times  Z'_{ab}&\rightarrow&\C\notag\\
( {z'}_1, z_2')&\mapsto&\mathrm{F}'( {z'}_1, z_2'):=\langle  {z'}_1,Cz_2'\rangle
\end{eqnarray}
where the (linear) generator of variance $C: Z'_{ab}\rightarrow  Z_{ab}$ is nondegenerate with domain ${\mathrm{D}_C}= Z'_{ab}$.  On the dual space $Z_{ab}$, construct an associated \emph{closed} sesquilinear form
\begin{eqnarray}
\mathrm{F}: Z_{ab}\times  Z_{ab}&\rightarrow&\C\notag\\
( z_1, z_2)&\mapsto&\mathrm{F}( z_1, z_2)-\mathrm{B}(z_1,z_2)=-\langle Dz_1, z_2\rangle=:\mathrm{F}_{\mathrm{B}}( z_1, z_2)
\end{eqnarray}
 where $D: Z_{ab}\rightarrow  Z'_{ab}$ is linear and $\mathrm{B}(z_1,z_2)$ is a sesquilinear boundary form.

 Let $\{z_{cr}\}$ be the set of critical paths of $\mathrm{F}_\mathrm{B}(z)$, that is $Dz_{cr}=0$ with boundary conditions $z_{cr}(\ti_{a,b})=c_{a,b}$, and put $Z_{\hat{z}_{cr}}:= Z_{ab}\backslash\mathrm{Ker}(D)$. Then, \emph{restricting to this factor space}, we require
\begin{eqnarray}
DC&=&\mathrm{Id}_{ Z'_{\hat{z}_{cr}}}\notag\\
CD&=&\mathrm{Id}_{ Z_{\hat{z}_{cr}}}\;,
\end{eqnarray}
and so in this sense $\mathrm{F}'$ and $\mathrm{F}$ are dual forms on $ Z_{\hat{z}_{cr}}$ modulo a boundary form. Further, on $Z_{0}$ (the space of path with $c_a=c_b=0$), any $z\in Z_{0}$ can be reached from a given $z_{cr}$ by $z(\ti)=z_{cr}(\ti)+(Cz')(\ti)$ for all $z'\in Z'_{\hat{z}_{cr}}$, and for each $z_{cr}\in\{\z_{cr}\}$ there is a copy of $Z_{\hat{z}_{cr}}$ in $Z_{0}$. Let $\{\bar{z}\}$ be the set of ``mean'' paths defined to be the stationary points of $\Gamma(z):=\mathrm{F}'(z')-\langle z', z\rangle$ at $z'=0$ with boundary conditions $c_B(\bar{z}(\ti_a),\dot{\bar{z}}(\ti_a))=c_{a,b}$. Since we are dealing with quadratic forms, it follows that $\mathrm{F}_\mathrm{B}(z_{cr})=\Gamma(\bar{z})$.

Decompose $\mathrm{F}_\mathrm{B}$ into Hermitian and skew-Hermitian parts according to $\mathrm{F}_\mathrm{B}=\mathrm{\mathrm{Q}}_\mathrm{B}+\Omega_\mathrm{B}$ where
\begin{eqnarray}
\mathrm{\mathrm{Q}}_\mathrm{B}( z_1, z_2):= -\tfrac{1}{2}\left\{\langle  Dz_1,z_2\rangle +\langle
 Dz_2,z_1\rangle\right\}= -\tfrac{1}{2}\left\langle  (D+D^\dag)z_1,z_2\right\rangle
 =:-\tfrac{1}{2}\left\langle  Qz_1,z_2\right\rangle
 \end{eqnarray}
 and
 \begin{eqnarray}
\Omega_\mathrm{B}( z_1, z_2)
 :=-\tfrac{1}{2}\left\{\langle  Dz_1,z_2\rangle -\langle
 Dz_2,z_1\rangle\right\}=-\tfrac{1}{2}\left\langle  (D-D^\dag)z_1,z_2\right\rangle
 =:-\tfrac{1}{2}\left\langle  \mathit{\Omega}z_1,z_2\right\rangle\;.
 \end{eqnarray}
Observe $\mathrm{Q}_\mathrm{B}(z_1,z_2)=\mathrm{Q}_\mathrm{B}(z_2,z_1)^\ast$ but $\Omega_\mathrm{B}(z_1,z_2)=-\Omega_\mathrm{B}(z_2,z_1)^\ast$. To make contact with QM, use $\mathrm{Q}_\mathrm{B}(z,z)$ (resp. $i\Omega_\mathrm{B}(z,z)$) to define a norm on $Z_{\hat{z}_{cr}}$ in the usual way, then complete $Z_{\hat{z}_{cr}}$ with respect to this norm to get the Hilbert space $\mathcal{H}_{\mathrm{Q}_\mathrm{B}}$ (resp. $\mathcal{H}_{\Omega_\mathrm{B}}$) of paths possessing boundary conditions encoded in the boundary term $\mathrm{B}(\cdot,\cdot)$.

\subsection{Gaussian integrators}\label{Gaussian integrators}
Consider the space $Z_0$ of piece-wise continuous paths $z:([\ti_a,\ti_b]=:\mathbb{T}\subseteq\R)\rightarrow \C^m$ satisfying \emph{vanishing} boundary conditions $c_B(z(\ti_a),\dot{z}(\ti_a))=0$ and $c_B(z(\ti_b),\dot{z}(\ti_b))=0$.  Then $z-\bar{z}$ will be paths in $Z_{ab}$. Gaussian integrators are constructed from Hermitian quadratic forms $\mathrm{Q}_\mathrm{B}$ on $Z_0$ ``shifted'' by $\bar{z}$.\footnote{Altough $z$ now has vanishing boundary conditions, the ``mean'' paths here have boundary conditions $c_B(\bar{z}(\ti_a),\dot{\bar{z}}(\ti_a))=c_{a,b}$, but they do not necessarily have to be critical points of $\Gamma(z)$: They can be any paths with suitable boundary conditions, e.g. critical points of $\mathrm{F}_{\mathrm{B}}$.}
\begin{definition}\label{Gaussian}
A family of Gaussian integrators $\mathcal{D}_\Lambda\omega_{\zz,\mathrm{Q}_\mathrm{B}}( z)$ is characterized
by
\begin{eqnarray}\label{Gaussian definition}
&&\Theta_{\zz,\mathrm{Q}_\mathrm{B}}( z, z')=e^{2\pi i \langle  z',z\rangle-(\pi/\s) \mathrm{Q}_\mathrm{B}( z-\zz)}\notag\\
&&\mathrm{Z}_{\zz,\mathrm{W}_\mathrm{B}}( z')=e^{2\pi i \langle  z',\zz\rangle}\,\mathrm{Det}_\lambda (\s{\mathrm{W}_\mathrm{B}})^{1/2}e^{-\pi\s \mathrm{W}_\mathrm{B}( z')}
\end{eqnarray}
where $\mathrm{W}_\mathrm{B}(z_1',z_2'):=-2\langle z_1',Cz_2'\rangle$ is inverse to $\mathrm{Q}_\mathrm{B}(z_1,z_2)$ according to $DC:=\mathrm{Id}_{ Z'_{\hat{z}_{cr}}}$ and $CD:=\mathrm{Id}_{Z_{\hat{z}_{cr}}}$. Explicitly,
\begin{equation}
\frac{\delta^2\mathrm{Q}_\mathrm{B}}{\delta z(\ti)^i\delta z(\ti)^j}\,\delta^{jk}\,\frac{\delta^2\mathrm{W}_\mathrm{B}}{\delta z'(\ti')^k\delta z'(\ti')^l}=\delta(\ti-\ti')\delta_{il}
\end{equation}
with $i,j,k,l\in\{1,\ldots,m\}$. The parameter $\s\in\C_+\cong\R_+\times i\R$ where $\R_+$ is the group of positive-definite reals, and the functional determinant is
assumed to be well-defined/regularized so that $\mathrm{Det}_\lambda(\cdot)$ coincides with $\mathrm{det}(\cdot)$ for a given $\lambda$ (up to a possible phase).

The Gaussian integrator family is defined in terms of the primitive integrator family $\mathcal{D}_\Lambda z$;
\begin{equation}\label{primative}
\mathcal{D}_\lambda\omega_{\zz,\mathrm{Q}_\mathrm{B}}( z)
:=e^{-(\pi/\s) \mathrm{Q}_\mathrm{B}( z-\zz)}\mathcal{D}_\lambda z
\end{equation}
where $\mathcal{D}_\lambda z$ is characterized by zero mean and trivial covariance
\begin{eqnarray}
&&\Theta_{0,\mathrm{Id}}( z, z')=e^{2\pi i \langle
 z', z\rangle-(\pi/\s)\mathrm{Id}( z)}\notag\\
&&\mathrm{Z}_{0,\mathrm{Id}}( z')=\mathrm{Det}_\lambda (\s\,\mathrm{Id})^{1/2}\,e^{-\pi\s \;\mathrm{Id}( z')}
\end{eqnarray}
with $\mathrm{Id}( z)=-\frac{1}{2}\langle Id\, z,z\rangle$. Given these objects, we define \emph{(c.f. \cite[ch.~2]{CA/D-W3})}
\begin{equation}
\int_{Z_0}e^{2\pi i \langle  z',z\rangle}
\;\mathcal{D}_\lambda\omega_{\zz,\mathrm{Q}_\mathrm{B}}(z)
=\sum_{\{z_{cr}\}}\!\!\!\!\!\!\!\!\!\int\;\int_{Z_{\hat{z}_{cr}}}\Theta_{\zz,\mathrm{Q}_\mathrm{B}}( z, z')\;\mathcal{D}_\lambda z
:=\mathrm{Z}_{\zz,\mathrm{W}_\mathrm{B}}( z')\;.
\end{equation}
\end{definition}
There is an obvious restriction $(\s^{-1}\mathrm{Q}_{\mathrm{B}}(z)|_{G_\lambda})\in\C_+$ ensuring integrable $\Theta_{\zz,\mathrm{Q}_\mathrm{B}}( z, z')$.

Observe the well-known Fourier transform interpretation between the left-hand and right-hand sides. Because $Z_0$ is an abelian group, $\mathcal{D}_\lambda z$ is translation invariant so we can write
\begin{align}
e^{-2\pi i \langle  z',\bar{z}\rangle}\,\mathrm{Z}_{\zz,\mathrm{W}_\mathrm{B}}( z')
&=\int_{Z_0}e^{2\pi i \langle  z',(z-\bar{z})\rangle}e^{-(\pi/\s) \mathrm{Q}_\mathrm{B}( z-\zz)}
\;\mathcal{D}_\lambda(z-\bar{z})\notag\\
&=\int_{Z_0}e^{2\pi i \langle  z',\tilde{z}\rangle}e^{-(\pi/\s) \mathrm{Q}_\mathrm{B}(\tilde{z})}
\;\mathcal{D}_\lambda\tilde{z}\notag\\
&=\int_{Z_0}e^{2\pi i \langle  z',\tilde{z}\rangle}
\;\mathcal{D}_\lambda\omega_{0,\mathrm{Q}_\mathrm{B}}(\tilde{z})\notag\\
&=\mathrm{Z}_{0,\mathrm{W}_\mathrm{B}}( z')\;.
\end{align}
Evidently, $\mathcal{F}(\omega_{\zz,\mathrm{Q}_\mathrm{B}}(z))=e^{2\pi i \langle  z',\bar{z}\rangle}\mathcal{F}(\omega_{0,\mathrm{Q}_\mathrm{B}}(z))$ where $\mathcal{F}(\cdot)$ denotes the functional Fourier transform with respect to $z\in Z_0$.

\begin{remark}
The resemblance between the functional form of $\mathrm{Z}_{\zz,\mathrm{W}_\mathrm{B}}(z')$ and the exponential multiplying the primitive integrator in \emph{(\ref{primative})} motivates the standard practice of
defining the effective action.

Start with the characteristic functional of the normalized Gaussian
\begin{equation}
\mathcal{Z}_{\zz,\mathrm{W}_\mathrm{B}}(z')
:=\frac{1}{{\mathrm{Z}_{\zz,\mathrm{W}_\mathrm{B}}(0)}}
\mathcal{F}\left(\omega_{\zz,\mathrm{Q}_{\mathrm{B}}}(z)\right)
=e^{2\pi i\langle z',\zz\rangle-\pi\s \mathrm{W}_\mathrm{B}(z')}
=:-e^{\pi \s\Gamma'_{\zz}(z')}\;.
\end{equation}
Then,
\begin{eqnarray}
\frac{1}{2\pi i}\frac{\delta}{\delta z'(\ti)}\,\ln\mathcal{Z}_{\zz,\mathrm{W}_\mathrm{B}}(z')
 &=&\left(\zz(\ti)-\frac{\s}{2i}\frac{\delta \mathrm{W}_\mathrm{B}(z')}{\delta z'(\ti)}\right)\notag\\
&=&\frac{1}{\mathcal{Z}_{\zz,\mathrm{W}_\mathrm{B}}(z')}\frac{\delta}{\delta z'(\ti)}
\int_{Z_0}e^{2\pi i\langle z',z\rangle}\;\mathcal{D}_\lambda\omega_{\zz,\mathrm{Q}_{\mathrm{B}}}(z)\notag\\
&=:& \mathbf{z}_{z'}(\ti)\;.
\end{eqnarray}
From the second equality it follows that $\mathbf{z}_{z'}:\mathbb{T}\rightarrow \C^m$ is an $L^{2,1}$ function with $\mathbf{z}_{z'}(\ti_{a,b})=c_{a,b}$ for all $z'\in X'_0$; also $\mathbf{z}_{z'_{cr}}(\ti)=\zz(\ti)$ is a ``mean'' path where $z'_{cr}$ is a critical point of $\mathrm{W}_\mathrm{B}(z')$.

Define the effective action function $\Gamma_{z'}$ evaluated at $\mathbf{z}_{z'}\in X_{\hat{z}_{cr}}$ by
\begin{equation}\label{effective action}
(\pi/\s)\Gamma_{z'}(\mathbf{z}_{z'})
:=\pi\s \mathrm{W}_\mathrm{B}(z')-2\pi i\langle z',\mathbf{z}_{z'}\rangle
=\pi\s\Gamma'_{\mathbf{z}_{z'}}(z')\;.
\end{equation}
The functional Fourier transform of $\frac{1}
{\mathrm{Z}_{\mathbf{z}_{z'},\mathrm{W}_\mathrm{B}}(0)}\omega_{\mathbf{z}_{z'},\mathrm{Q}_{\mathrm{B}}}(z)$ is the exponentiated effective action. Explicitly,
\begin{eqnarray}
\mathcal{Z}_{\mathbf{z}_{z'},\mathrm{W}_\mathrm{B}}(z')
&=&\frac{1}{{\mathrm{Z}_{\mathbf{z}_{z'},\mathrm{W}_\mathrm{B}}(0)}}
\;\int_{Z_0}e^{2\pi i\langle z',z\rangle-(\pi/\s) \mathrm{Q}_\mathrm{B}(z-\mathbf{z}_{z'})} \;\mathcal{D}_\lambda z\notag\\
 &=&e^{2\pi i\langle z',\mathbf{z}_{z'}\rangle}
 \,\mathcal{Z}_{0,\mathrm{W}_\mathrm{B}}(z')\notag\\
 &=&e^{-(\pi/\s)\Gamma_{z'}(\mathbf{z}_{z'})}\;.
\end{eqnarray}
Notice that, since the sesquilinear form $\mathrm{F}$ is quadratic,  the functional integral is easily evaluated once all $\zz$ are known because $\Gamma_{z'}(\mathbf{z}_{z'})|_{z'=z'_{cr}}=\Gamma_{z'_{cr}}(\bar{z})=\mathrm{Q}_{\mathrm{B}}(\bar{z})
=\mathrm{Q}_{\mathrm{B}}(z_{cr})$.  In this quadratic case, the formulation of effective action is fairly trivial and unproductive.

However, Gaussian integrators can be readily generalized to non-Gaussian integrators based on non-quadratic action functionals $\mathrm{S}_{\mathrm{B}}=\mathrm{Q}_{\mathrm{B}}+\mathrm{V}_{\mathrm{B}}:Z_0\times Z_0\rightarrow\C$ --- in which case the effective action becomes a useful tool. Define ${\mathrm{Z}}_{\zz,\mathrm{S}_{\mathrm{B}}'}(z'):=\int_{X_0}e^{2\pi i\langle z',z\rangle}e^{-(\pi/\s)V_{\mathrm{B}}(z-\mathbf{z}_{z'})}
\;\mathcal{D}\omega_{\zz,\mathrm{Q}_{\mathrm{B}}}(z)$ and ${\mathcal{Z}}_{\zz,\mathrm{S}_{\mathrm{B}}'}(z')
:=\frac{1}{\mathrm{Z}_{\zz,\mathrm{S}_{\mathrm{B}}'}(0)}
\mathcal{F}\left(e^{-(\pi/\s)V_{\mathrm{B}}(z-\mathbf{z}_{z'})}\,\omega_{\zz,\mathrm{Q}_{\mathrm{B}}}(z)\right)$. Then define
\begin{equation}
{\mathcal{Z}}_{\zz,\mathrm{S}_{\mathrm{B}}'}(z')
:=\frac{1}{{\mathrm{Z}_{\zz,\mathrm{S}_{\mathrm{B}}'}(0)}}\int_{Z_0}e^{2\pi i\langle z',z\rangle}e^{-(\pi/\s)V_{\mathrm{B}}(z-\mathbf{z}_{z'})}\;\mathcal{D}_\lambda\omega_{\zz,\mathrm{Q}_{\mathrm{B}}}(z)
=:e^{(\pi/\s)\widetilde{\Gamma}'_{\zz}(z')}
=:e^{2\pi i\langle z',\zz\rangle-\pi\s \,\mathrm{S}_{\mathrm{B}}'(z')}\;,
\end{equation}
and
\begin{equation}
(\pi/\s)\widetilde{\Gamma}_{z'}(\mathbf{z}_{z'})
:=\pi\s\,\mathrm{S}_{\mathrm{B}}'(z')-2\pi i\langle z',\mathbf{z}_{z'}\rangle\;.
\end{equation}
Now, however, $\widetilde{\Gamma}_{z'}(\mathbf{z}_{z'})|_{z'=z'_{cr}}\neq\mathrm{S}_{\mathrm{B}}(z_{cr})$  (where now $z'_{cr}$ is a critical point of $\mathrm{S}_{\mathrm{B}}'$ and $z_{cr}$ is a critical point of $\mathrm{S}_{\mathrm{B}}$) because $\mathrm{S}_{\mathrm{B}}'$ is no longer a simple bilinear form. Nevertheless, we still get
\begin{equation}
{\mathcal{Z}}_{\mathbf{z}_{z'},\mathrm{S}_{\mathrm{B}}'}(z')
=e^{2\pi i\langle z',\mathbf{z}_{z'}\rangle}
\,{\mathcal{Z}}_{0,\mathrm{S}_{\mathrm{B}}'}(z')
 =e^{(\pi/\s)\widetilde{\Gamma}_{z'}(\mathbf{z}_{z'})}\;.
\end{equation}
 As is well-known, ${\mathcal{Z}}_{0,\mathrm{S}_{\mathrm{B}}'}(z')$ is generally not computable in closed form, and one usually resorts to expansions around $\mathrm{Q}_{\mathrm{B}}$.
\end{remark}

\begin{remark}
The definition of functional integral allows to take limits of Gaussian integral operators with respect to the parameter $\s$ when the limits exist for the finite-dimensional integrals. Accordingly, one can define an integrator analog of the Dirac measure;
\begin{eqnarray}\label{dirac measure}
\lim_{|\s|\rightarrow0}\frac{1}{\mathrm{Z}_{\zz,\mathrm{W}_\mathrm{B}}(0)}\int_{Z_0}\Theta_{\zz,\mathrm{Q}_\mathrm{B}}( z, z')\mathcal{D}_\lambda z
&=:&\int_{Z_0}e^{2\pi i \langle  z',z\rangle}\delta(z-\zz)\mathcal{D}_\lambda z\notag\\
&=:&\int_{Z_0}e^{2\pi i \langle  z', z\rangle}\mathcal{D}_\lambda\delta_{\bar{z}}(z)\notag\\
&=&\lim_{|\s|\rightarrow0}e^{2\pi i\langle z',\bar{z}\rangle}e^{-\pi\s \mathrm{W}_\mathrm{B}(z')}\notag\\
&=&e^{2\pi i\langle z',\bar{z}\rangle}\;.
\end{eqnarray}
This definition makes sense because: i)  the ratio $\Theta_{\zz,\mathrm{Q}_\mathrm{B}}( z, 0)/\mathrm{Z}_{\zz,\mathrm{W}_\mathrm{B}}(0)=\Theta_{\zz,\mathrm{Q}_\mathrm{B}}( z, 0)/\mathrm{Z}_{\zz,\mathrm{Q}^{-1}_\mathrm{B}}(0)$ localizes to a Gaussian distribution which tends to a delta function as $|\s|\rightarrow0$; and ii) it is consistent with the finite-dimensional definition.

It is not surprising that the Dirac integrator behaves as one expects under a linear map $\mathrm{L}:Z_0\rightarrow Z_0$ by $z\mapsto \mathrm{L}z$ with $\mathrm{Det}\, \mathrm{L}'\neq0$ \emph{(\cite[pg. 387]{LA2})};
\begin{equation}\label{delta integrator}
\int_{Z_0}F(z)\mathcal{D}_\lambda\delta_{\bar{z}}(\mathrm{L}(z))
=\sum_{z_0}\left(\mathrm{Det}_\lambda \mathrm{L}'_{z_0}\right)^{-1}F(z_0)
\end{equation}
given that $(\mathrm{L}(z_0)-\bar{z})=0$, and $\mathrm{Det}_\lambda \mathrm{L}'_{z_0}\neq0$ for all $z_0\in\mathrm{Ker}(\mathrm{L}-\bar{z}\mathrm{Id})$.

On the other hand, for $|\s|\rightarrow\infty$,
\begin{eqnarray}
\lim_{|\s|\rightarrow\infty}\int_{Z_0}\Theta_{\zz,\mathrm{Q}_\mathrm{B}}( z, z')\mathcal{D}_\lambda z
&=:&\int_{Z_0}e^{2\pi i \langle  z',z\rangle}\mathcal{D}_\lambda z\notag\\
&=&\lim_{|\s|\rightarrow\infty}\,\mathrm{Det}_\lambda (\s{\mathrm{W}_\mathrm{B}})^{1/2}e^{-\pi\s \mathrm{W}_\mathrm{B}( z')}\notag\\
&=&\lim_{|\tilde{\s}|\rightarrow0}\,
\mathrm{Det}_\lambda ({\mathrm{Q}^{-1}_\mathrm{B}}/\tilde{\s})^{1/2}e^{-(\pi/\tilde{\s}) \mathrm{Q}^{-1}_\mathrm{B}( z')}\notag\\
&=:&\delta(z')\;.
\end{eqnarray}
Again, the definition makes sense for the same reasons --- it is essentially steepest descent. But notice the mismatch in normalization between the two cases.
\end{remark}

This subsection has presented the construction of Gaussian integrators in terms of continuous paths in order to introduce the concepts in a familiar context. But from Definition \ref{int-def} it is clear that our family of Gaussian integrators can be constructed on any abelian group. So henceforth $Z_0$ will refer to the underlying abelian group of some topological Banach space.

\subsection{Skew-Gaussian integrators}\label{Skew-Gaussian integrators}
Use the Hermitian form $\mathrm{Id}_\mathrm{B}$ to define an inner product on some topological Banach space $Z_0$
\begin{equation}
\mathrm{Id}_\mathrm{B}(z_1,z_2)=-\tfrac{1}{2}\langle Id\, z_1, z_2\rangle=:( z_1| z_2)_{Z_0}\;.
\end{equation}
 Employ this inner product and the complex structure $\mathrm{J}$ on $Z_0$ to define the associated map $J:Z_0\rightarrow Z'_0$ by $-\frac{1}{2}\langle J z_1, z_2\rangle:=(\mathrm{J} z_1| z_2)_{Z_0}$. Similarly, the skew form $\Omega_{\mathrm{B}}$ defines an associated structure $\Omega$ on $Z_0$ by
 \begin{equation}
\Omega_{\mathrm{B}}(z_1,z_2)=-\tfrac{1}{2}\langle \mathit{\Omega} z_1, z_2\rangle=:(\Omega z_1| z_2)_{Z_0}\;.
\end{equation}

 Skew-Gaussian integrators are constructed from $\Omega_\mathrm{B}$ provided that the form is $\mathrm{J}$-skew, that is
 \begin{equation}
\Omega\mathrm{J}+\mathrm{J}^\dag{\Omega}=[\Omega,\mathrm{J}]=0\;.
\end{equation}
To emphasize the skew nature of $\Omega_\mathrm{B}$, we will change notation $z\rightarrow\eta$ in this subsection but stress that $\eta$ is \emph{not} Grassmann: The notation $\eta$ is only meant to remind of the underlying skew symmetry. That being said, the pair $(\eta,\mathcal{D}\eta)$ models $1$-forms in the skew-Gaussian context in contrast to the model of vectors by $(z,\mathcal{D}z)$ in the Gaussian context.

\begin{definition}
 Given a $\mathrm{J}$-skew symplectic form $\Omega$ on $Z_{\hat{\eta}_{cr}}$, a family of skew-Gaussian integrators
$\mathcal{D}\omega_{\bar{\eta},\Omega_\mathrm{B}}(\eta)$ is
characterized by
\begin{eqnarray}\label{Gaussian definition}
&&\Theta_{\bar{\eta},\Omega_\mathrm{B}}( \eta,\eta')=e^{2\pi i \langle  \eta',\eta\rangle -\pi\s \,\Omega_\mathrm{B}( \eta-\bar{\eta})}\notag\\
&&\mathrm{Z}_{\bar{\eta},\mathrm{M}_\mathrm{B}}( \eta')=e^{2\pi i \langle  \eta',\bar{\eta}\rangle}\,\mathrm{Pf}_\lambda(\s\,{\mathrm{M}_\mathrm{B}})^{-1}
e^{-(\pi/\s) \mathrm{M}_\mathrm{B}( \eta')}
\end{eqnarray}
where

\begin{equation}
\Omega_{\mathrm{B}}(\eta):=i\Omega_{\mathrm{B}}(\eta,\eta)
=i(\Omega\eta|\eta)_{Z_0}=(\Omega\eta|\mathrm{J}\eta)_{Z_0}
\end{equation}
and
\begin{equation}
\mathrm{M}_\mathrm{B}(\eta'):=i( \mathrm{M}\eta'|\eta')_{Z'_0}\;.
\end{equation}
Parallel to the Gaussian case, $\mathrm{M}_\mathrm{B}$ and $\Omega_\mathrm{B}$ are inverses on $Z_0\backslash\mathrm{Ker}(\OO)$; $\OO{M}:=\mathrm{Id}_{ Z'_{\hat{\eta}_{cr}}}$ and ${M}\OO:=\mathrm{Id}_{Z_{\hat{\eta}_{cr}}}$.\footnote{By definition, $\Omega$ is non-degenerate so $\OO$ is invertible on $Z_0$.} The restriction $(\s\,\Omega_\mathrm{B}(\eta)|_{G_\lambda})\in\C_+$ ensures integrability, and the functional Pfaffian is assumed well defined so that $\mathrm{Pf}_\lambda(\cdot)=\mathrm{pf}(\cdot)$  for a given $\lambda$ (up to a possible phase).

This integrator family is defined in terms of the primitive
skew-Gaussian integrator family $\mathcal{D}_\Lambda\eta$ according to
\begin{equation}
\mathcal{D}_\lambda\omega_{\bar{\eta},\Omega_\mathrm{B}}(\eta)
:=e^{-\pi\s \Omega_\mathrm{B}( \eta-\bar{\eta})}\mathcal{D}_\lambda\eta
\end{equation}
where $\mathcal{D}_\lambda\eta$ is characterized by
\begin{eqnarray}
&&\Theta(\eta,\eta')=e^{2\pi i \langle
\eta',\eta\rangle-\pi \s\mathrm{Id}_\mathrm{B}(\eta)}\notag\\
&&\mathrm{Z}(\eta')=\mathrm{Pf}_\lambda(\mathrm{Id}_\mathrm{B}/\s)\,e^{(-\pi /\s) \;\mathrm{Id}_\mathrm{B}(\eta')}\;.
\end{eqnarray}
Here $\mathrm{Id}_\mathrm{B}(\eta):=i(\mathrm{J}\eta|\eta)_{Z_0}$ and $\mathrm{Id}_\mathrm{B}(\eta'):=i(\mathrm{J}'\eta'|\eta')_{Z'_0}$. Then
\begin{equation}
 \int_{Z_0}e^{2\pi i \langle
\eta',\eta\rangle}\mathcal{D}_\lambda\omega_{\bar{\eta},\Omega_\mathrm{B}}(\eta)
=\sum_{\{\eta_{cr}\}}\!\!\!\!\!\!\!\!\!\int\;\int_{Z_{\hat{\eta}_{cr}}}
\Theta_{\bar{\eta},\Omega_\mathrm{B}}( \eta, \eta')\;\mathcal{D}_\lambda \eta
:=\mathrm{Z}_{\bar{\eta},\mathrm{M}_\mathrm{B}}( \eta')\;.
\end{equation}
\end{definition}

Because $\Omega_{\mathrm{B}}$ is skew-symplectic on $Z_0\backslash\mathrm{Ker}(\OO)$, the skew-Gaussian functional integral $\sum_{\{\eta_{cr}\}}\int_{Z_{\hat{\eta}_{cr}}}\mathcal{D}_\lambda\omega_{\bar{\eta},\Omega_\mathrm{B}}(\eta)
\sim\mathrm{Pf}_\lambda(\Omega/\s)$  can be heuristically interpreted as a sum/integral over the set $\{\eta_{cr}\}$ of symplectomorphism-invariant \emph{top-forms} over $Z_0$ viewed as the phase space of an infinite dimensional dynamical system. From a mathematics perspective, one could rightly claim this characterization is merely a change of notation from the Berezin Gaussian integral over Grassmann variables. But, physically, it transfers the duty of non-commuting dynamical variables to the skew-Gaussian integrators; which turns out to be a useful shift in perspective.

In contrast to the Gaussian case, the \emph{opposite}\footnote{Compare the scaling parameter $\mathrm{s}$ between the Gaussian and skew-Gaussian definitions: The rationale for them being inverse to each other will be explained at the end of this subsection.} extreme values of $\s$ lead to
\begin{eqnarray}
\lim_{|\s|\rightarrow\infty}\frac{1}{\mathrm{Z}_{\bar{\eta},\mathrm{M}_\mathrm{B}}(0)}
\int_{Z_0}\Theta_{\bar{\eta},\Omega_\mathrm{B}}( \eta,\eta')\mathcal{D}_\lambda \eta
&=:&\int_{Z_0}e^{2\pi i \langle  \eta', \eta\rangle}\delta(\eta-\bar{\eta})\mathcal{D}_\lambda \eta\notag\\
&=&\lim_{|\s|\rightarrow\infty}e^{2\pi i \langle  \eta',\bar{\eta}\rangle}e^{-(\pi/\s) \mathrm{M}_\mathrm{B}( \eta')}\notag\\
&=&e^{2\pi i \langle  \eta',\bar{\eta}\rangle}\;,
\end{eqnarray}
and
\begin{eqnarray}\label{symplectic delta functional}
\lim_{|\s|\rightarrow0}\int_{Z_0}\Theta_{\bar{\eta},\Omega_\mathrm{B}}( \eta,\eta')\mathcal{D}_\lambda \eta
&=&\int_{Z_0}e^{2\pi i \langle  \eta',\eta\rangle}\mathcal{D}_\lambda \eta\notag\\
&=&\lim_{|\s|\rightarrow0}
\,\mathrm{Pf}_\lambda({\omega_\mathrm{B}}/\s)e^{-(\pi/\s) \Omega_\mathrm{B}( \eta')}\notag\\
&=:&\delta(\eta')\;.
\end{eqnarray}

Skew-Gaussian integrators enable functional integration on the Banach space of forms as exemplified by the following.

\begin{application}\emph{Mathai-Quillen Thom class representative:}

Suppose $Z_0$ arises as the complexified vector space of forms $\Lambda ^\ast(\M)$ on some given manifold $\M$ and $\OO={\mathrm{D}'} J\mathrm{D}^\dag$ where $\mathrm{D}:Z_0\rightarrow Z_0$ is a linear operator. Its transpose is determined by $\langle{\mathrm{D}'}\eta',\eta\rangle:=\langle \eta',{\mathrm{D}}\eta\rangle$. (Below we will not distinguish between $\mathrm{D}'(or\, \mathrm{C}')$ and $\mathrm{D}(or\, \mathrm{C})$ unless to avoid confusion). Typically, $J\mathrm{D}$ is a covariant exterior differential associated with a connection on a principal fiber bundle of interest or an equivariant exterior differential.

Recall ${{J}}:Z_0\rightarrow Z'_0$ is canonically associated with the complex structure $\mathrm{J}$ such that ${J'}J=-\mathrm{Id}_{Z_0}$ and $J^\dag=-J$. Now, on $Z_0\backslash\mathrm{Ker}(\OO)$, $\mathrm{M}={\Omega}^{-1}$ implies $M= {\mathrm{C}}^\dag {J}^\dag{\mathrm{C}}$ and
\begin{eqnarray}\label{psi'}
\mathrm{M}_\mathrm{B}(\eta'_1,\eta'_2)
 &=&-\frac{1}{2}\left\{\langle \eta'_1,{{\mathrm{C}}^\dag} {J}'^\dag{\mathrm{C}}\eta'_2\rangle-\langle {\eta}'_1,{{\mathrm{C}}^\dag} {{J}'}{\mathrm{C}}\eta'_2\rangle \right\}\notag\\
 &=&\langle {{\mathrm{C}}}{\eta'}_1,{J}'{{\mathrm{C}}}\eta'_2\rangle\notag\\
 &=:&{\mathrm{M}}_\mathrm{B}(\psi'_1,\psi'_2)
 \end{eqnarray}
where we have defined $\psi':={{\mathrm{C}}}\eta'$ with ${\mathrm{D}}{\mathrm{C}}:=\mathrm{Id}_{Z'_{\hat{\eta}_{cr}}}$.\footnote{As $ z=\bar{ z}+Cz'$ for all $ z'\in Z'_{\hat{\eta}_{cr}}$, one can view $\psi'$ as a fluctuation $\eta-\bar{\eta}=C\eta'=J'\mathrm{C}\eta'=J'\psi'$, or alternatively as $\psi'\in H^1(Z'_0)$.} Restricting to $Z'_{\hat{\eta}_{cr}}$ yields a Hermitian inner product on $Z'_{\hat{\eta}_{cr}}\times Z'_{\hat{\eta}_{cr}}$ given by
 \begin{equation}
{\mathrm{M}}_\mathrm{B}(\psi')|_{Z'_{\hat{\eta}_{cr}}}
:=i{\mathrm{M}}_\mathrm{B}(\psi',\psi')|_{Z'_{\hat{\eta}_{cr}}} =i(\mathrm{J}'\psi'|\psi')_{Z'_{\hat{\eta}_{cr}}}=:|\psi'|^2\;.
\end{equation}

The corresponding skew-Gaussian functional integral encodes a useful Fourier duality between $\eta$ and
$\psi'$ that generates a Thom class representative;
\begin{equation}\label{duality}
\int_{Z_{\hat{\eta}_{cr}}}e^{2\pi i \langle
{\mathrm{D}}\psi',\eta\rangle-\pi\s\Omega_\mathrm{B}(\eta)}
\,\mathcal{D}_\lambda\eta =\mathrm{Pf}_\lambda(\Omega_\mathrm{B}/\s )e^{-(\pi /\s)|\psi'|^2}\;.
\end{equation}
Consider the action functional contained in \emph{(\ref{duality})},
\begin{equation}
\mathrm{S}(\eta,\psi'):=2\pi i \langle{\mathrm{D}}\psi',\eta\rangle-\pi\s \Omega_\mathrm{B}(\eta)+(\pi/\s)
|\psi'|^2\;.
\end{equation}
Using $\delta\mathrm{D}=\mathrm{D}\delta$ and the skew symmetry of $\OO$, it follows that $\delta \mathrm{S}(\eta,\psi')=0$ for the variations
\begin{eqnarray}
\delta\eta&=&\mathrm{J}'\psi'\notag\\
\delta\psi'&=&\s J'{\mathrm{D}}\psi'\;.
\end{eqnarray}
 Notice $\delta\eta\in Z'_{{\hat{\eta}_{cr}}}$ while $\delta\psi'\in Z_{{\hat{\eta}_{cr}}}$. This SUSY-like symmetry owes its existence to the J-skew symmetry of $\Omega_\mathrm{B}$, the $\mathbb{Z}_2$ grading induced by the complex structure on $Z_0$, and the duality between $Z_0$ and $Z_0'$. They induce morphisms $(\mathit{\Omega},\mathit{\Omega}'):Z_0\times Z'_0\rightarrow Z_0\times Z'_0$ and $(J,J'):Z_0\times Z'_0\rightarrow Z_0\times Z'_0$ satisfying $[\mathit{\Omega},J]=0$ and $[\mathit{\Omega}',J']=0$  that will feature prominently later.

 To verify that $\int_{Z_{\hat{\eta}_{cr}}}\exp[\mathrm{S}(\eta,\psi')]\mathcal{D}\eta$ is indeed a Thom class representative when $J{\mathrm{D}}$ is a covariant derivative, note that: i) it is clearly horizontal and invariant under the rotational symmetry of $(\cdot \,|\cdot\,)_{Z'_{\hat{z}_{cr}}}$, ii) the action functional is $\delta$-closed, but introducing an auxiliary map that is Gaussian-dual to $\psi'$ in the usual way renders it $\delta$-exact,\footnote{This boils down to re-writing $e^{-(\pi /\s)|\psi'|^2}$ as a Gaussian functional integral for the dual pair $\langle \psi',B\rangle$ over the dual space $Z_{{\hat{\eta}_{cr}}}$.} and iii) normalization is enforced by \emph{(\ref{symplectic normalization})} below.

The topological and cohomological aspects of \emph{(\ref{duality})}  are well studied and understood --- usually in the context of supersymmetric QM/QFT and Berezin functional integrals \emph{(see e.g. \cite{BT, SZ})}. It is noteworthy that the complexified vector space of $p$-forms and the associated skew-Gaussian integrators are also germane in this context.\footnote{Of course this is no accident. A Clifford algebra obtains if one defines a symmetric product  on $Z_0$ by polarization of $\Omega_{\mathrm{B}}(\eta)$ according to $\eta_1\eta_2+\eta_2\eta_1:=\Omega_{\mathrm{B}}(\eta_1+\eta_2)
-\Omega_{\mathrm{B}}(\eta_1)
-\Omega_{\mathrm{B}}(\eta_2)$. This leads directly to the Grassmann/Berezin picture for any Lagrangian (or isotropic) subspace of ${Z}_0$ since then $\Omega_{\mathrm{B}}(\cdot)$ vanishes on such a subspace. In other words, one can view anticommuting degrees of freedom as elements of the algebra induced from restricting a symplectic form to a Lagrangian subspace modulo symplectomorphisms (which, in particular, records dynamics). Also, note that variation of $\mathrm{S}(\eta,\psi')$ with respect to $\eta$ yields $\mathrm{D}\psi'=-i\s\mathrm{D}^\dag J\mathrm{D}\eta$ so on shell $\{\delta^\dag,\delta\}\eta=2\s^2\mathit{J'\Omega}\eta$. Consequently $\delta$ certainly looks like a supersymmetry, but in general $\psi'\in Z'_{{\hat{\eta}_{cr}}}$ and $\eta\in Z_0$ are not exactly topological duals so we call $\delta$ a `SUSY-like' symmetry.}

In order to illustrate how all this works in detail, we follow a standard example and use it to express the Euler characteristic of a smooth compact real Riemannian manifold  $\M$. Its based loop space $L_{\m_a}\M$  consists of smooth loops $\gamma:\mathbb{T}\rightarrow\M$ with $\gamma(\ti_a)=\gamma(\ti_b)=\m_a$ and $\m_a\in\M$. It is convenient to parametrize $(L_{\m_a}\M)^\C$ by the Banach space $T_\gamma(L_{\m_a}\M)^\C$ via the exponential map $\mathrm{Exp}:T_\gamma(L_{\m_a}\M)^\C\rightarrow(L_{\m_a}\M)^\C$. \emph{(see appx. \ref{approaches})}

Let $Z_\gamma$ be the underlying abelian group of $T_\gamma(L_{\m_a}\M)^\C$. Since $Z_\gamma\ni\eta_\gamma:T\mathbb{T}\rightarrow T_\gamma\M^\C$, the parametrizing space $Z_\gamma$ is isomorphic to the Banach space of sections $\Gamma(\gamma^\ast T\M^\C)$. It follows that the domain of the skew-Gaussian functional integral is the underlying abelian group of the Banach space of sections $\Gamma(\gamma^\ast T\M^\C)$  such that $\eta_\gamma(\ti)\in  T_{\gamma(\ti)}\M^\C$.

The complex structure $\mathrm{J}$ on $T\M^\C$ pulls back to a complex structure on $T_\gamma\M^\C$ allowing the decomposition $T_{\gamma(\ti)}\M^\C=T_{\gamma(\ti)}\M^+\oplus T_{\gamma(\ti)}\M^-$ for all $\ti\in\mathbb{T}$. This in turn induces  the decomposition of $Z_\gamma=Z^+_\gamma\oplus Z^-_\gamma$ into holomorphic and anti-holomorphic vector-valued $1$-forms which allows to fix boundary conditions such that $\eta_\gamma(\ti_a)=\eta_\gamma(\ti_b)=(\varsigma_\gamma^+,0)$.

 Choose the linear first-order operator ${\mathrm{D}}=\nabla_{\ti}$ where $\nabla$ is the Levi-Civita connection on the complex frame bundle $F\M^\C$ and $\nabla_\ti$ denotes the pull-back of $\nabla$ by $\gamma$. The action of $\nabla$ on a holomorphic frame $\{\mathrm{e}_\alpha\}$ in $T\M^\C$ is given by $\nabla \mathrm{e}_\alpha=\mathrm{e}_\beta\otimes\omega^\beta_{\;\;\alpha}$ where $\omega^\beta_{\;\;\alpha}$ are the local connection $(1,0)$-forms on $U_i\subset \M$. Likewise,
 \begin{equation}
 J\nabla^\dag\mathrm{e}^\alpha=J\mathrm{e}^\beta\otimes\omega_{\;\;\beta}^{\alpha}
 =i\mathrm{e}_\beta\otimes\omega^{\alpha\beta}\;.
\end{equation}
This implies (with $\mathrm{R}^{\alpha\beta}$ the curvature $(1,1)$-forms)
\begin{equation}
\nabla J\nabla^\dag\mathrm{e}^\alpha=i\mathrm{e}_\beta\otimes \mathrm{R}^{\alpha\beta}\;.
\end{equation}

 Hence, in a local trivialization with
 \begin{equation}
\nabla\eta_\gamma(\ti)=(\gamma(\ti),\Bold{v}_{\gamma(\ti)}(\ti))
\in T_{\gamma(\ti)}\M^+\otimes T^\ast_{\gamma(\ti)}\M^+\;,
\end{equation}
the symplectic form is represented by the matrix of the components of curvature $(1,1)$-forms; that is
$\Omega_\mathrm{B}(\eta)={v}^\ast_\alpha \mathrm{R}^{\alpha\beta}v_\beta$ (where we have suppressed the $1$-form indices).
The explicit action functional in this case is
\begin{equation}
\int_{\mathbb{T}}\left[2\pi i\,v^\ast_\alpha\nabla_\ti\psi'^\alpha
-\pi\s \,{{v}^\ast_\alpha} \mathrm{R}^{\alpha\beta}v_\beta
+\pi/\s(\psi'|\psi')\right]_{\gamma(\ti)}\,d\ti\;.
\end{equation}

Now, since
\begin{equation}
 D\eta_{cr}=J\mathrm{D}{\eta_{cr}}=\nabla{\eta_{cr}}=0
\end{equation}
 by definition, the boundary conditions on $\eta_\gamma$ imply
  \begin{equation}
{\eta_{cr}}_\gamma(\ti)=(\m_a,0)\;\;\forall\ti\in\mathbb{T}\;.
\end{equation}
That is, $(\nabla J^\dag \nabla^\dag){\eta_{cr}}_\gamma=0$ implies that $[{\eta_{cr}}_\gamma(\ti),\eta_\gamma(\ti)]$  is horizontal for all $\eta_\gamma(\ti)$ (with $[\cdot,\cdot]$ the Lie bracket) and therefore
\begin{equation}
{\Bold{v}_{cr}}_{\gamma(\ti)}(\ti)=0 \;\;\forall \ti\in\mathbb{T}\;.
\end{equation}
Intuitively, ${\eta_{cr}}_\gamma$ is just the trivial zero mode. This holds for every based loop, and so the set $\{\eta_{cr}:J\mathrm{D}\eta_{cr}=0\}$ coincides with the zero section $\Gamma_0(T\M^+)\cong\M$. It follows that loop space $L\M$ can be parametrized by the disjoint union $X_{a}:=\bigsqcup_{\m_a}Z^+_\gamma$.

Finally, fix $\s=1$ and momentarily restrict to a single $Z^+_\gamma$. Thereby obtain the Euler class representative \`{a} la Mathai-Quillen
\begin{eqnarray}\label{Mathai-Quillen}
\int_{Z^+_\gamma}e^{2\pi i \langle{\mathrm{D}}\psi',\eta\rangle-\pi \Omega_\mathrm{B}(\eta)+\pi
(\psi'|\psi')}
\,\mathcal{D}_\lambda\eta
=:\int_{Z^+_\gamma}e^{-\pi\mathrm{F}_{\mathrm{B}}(\eta,\psi')}
\,\mathcal{D}_\lambda\eta
 =\mathrm{Pf}_{\psi',\lambda}
(\Omega_\mathrm{B})
\end{eqnarray}
where a normalization/regularization for $\mathrm{Pf}_\lambda$ must be fixed for a given $\psi'$ by suitable choice of $\lambda$ and $\mathrm{F}_{\mathrm{B}}:Z^+_\gamma\times {Z^+_\gamma}'\rightarrow\C$ by (using $\langle{\mathrm{D}}'\psi',\eta\rangle=\langle{\psi',\mathrm{D}}\eta\rangle$)
\begin{equation}\label{supersymmetric form}
\mathrm{F}_{\mathrm{B}}(\eta,\psi')=\big(\eta,\psi'\big)
\left(
  \begin{array}{cc}
\Omega\mathrm{J} & J'\mathrm{D}' \\
    J\mathrm{D} & -\mathrm{Id} \\
  \end{array}
\right)\left(
         \begin{array}{c}
           \eta \\
           \psi' \\
         \end{array}
       \right)
\;.
\end{equation}

In particular, for $\psi'=\mathrm{C}\eta'=0$ so that $\langle \mathrm{D}\psi',\eta\rangle=0$, the set $\{\eta_{cr}:J\mathrm{D}\eta_{cr}=0\}$ yields an integration over the full space $X_a$; reproducing the well-known functional integral representation of the Euler characteristic
\begin{equation}
\int_{X_a}e^{-\pi
\Omega_\mathrm{B}(\eta)}
\,\mathcal{D}_\lambda\eta
=\int\!\!\!\!\!\!\!\!\sum_{\{\eta_{cr}\}}\int_{Z^+_\gamma}e^{-\pi
\Omega_\mathrm{B}(\eta)}
\,\mathcal{D}_\lambda\eta =\int_\M\mathrm{pf}(\mathrm{R})=\chi(\M)\;.
\end{equation}
Alternatively, re-instate the parameter $\s$ and take the limit $|\s|\rightarrow0$ along the imaginary axis. In this case, if $\psi'$ is generic, the left-hand side of \emph{(\ref{Mathai-Quillen})} localizes onto the zero-locus of $\psi'$ according to \emph{(\ref{symplectic delta functional})} and \emph{(\ref{psi'})}.\footnote{The delta functional $\delta(\psi')$ represents a complex quantity whose phase at an isolated zero $\m_0$ in this case is normalized/taken to be $i^{\mathrm{Ind}(\psi',\m_0)}$.}

Note that we can arrange to have $d Z_{\bar{\eta},\mathrm{M}_\mathrm{B}}( \eta')/d\s=0$ simply by scaling $\eta'\rightarrow \s^{1/2}\eta'$ and, to maintain the topological duality, scaling  $\eta\rightarrow \s^{-1/2}\eta$. Since $Z_{\bar{\eta},\mathrm{M}_\mathrm{B}}( \eta')$ doesn't depend on $\s$, the two cases,  \emph{($\s=1$ and $\psi'=0$)} versus \emph{($|\s|\rightarrow 0$ and $\psi'$ generic)}, yield the well-known equivalence between the Gauss-Bonnet versus Poincar\'{e}-Hopf representations of the Euler characteristic.

To finish the story, recall \emph{(\ref{manifold integral})}. Then, by definition
\begin{eqnarray}
\int_{L\M}\mathrm{F}(m)\mathcal{D}_\lambda m
:=\int_{Z_a}e^{-\pi
\Omega_\mathrm{B}(\eta)}
\,\mathcal{D}_\lambda\eta
&=&\int_\M\int_{Z^+_\gamma}e^{-\pi\Omega_\mathrm{B}(\eta)}\,\mathcal{D}_\lambda\eta\notag\\
&=&\int_\M\mathrm{pf}
(\mathrm{R})=\chi(\M)
\end{eqnarray}
as long as $\mathrm{F}$ and $\mathcal{D}m$ are chosen to satisfy $\mathrm{F}(P(\eta))\,\mathcal{D}(P(\eta))=e^{-\pi
\Omega_\mathrm{B}(\eta)}
\,\mathcal{D}\eta$. In essence, the functional integral is realized by an entire family of integrals $\int_{Z^+_\gamma}e^{-\pi\Omega_\mathrm{B}(\eta)}\,\mathcal{D}_\lambda\eta:=\mathrm{pf}
(\mathrm{R})$ indexed by the points of $\M$ --- which demonstrates yet again the localization theme.
\end{application}

Before moving on, we digress for a moment to indicate why characterizing a skew-Gaussian
 integrator by $\mathrm{Pf}(\s\,\mathrm{M}_\mathrm{B})^{-1}\sim\mathrm{Pf}(\Omega_\mathrm{B}/\s)$ is justified --- after all, it seems a bit contrived to invert the Pfaffian.

 Suppose $Z_0$ is endowed with Hermitian form $\mathrm{Q}$ and complex structure $\mathrm{J}$. Let $\Omega$ be the associated K\"{a}hler form. From the topological duality, $\mathrm{Q}$, $\mathrm{J}$, and $\Omega$ extend to $Z_0\times Z'_0$ which is $\mathbb{Z}_2$-graded and decomposes as
  \begin{eqnarray}
  Z_0\times Z'_0=\bigoplus_\pm\left[\left(Z^\pm_0\times Z'^\pm_0\right)\oplus \left(Z^\pm_0\times Z'^\mp_0\right)\right]
  =:W^{\mathbf{e}}_0\oplus W^{\mathbf{o}}_0=:W_0\;.
\end{eqnarray}
 We can characterize an invariant  integrator $\mathcal{D}(w^{\mathbf{e}},w^{\mathbf{o}})$ on $Z_0\times Z'_0$ by
\begin{equation}
\int_{Z_0\times Z'_0}\Theta_{(\bar{w}^{\mathbf{e}},\bar{w}^{\mathbf{o}}),\mathrm{F}_\mathrm{B}}(w^{\mathbf{e}},w^{\mathbf{o}})\,\mathcal{D}_\lambda(w^{\mathbf{e}},w^{\mathbf{o}})
:=1
\end{equation}
with  $w^{\mathbf{e},\mathbf{o}}\in W^{\mathbf{e},\mathbf{o}}_0$, and a block-diagonal sesquilinear form on ${W}_0\times W_0$
\begin{equation}
\mathrm{F}_\mathrm{B}(w^{\mathbf{e}},w^{\mathbf{o}})=\frac{1}{\s}\mathrm{Q}(w^{\mathbf{e}}_1,w^{\mathbf{e}}_2)+\s\,\Omega(w^{\mathbf{o}}_1,w^{\mathbf{o}}_2)
\end{equation}
where $\Omega(w^{\mathbf{o}}_1,w^{\mathbf{o}}_2)=\mathrm{Q}(\mathrm{J}w^{\mathbf{o}}_1,w^{\mathbf{o}}_2)$ is skew-Hermitian.

As long as $w^{\mathbf{e}}$ and $w^{\mathbf{o}}$ are independent we have $\mathcal{D}(w^{\mathbf{e}},w^{\mathbf{o}})=\mathcal{D}w^{\mathbf{e}}\mathcal{D}w^{\mathbf{o}}$. Then the characterization of skew-Gaussian integrators follows from the prescribed normalization (and functional Fubini);
\begin{eqnarray}\label{symplectic normalization}
1\equiv\int_{Z_0\times Z'_0}\Theta(w^{\mathbf{e}},w^{\mathbf{o}})\,\mathcal{D}_\lambda(w^{\mathbf{e}},w^{\mathbf{o}})
&=&\int_{W^{\mathbf{o}}_0}\left[\int_{W^{\mathbf{e}}_0}\Theta(w^{\mathbf{e}},w^{\mathbf{o}})
\,\mathcal{D}_\lambda w^{\mathbf{e}}\right]\mathcal{D}_\lambda w^{\mathbf{o}}\notag\\
&=&\int_{W^{\mathbf{o}}_0}\mathrm{Det} ({\mathrm{Q}_\mathrm{B}}/\s)^{-1/2}\,e^{-\pi \s\, \Omega_\mathrm{B}( w^{\mathbf{o}}-\bar{w}^{\mathbf{o}})}\,\mathcal{D}_\lambda w^{\mathbf{o}}\;.
\end{eqnarray}
 As $Z_0$ and $Z_0'$ are dual symplectic vector spaces, the primitive integrator on $Z_0\times Z'_0$ should be invariant under a change of integration variable;  implying $\mathcal{D}_\lambda \widetilde{w}^{\mathbf{e}} \mathcal{D}_\lambda \widetilde{w}^{\mathbf{o}}= \mathcal{D}_\lambda {w^{\mathbf{e}}} \mathcal{D}_\lambda {w^{\mathbf{o}}}$. Now suppose the change of variable renders $\mathrm{Q}_\mathrm{B}\mapsto\mathrm{Id}_\mathrm{B}$, then $\mathcal{D}_\lambda \widetilde{w}^{\mathbf{e}}\sim\mathrm{Det}_\lambda ({\s\,\mathrm{W}_\mathrm{B}})^{1/2}\,\mathcal{D}_\lambda w^{\mathbf{e}}$.\footnote{For a general change of variable, say $\mathrm{Q}_\mathrm{B}\mapsto\widetilde{\mathrm{Q}}_\mathrm{B}$, the two integrators are related through the ratio $\mathrm{Det}_\lambda \left(\frac{\mathrm{Q}_\mathrm{B}}{\widetilde{\mathrm{Q}}_\mathrm{B}}/\s\right)^{1/2}
 :=\s^{-\mathrm{ran}(\widetilde{\mathrm{Q}}_\mathrm{B})/2
 -\mathrm{ind}(\widetilde{\mathrm{Q}}_\mathrm{B})
 +\mathrm{ind}(\mathrm{Q}_\mathrm{B})}
 \left|\left[\mathrm{det}({\mathrm{Q}_\mathrm{B}})^{1/2}\right]/ [\mathrm{det} ({\widetilde{\mathrm{Q}}_\mathrm{B}})^{1/2}]\right|$. The same conclusion follows.} Hence we require $\mathcal{D}_\lambda \widetilde{w}^{\mathbf{o}}\sim \mathrm{Pf}_\lambda (\s\,{\mathrm{M}_\mathrm{B}})^{-1}\,\mathcal{D}_\lambda w^{\mathbf{o}}$, and $\mathcal{D}_\lambda w^{\mathbf{o}}$ is a top-form density of weight $-1$.

 Physically, this construction can be interpreted to mean that Gaussian integrators represent \emph{correlations} among complex degrees of freedom. Such correlations are characterized by symmetric bilinear forms (hence by covariance in the statistical sense). The topological duals in this case are related through inverse symmetric bilinear forms, i.e. variances in the statistical sense. On the other hand, skew-Gaussian integrators represent \emph{dynamical relations}  among complex degrees of freedom. These are characterized by skew-symmetric bilinear forms (hence by conjugation in the dynamical sense). So it is natural to impose an inverse symplectic relationship between the corresponding duals, and therefore the condition $\mathrm{Pf}(\mathrm{M})=\mathrm{Pf}({\Omega}^{-1})$. Note this accommodates symplectomorphism invariance among Lagrangian subspaces in phase space; which is the hallmark of a Hamiltonian dynamical system. The two integrator types are dual (by design) in the sense that $\mathcal{D}w^{\mathbf{e}}$ and $\mathcal{D}w^{\mathbf{o}}$ transform inversely under a change of variable in congruity with covariance v.s contravariance in finite-dimensions. However, $\mathcal{D}w^{\mathbf{o}}$ depends on the underlying symplectic structure on $W_0$ while $\mathcal{D}w^{\mathbf{e}}$ depends on  both the symplectic \emph{and} complex structures.

 Reversing integration order allows to define a functional Liouville integrator
\begin{eqnarray}
\int_{W^{\mathbf{e}}_0}\left[\int_{W^{\mathbf{o}}_0}\Theta(w^{\mathbf{e}},w^{\mathbf{o}})\,\mathcal{D}_\lambda w^{\mathbf{o}}\right]\mathcal{D}_\lambda w^{\mathbf{e}}
&=&\int_{W^{\mathbf{e}}_0}\mathrm{Pf} ({\Omega_\mathrm{B}}/\s)e^{-(\pi/\s) \mathrm{Q}_\mathrm{B}( w^{\mathbf{e}}-\bar{w}^{\mathbf{e}})}\,\mathcal{D}_\lambda w^{\mathbf{e}}\notag\\
&=:&\int_{W^{\mathbf{e}}_0}\mathcal{D}_\lambda \ell_{\bar{w}^{\mathbf{e}},\,\Omega_\mathrm{B},\,\mathrm{Q}_\mathrm{B}}(w^{\mathbf{e}})\;.
\end{eqnarray}
Since $Z_0\times Z'_0$ models a supersymplectic manifold, this Liouville integrator clearly has application in the context of phase space path integrals if $w^{\mathbf{e}}$ is restricted to a Lagrangian subspace of $W_0^{\mathbf{e}}$.

\begin{remark}
 Skew-Gaussian integrators possess two key properties: i) they provide access to Pfaffian-type generating functions without invoking Berezin integration, and ii) like the Gaussian case, they allow to define a delta functional on $W_0^{\mathbf{o}}$ but in the {inverse} $|\s|$ scaling limit. In consequence, a delta functional localization cannot be achieved  on $Z_0\times Z'_0$ simultaneously using Gaussian and skew-Gaussian integrators since their respective delta functionals require inverse limits for $|\s|$. Meanwhile, the limit $|\s|\rightarrow 0$ does yield delta functional localization on $W_0^{\mathbf{e}}$ while $|\s|\rightarrow\infty $ localizes on $W_0^{\mathbf{o}}$.

These properties, along with the physical interpretation of the integrators discussed above, suggest that in the context of quantum physics one can interpret $Z_0$ and $Z'_0$ as describing two Fourier-dual vector spaces underlying a Bargmann-Fock space of {complex} degrees of freedom; while the decomposition $W_0^{\mathbf{e}}$ and $W_0^{\mathbf{o}}$ can be interpreted as describing a probabilistic  {and} a dynamical nature (respectively) simultaneously possessed by {complex} degrees of freedom. Mixing these attributes together on $W_0$ when $\mathrm{F}_\mathrm{B}(w^{\mathbf{e}},w^{\mathbf{o}})$ is supersymmetric and $\eta,\psi$ represent balanced degrees of freedom leads to supersymmetry --- but without the compulsion to introduce superpartners into the physical arena.\footnote{Clearly one could interpret the constituents of $W_0$ as superpartners. However, to the extent that $Z_0$ models already a symplectic manifold via its complex structure and the Liouville integrators characterize quantum systems, it seems only one side of the symmetric/skew-symmetric split contributes to the physical Hilbert space. It goes counter to experiment (at least up to now) to build a realistic physical model whose transition amplitudes include both sides at once.}
\end{remark}

Everything in this section can be promoted to non-abelian topological linear Lie groups. In this case, one would identify $Z_0$ and $Z_0'$ with the Lie algebra and dual Lie algebra respectively of the non-abelian group and use $\mathrm{Exp}$ or $\mathrm{Dev}$ (as described in appx. \ref{approaches}) to parametrize the group. The non-trivial Lie bracket would allow development of Lie derivatives, exterior and interior products, etc.
\section{Non-Gaussian integrators}
One of the main themes we continue to emphasize is that the proposed definition of functional integral gives rise to a perspective that spurs developing new types of functional integrals beyond the typical Gaussian case. This section highlights four non-Gaussian integrators.

\subsection{Gamma integrators}
\begin{definition}
Let $T_a$ be the space of continuous pointed maps
$\tau:(\mathbb{T},\ti_a)\rightarrow(\C^\times,1)$ where $\mathbb{T}=[\ti_a,\ti_b]\subset\R$ and $\C^\times:=\C\backslash\{0\}\cong\R_+\times \mathrm{S}^1$. $T_a$ is an abelian topological group under point-wise multiplication as $\C^\times$ is the multiplicative group of (non-zero) complex numbers. Let $\beta'$ be a fixed element in the dual $\T\equiv\mathrm{Hom}_C(\Ta,\C)$. A gamma family of integrators
$\mathcal{D}_\Lambda\gamma_{\alpha,\beta'}(\tau)$ on ${T_a}$ is characterized
by
\begin{eqnarray}
&&\Theta_{\alpha,\beta'}(\tau,\tau')= e^{ i\dual-\langle\beta',\tau\rangle}\,\tau^\alpha \notag\\
&&\mathrm{Z}_{\alpha,\beta'}(\tau')
=({\beta'}-i{\tau'})_\lambda^{-\alpha}
\end{eqnarray}
where $\alpha\in\C_+$ and $\tau^\alpha$ is defined point-wise by $\tau^\alpha(\ti):=e^{\alpha\log\tau(\ti)}$ with the principal value prescription for $\log\tau(\ti)$.

The gamma integrator family is defined in terms of the primitive
integrator $\mathcal{D}_\lambda\tau$ by
\begin{equation}\label{gamma}
\mathcal{D}_\lambda\gamma_{\alpha,\beta'}(\tau) :=
e^{-\langle\beta',\tau\rangle}\tau^{\alpha}\, \mathcal{D}_\lambda\tau
\end{equation}
where $\mathcal{D}_\lambda\tau$ is
characterized by
\begin{eqnarray}
&&\Theta_{0, Id'}(\tau,\tau')=\exp\{ i\dual-\langle Id',\tau\rangle\}\notag\\
&&\mathrm{Z}_{0, Id'}(\tau')=\Gamma_\lambda(0)\;,
\end{eqnarray}
and implicit in $\Gamma_\lambda(0)$ is a regularization, e.g. choose $\lambda$ so that $\Gamma_\lambda(0)\stackrel{\lambda}{\rightarrow}\lim_{\alpha\rightarrow0}\frac{1}{\Gamma(\alpha)}
\int\tau^\alpha\, \mathcal{D}_\lambda\tau$.
\end{definition}

 In applications, one often imposes a bound on $\tau(\ti)$; say $|\tau(\ti)|\leq|c|$ for all $\ti\in[\ti_a,\ti_b]$ and for some finite constant $c\in\C^\times$. The obvious tool to enforce this constraint is the functional analog of Heaviside; yielding a `cut-off' gamma family that generalizes the previous definition but reduces to it as the cutoff $|c|\rightarrow\infty$:

\begin{definition}
Let $\Ta$ be the space of continuous pointed maps
$\tau:(\mathbb{T},\ti_a)\rightarrow(\C^\times,1)$. Let $\beta'$ be a fixed element in the dual $\T$  and fix a fiducial $\tau_c\in\Ta$ such that $\langle\beta',\tau_c\rangle=c\in\C^\times$. A lower gamma family of integrators
$\mathcal{D}_\Lambda\gamma_{\alpha,\beta',c}(\tau)$ on $T_a$ is characterized
by
\begin{eqnarray}
&&\Theta_{\alpha,\beta'}(\tau,\tau')= e^{ i\dual-\langle\beta',\tau\rangle}\,\tau^\alpha \notag\\
&&\mathrm{Z}_{\alpha,\beta',c}(\tau')
=\gamma\left(\alpha,c\right)(\beta'-i{\tau'})_\lambda^{-\alpha}
\end{eqnarray}
where  $\gamma\left(\alpha,c\right)$ is
the lower incomplete gamma function given by
\begin{equation}
\gamma\left(\alpha,c\right)
=\Gamma(\alpha)e^{-c}\sum_{n=0}^\infty
\frac{(c)^{\alpha+n}}{\Gamma(\alpha+n+1)}\;.
\end{equation}
This renders a gamma-type functional integral;
\begin{equation}
\int_{T_a}e^{ i\dual}\;\mathcal{D}_\lambda\gamma_{\alpha,\beta',c}(\tau)
=\mathrm{Z}_{\alpha,\beta',c}(\tau')\;.
\end{equation}
An upper gamma family of integrators
$\mathcal{D}_\Lambda\Gamma_{\alpha,\beta',c}(\tau)$ is defined
similarly where
\begin{equation}
\Gamma\left(\alpha,c\right)
=\Gamma(\alpha)-\gamma\left(\alpha,c\right)
\end{equation}
is the upper incomplete gamma function.
\end{definition}

Using this notion, the fiducial gamma integrator $\mathcal{D}_\lambda\tau$ is
$\mathcal{D}_\lambda\gamma_{0, Id',\infty}(\tau)$. By definition it is
normalized according to
\begin{equation}
\frac{1}{\Gamma(0)}\int_{T_a}\mathcal{D}_\lambda\gamma_{0, Id',\infty}(\tau):=1
=:\frac{1}{\Gamma(0)}\int_{T_a}\mathcal{D}_\lambda\Gamma_{0, Id',0}(\tau)\;,
\end{equation}
but the other family members yield
\begin{equation}\label{gamma normalization}
\frac{1}{\Gamma(\alpha)}\int_{T_a}\mathcal{D}_\lambda\gamma_{\alpha,\beta',\infty}(\tau)
={\beta'}_\lambda^{-\alpha}
 =\frac{1}{\Gamma(\alpha)}\int_{T_a}\mathcal{D}_\lambda\Gamma_{\alpha,\beta',0}(\tau)\;.
\end{equation}

\subsubsection{Distributionals}
As we have stressed, an important aspect of the proposed scheme is localization in function spaces. The aim of this subsection is to develop tools to effect localization on the dual  $T_a'$.

As motivation, consider the lower gamma integrator and restrict to $\tau$ that is real-valued, i.e. $\tau:(\mathbb{T},\ti_a)\rightarrow(\R^\times,1)$. Put
$\tau'=0$ and $\alpha=1$, and implement the  localization by $\lambda_{\R^\times}:\Ta\rightarrow\R^\times$. Choose $\beta'$ such that $\langle\beta',\tau\rangle=\langle\omega\delta_{\ti},\tau\rangle={\omega t}$ with $t\in\R^\times$ and $\omega\in i\R$. Then,
\begin{equation}
\int_{\Ta}\mathcal{D}_\lambda\gamma_{1,\beta'}(\tau)\stackrel{\lambda_{\R^\times}}{\longrightarrow}
\int_{\R^\times}e^{-\langle\beta',\tau(t)\rangle}\tau(t)\,d\log\tau(t)
=\int_{\R^\times}e^{-{\omega}t}\;dt=2\pi\delta(|\omega|)
\end{equation}
with the integral over $\R^\times$ understood as a two-sided Laplace transform (or an inverse Laplace transform after $\omega t\rightarrow i|\omega|t$). On the other hand,
\begin{equation}
\int_{T_a} \mathcal{D}_\lambda\gamma_{1, \beta',c}(\tau)
:=\gamma(1,c) {\beta'}^{-1}_\lambda
=(1-e^{- c}) {\beta'}^{-1}_\lambda\;,
\end{equation}
and so the integrator $\mathcal{D}_\lambda\gamma_{1,\beta'}(\tau)\equiv\mathcal{D}_\lambda{\gamma}_{1,\beta',\infty}(\tau)$ can be understood as a limit;
\begin{equation}
\int_{T_a}\mathcal{D}_\lambda{\gamma}_{1,\beta',\infty}( \tau): =\lim_{
|c|\rightarrow\infty}
\int_{T_a}\mathcal{D}_\lambda{\gamma}_{1,\beta',c}( \tau)\;.
\end{equation}

Consequently, when $c$ is strictly imaginary,
$\mathcal{D}_\lambda\gamma_{1,\beta',\infty}(\tau)$ can be interpreted as
the functional analog of a two-sided (or inverse) Laplace transform implying
\begin{equation}
\int_{T_a} \mathcal{D}_\lambda\gamma_{1, \beta',\infty}(\tau) =\lim_{
c\rightarrow \pm i\infty} (1-e^{- c}){\beta'}^{-1}_\lambda\;;
\end{equation}
which formally vanishes on average unless ${\beta'}^{-1}_\lambda$ diverges.\footnote{To be precise, we mean that $\langle{\beta'}^{-1}_\lambda,\tau|_{G_\lambda}\rangle
\rightarrow\infty$ for all $\tau\in T_a$ with $\lambda:T_a\rightarrow G_\lambda$.} This can be interpreted as the functional analog of a delta function as our motivation in the previous paragraph suggested. In particular, this integrator can be used to localize onto the kernel of $\beta'$.

Conversely, if $c$ is strictly positive-real,  then
\begin{equation}
\int_{T_a} \mathcal{D}_\lambda\gamma_{1, \beta',\infty}(\tau)
:=\lim_{c\rightarrow\infty}
\gamma(1,c){\beta'}^{-1}_\lambda
=\lim_{c\rightarrow\infty} (1-e^{- c}){\beta'}^{-1}_\lambda\;,
\end{equation}
which we interpret as a principal value.

These observations suggest the definition:
\begin{definition}\label{def 4.4}
Suppose $\langle \beta',\tau\rangle\in i\R$ and ${\beta'}^{-1}_\lambda$ diverges. A delta functional on $T_a'$ is defined by
\begin{equation}
\delta_\lambda(\beta')
:=\frac{1}{\Gamma(1)}\int_{T_a}\mathcal{D}_\lambda{\gamma}_{1,\beta'}(\tau)\;.
\end{equation}
If instead $\langle \beta',\tau\rangle\in\R_+$\;,
\begin{equation}
\mathrm{Pv}_\lambda(\beta'^{-1})
:=\frac{1}{\Gamma(1)}\int_{T_a}\mathcal{D}_\lambda{\gamma}_{1,\beta'}(\tau)\;.
\end{equation}
\end{definition}
Together these indicate the heuristic
\begin{equation}
 {\beta'^{-1}}=\mathrm{Pv}(\beta'^{-1})+i\delta(\beta')
\end{equation}
when $\langle \beta',\tau\rangle\in \C_+\cong \R_+\times i\R$. That is, given a choice of $\lambda$, we have
\begin{eqnarray}
\beta'^{-1}:T_a&\rightarrow&\C\notag\\
\widetilde{\tau}&\mapsto&\langle\int_{T_a}\mathcal{D}_\lambda{\gamma}_{1,\beta'}(\tau),\widetilde{\tau}\rangle\notag\\
&=:&\langle\mathrm{Pv}_\lambda(\beta'^{-1}),\widetilde{\tau}\rangle
+i\langle\delta_\lambda(\beta'),\widetilde{\tau}\rangle\;.
\end{eqnarray}

Remark that Definition \ref{def 4.4} suggests the
characterization
\begin{equation}
\delta_\lambda^{(\alpha-1)'}(\beta')
=\frac{i^{\alpha-1}}{\Gamma(\alpha)}\int_{T_a}\mathcal{D}_\lambda{\gamma}_{\alpha,\beta'}(\tau)
\end{equation}
when $\langle \beta',\tau\rangle\in i\R$ and ${\beta'}^{-1}_\lambda$ diverges.
 The characterization is ``good" in the sense that $\delta_\lambda(\beta')$ reduces to the usual Dirac delta function under $\lambda:T_a\rightarrow (i\mathbb{R}^\times)^n$ for any $n$ where $\langle\beta',\tau\rangle=\Bold{\omega}\cdot\Bold{t}$ with $\Bold{\omega}\in i\R^n$ and $\Bold{t}\in(\mathbb{R}^\times)^n$. So for $\alpha=m\geq1$ with $m\in\N$ we have
 \begin{equation}
 \delta_\lambda^{(m-1)'}(\beta')=i^{m-1}\frac{\Gamma(m-1)}{\Gamma(m)}
 \int_{T_a}\frac{\delta^m}{\delta \beta'^m}\mathcal{D}{\gamma}_{0,\beta'}(\tau)\;.
 \end{equation}
It appears that gamma integrators and their associated functional
integrals might be used as a basis to build up a theory of what might be
called `distributionals', but of course much work is required to develop and verify such a concept.

\begin{remark}
Delta functionals  defined in terms of gamma-type integrators are important for imposing constraints that lead to certain types of localization. Notice that they can be interpreted as the functional analog of the inverse Laplace transform of the identity map, and the duality allows them to be transferred to $T_a$. On the other hand, as remarked in the previous section, the notion of delta functionals can also be formulated using Gaussian-type integrators \emph{\cite{LA4}}. Again, duality --- but this time  Fourier duality --- allows them to be transferred between dual spaces. But there is a big difference between the two. How is one to know which type of delta functional is appropriate in a given application?

The answer proposed in \emph{\cite{LA1}} is based on analogy with Bayesian inference in probability theory. In essence, the type of delta functional depends on the integrator family characterizing the function space of interest. For example, if the function space is $T_a$, then the gamma-type delta functional is indicated. However, if the function space is a Banach space characterized by a Gaussian integrator family, \textbf{both} types of delta functional are required in general. Specifically, one uses a gamma-type delta functional to localize the  mean and a Gaussian-type to localize the  covariance. The latter corresponds to the Faddeev-Popov method successfully utilized in QFT (recall \emph{(\ref{delta integrator})}).

It is known that the Faddeev-Popov method is not appropriate for all types of
localization: In particular, it is not applicable to fixed energy
  path integrals or paths with fixed boundary conditions. But these types of constraints localize the mean of a Gaussian
   and should therefore be implemented with gamma-type delta functionals \emph{\cite{LA1,LA2}}.
\end{remark}

\subsection{Poisson integrators}
Restrict the gamma integrator to integers, i.e. $\alpha=n\in\mathbb{N}$. Take the lower gamma integrator and regularize by replacing
$\gamma(n,c)$ with the regularized lower
incomplete gamma function
\begin{equation}
P(n,c):=\gamma(n,c)/\Gamma(n)\;.
\end{equation}
Likewise for the upper incomplete gamma function
\begin{equation}
\widehat{P}(n,c):=\Gamma(n,c)/\Gamma(n)\;.
\end{equation}

Note that, for $N\in
Pois(c)$ a Poisson random variable, we have
\begin{equation}
Pr(N<n)=\sum_{k<n}e^{-c}\frac{(c)^k}{k!}=\widehat{P}(n,c)\;.
\end{equation}
Hence,
\begin{equation}
Pr(N\geq n)=\sum_{k=n}^\infty
e^{-c}\frac{(c)^k}{k!}=P(n,c)
=\frac{1}{\Gamma(n)}\int_{T_a}
\mathcal{D}_\lambda\gamma_{n, Id',c}(\tau)
\end{equation}
which, in particular, implies
\begin{equation}
\frac{1}{\Gamma(0)}\int_{T_a}\;\mathcal{D}_\lambda\gamma_{0, Id',c}(\tau)
=\sum_{k=0}^\infty e^{-c}\frac{(c)^k}{k!}\;.
\end{equation}
On the other hand,
\begin{equation}
e^{-c}\frac{(c)^k}{k!}=\frac{e^{-c}}{k!}
\int_0^{c}\cdots\int_0^{c} \;d\tau_1,\ldots,d\tau_k\;.
\end{equation}

Evidently, the Poisson distribution is closely related to the
restricted gamma integrator which motivates the following
definition:
\begin{definition}\label{dirac}
Let $\Ta$ be the space of continuous pointed maps
$\tau:(\mathbb{T},\ti_a)\rightarrow(\C^\times,1)$
endowed with a lower gamma family of integrators. Let $\alpha=n\in\mathbb{N}$ and
$\langle \beta',\tau_c\rangle=c\in\C^\times$ for some fiducial $\tau_c\in T_a$. The Poisson integrator family
$\mathcal{D}_\Lambda\pi_{n,\beta',c}(\tau)$ is characterized by
\begin{eqnarray}
&&\Theta_{n,\beta'}(\tau,\tau')=e^{i\dual-\langle \beta',\tau\rangle} \tau^n\notag\\
&&\mathrm{Z}_{n, \beta',c}(\tau') = P\left(n, c\right)\left( \beta'-i{\tau'}\right)_\lambda^{-n}\;.
\end{eqnarray}
The Poisson family is defined in terms of the primitive
integrator $\mathcal{D}_\lambda\tau$ by
\begin{equation}
\mathcal{D}_\lambda\pi_{n,\beta',c}(\tau) :=
e^{-\langle \beta',\tau\rangle}\tau^n\,
\mathcal{D}_\lambda\tau\;.
\end{equation}
\end{definition}
Note the normalization of the fiducial Poisson integrator
\begin{equation}
\int_{T_a}\;\mathcal{D}_\lambda\pi_{0,Id',c}(\tau) :=1\;,
\end{equation}
and the rest of the family
\begin{equation}
\int_{T_a}\;\mathcal{D}_\lambda\pi_{n,Id',c}(\tau)
=P(n, c)\;.
\end{equation}
Poisson functional integrals are useful tools for solving systems of first-order differential equations.

\subsection{Matrix gamma integrators}
The target manifold for gamma integrators can be readily generalized to higher dimensions. The same can be done for Gaussian integrators. This brings \emph{matrix} functional integrals into the fold --- complex Wishart for gamma and random matrix models for Gaussian. We'll do it here for gamma integrators restricted to positive-definite Hermitian matrices.

\begin{definition}
Let $\Bold{T}_a$ denote a topological space of continuous matrix-valued pointed maps
$\Bold{\tau}:(\mathbb{T},\ti_a)\rightarrow(V_H,Id)$ where $V_H$ is the vector space of $n\times n$ positive-definite Hermitian matrices over $\R$. $\Bold{T}_a$ is a non-abelian topological group under point-wise multiplication. Let $\Bold{\beta}'\in\mathrm{Hom}_C(\Bold{T}_a,\C^{n\times n})$ be a fixed element such that $\langle{\beta}',\Bold{\tau}\rangle=\mathrm{tr}\,\Bold{\beta}'(\Bold{\tau})$. A matrix gamma family of integrators
$\mathcal{D}_\Lambda\gamma_{\alpha,\Bold{\beta}'}(\Bold{\tau})$ on $\Bold{T}_a$ is characterized
by
\begin{eqnarray}
&&\Theta_{\alpha,\Bold{\beta}'}(\Bold{\tau},\Bold{\tau}')
=e^{i\,\mathrm{tr}\,\Bold{\tau}'(\Bold{\tau})
-\mathrm{tr}\,\Bold{\beta}'(\Bold{\tau})}\,\mathrm{det}(\Bold{\tau}^\alpha) \notag\\
&&\mathrm{Z}_{\alpha,\Bold{\beta}'}(\Bold{\tau}')
=\mathrm{det}\,({\Bold{\beta}'}-i{\Bold{\tau}'})_\lambda^{-\alpha}
\end{eqnarray}
where $\Bold{\tau}'\in\mathrm{Hom}_C(\Bold{T}_a,\C^{n\times n})$, $\alpha\in\C_+$, and $\mathrm{det}(\Bold{\tau}^\alpha)$ is defined point-wise by
\begin{equation}
\mathrm{det}(\Bold{\tau}^\alpha)(\ti):=e^{\alpha\,\mathrm{tr}\log\Bold{\tau}(\ti)}\;.
\end{equation}
Note that $\mathrm{tr}\,\log\Bold{\tau}(\ti)$ (with principal value prescription) is \textbf{real} analytic here since $\Bold{\tau}(\ti)$ is Hermitian.
As before, the functional determinant $\mathrm{det}\,({\Bold{\beta}'}-i{\Bold{\tau}'})_\lambda^{-\alpha}$ is assumed to be well defined for a given $\lambda$.

The gamma integrator family is defined in terms of the primitive
integrator $\mathcal{D}_\lambda\Bold{\tau}$ by
\begin{equation}\label{gamma}
\mathcal{D}_\lambda\gamma_{\alpha,\Bold{\beta}'}(\Bold{\tau}) :=
e^{-\mathrm{tr}\,\Bold{\beta}'(\Bold{\tau})}\mathrm{det}(\Bold{\tau}^\alpha)\, \mathcal{D}_\lambda\Bold{\tau}
\end{equation}
where $\mathcal{D}_\lambda\Bold{\tau}$ is
characterized by
\begin{eqnarray}
&&\Theta_{0, \Bold{Id}'}(\Bold{\tau},\Bold{\tau}')=\exp\{ i\,\mathrm{tr}
\,\Bold{\tau}'(\Bold{\tau})-\mathrm{tr}\, \Bold{Id}'(\Bold{\tau})\}\notag\\
&&\mathrm{Z}_{0, \Bold{Id}'}(\Bold{\tau}')=\Gamma_n(0)_\lambda\;,
\end{eqnarray}
where
\begin{equation}\label{n-gamma}
\Gamma_n(\alpha)_\lambda:=\int_{\Bold{T}_a}e^{-\mathrm{tr}\,\Bold{\tau}}\,
\mathrm{det}(\Bold{\tau}^\alpha)\, \mathcal{D}_\lambda\Bold{\tau}
\end{equation}
and implicit in $\Gamma_n(0)_\lambda$ is a regularization.
\end{definition}
In applications, the archetypical localization turns out to be $\lambda:\Bold{T}_a\rightarrow V_H\subseteq\C^{n(n+1)/2}$; yielding in particular the complex Wishart integral for the special case $\Re(\alpha)=d/2$ with $d>n-1$ and $\mathcal{D}_\lambda\Bold{\tau}=(\det \Bold{\tau})^{-(n+1)/2}\prod_{1\leq i\leq j\leq n}d\tau_{i,j}$.

Analogous to the gamma integrator, consider a cut-off map $\Bold{\tau}_o$, and interpret it as a bound on the spectral radius $\rho(\Bold{\tau}(\ti))\leq\|\Bold{\tau}_o(\ti)^k\|^{1/k}$  for all $\ti\in[\ti_a,\ti_b]$ and all $k\in\mathbb{N}$. This leads to a lower matrix gamma integrator;
\begin{definition}
Let $\Bold{T}_a$ denote a topological space of continuous, matrix-valued pointed maps
$\Bold{\tau}:(\mathbb{T},\ti_a)\rightarrow V_H$. Fix a fiducial $\Bold{\tau}_o\in\Bold{T}_a$ and some $\Bold{\beta}'\in\mathrm{Hom}_C(\Bold{T}_a,\C^{n\times n})$ such that $\langle\beta',\Bold{\tau}_o\rangle=\mathrm{tr}\,\Bold{\beta}'(\Bold{\tau}_o)=r\in\R_+$. A lower matrix gamma family of integrators
$\mathcal{D}_\Lambda\gamma_{\alpha,\Bold{\beta}',r}(\Bold{\tau})$ on $\Bold{T}_a$ is characterized
by
\begin{eqnarray}
&&\Theta_{\alpha,\Bold{\beta}'}(\Bold{\tau},\Bold{\tau}')= e^{ i\,\mathrm{tr}\,\Bold{\tau}'(\Bold{\tau})-\mathrm{tr}\,\Bold{\beta}'(\Bold{\tau})}
\,\mathrm{det}(\Bold{\tau}^\alpha) \notag\\
&&\mathrm{Z}_{\alpha,\Bold{\beta}',r}(\Bold{\tau}')
=\gamma_n\left(\alpha,r\right)
\mathrm{det}\,({\Bold{\beta}'}-i{\Bold{\tau}'})_\lambda^{-\alpha}
\end{eqnarray}
where  $\gamma_n\left(\alpha,r\right)$ is
the lower incomplete gamma function associated with $\Gamma_n(\alpha)$ defined in \emph{(\ref{n-gamma})},
and again the functional determinant is assumed to be well defined by suitable choice of $\lambda$. The upper matrix gamma family
$\mathcal{D}_\Lambda\Gamma_{\alpha,\Bold{\beta}',r}(\Bold{\tau})$ is defined
similarly.
\end{definition}

\begin{remark}
Choosing $i\,\mathrm{tr}\,\Bold{\tau}'(\Bold{\tau})
-\mathrm{tr}\,\Bold{\beta}'(\Bold{\tau})=\mathrm{tr}(\sum_{k\neq0}a_k\Bold{\tau}^k-\Bold{\tau})$ models the generating functional of the Laguerre unitary ensemble studied in \emph{\cite{GGR}}. They calculate correlators for $k\in\mathbb{Z}\backslash\{0\}$ and relate them to certain Hurwitz numbers and Hodge integrals. This example and the next two applications highlight the combinatoric nature of gamma functional integrals.
\end{remark}

The next two applications use gamma, Poisson, and matrix gamma integrators to construct functional integral representations of certain \emph{average} prime and prime $\mathrm{k}$-tuple counting functions. We formulate the counting functions in the spirit of quantum-mechanical expectation values in the sense that they represent a sum over all `paths' with certain attributes. This approach leads to average single-prime counting functions that agree with their known number theory counterparts --- albeit from a quite different perspective. The average prime $\mathrm{k}$-tuple counting functions, though in agreement \emph{asymptotically} with the Hardy-Littlewood prime $\mathrm{k}$-tuple conjecture \cite{LH}, actually yield more accurate counts. (see appx. C)

\begin{application}\emph{A toy model for counting primes:}

 If one is willing to compare the pseudo-random occurrence of prime numbers and their predictable averages to quantum evolution, then analogy with quantum systems suggests that \textbf{counting} prime powers can be formulated in terms of an integral kernel/propagator associated with a gamma functional integral. This provides our starting hypothesis: The very nature of counting random prime-power events on the integer lattice dictates a Poisson process which can be represented by a gamma functional integral. All we need do is determine the relevant, perhaps non-homogeneous, scaling factor that parametrizes the process.

To see how to proceed, we first calculate the expected number of integers occurring up to some cut-off integer $ \mathsf{x}\in\R_+$  by defining a suitable $\alpha$-trace (to be defined) applied to the simple case of a homogeneous process. That is, we take $\beta'=-Id'$ in the lower gamma integral and fix the localization  by restricting the domain of  paths via the homomorphism $\lambda_{\R_+}:T_a\rightarrow\R_+$.  In this case, the functional integral can be explicitly evaluated and we get
\begin{eqnarray}\label{counting}
N( \mathsf{x})
:=\mathrm{tr}_\alpha\left[\int_{\R_+}
\;\mathcal{D}\gamma_{\alpha,-Id', \mathsf{x}}(\tau(\ti))\right]
&=&\mathrm{tr}_\alpha\bigg[\,(-1)^{\alpha}\gamma(\alpha, \mathsf{x})\,\bigg]\notag\\
&:=&\int_{\mathcal{C}}\frac{\Gamma(1-\alpha)}{2\pi
i}\,\bigg[\,(-1)^{\alpha}\gamma(\alpha, \mathsf{x})\,\bigg]\,d\alpha\notag\\
&=&\sum_{n=1}^\infty\frac{(-1)^{2n}}{(n-1)!}\,\gamma(n, \mathsf{x})\notag\\
&=&\sum_{n=1}^\infty P(n, \mathsf{x})= \mathsf{x}
\end{eqnarray}
where the contour $\mathcal{C}$ encircles the positive real axis. This motivates and justifies the definition of the $\alpha$-trace $\mathrm{tr}_\alpha$.\footnote{We could, of course, take $\beta'=+Id'$ and adjust the definition of $\mathrm{tr}_\alpha$.}

Now postulate that the prime-power counting function is the $\alpha$-trace of a
gamma process with unknown scale parameter due to the constraint
associated with counting only prime events on the integer lattice. According to \emph{\cite{LA1}}, the constrained functional
integral that represents the propagator can be formulated in terms of a
constrained functional that is integrable with respect to two
marginal gamma integrators. Accordingly, we posit the average number of \textbf{prime powers} $p^k$ up to some cut-off integer $\x$ is given by
\begin{eqnarray}
\overline{N_{p^k}( \x)}_\lambda
&=&\mathrm{tr}_{\alpha+1}\int_{{\Ta}\times C}\;
\mathcal{D}_\lambda\gamma_{\alpha,-Id', \x}(\tau)
\;\mathcal{D}_\lambda\gamma_{1,ic'(\tau),\infty}(c)\notag\\
&=&\mathrm{tr}_{\alpha+1}\int_{{\Ta}}\;
\mathcal{D}_\lambda\gamma_{\alpha,-Id',s( \x)}({\tau})
\end{eqnarray}
where $c'(\tau)$ represents the constraint and $s( \x)$ represents an unknown
possibly non-homogenous scaling factor (see \emph{\cite[\S~4.1]{LA1}}). The $\alpha$-trace has been shifted by one because the counting should begin with the second event (since primes start with $p=2$). Note that the constraint just imposes a scaling factor on the cut-off (compare with \emph{(\ref{counting})}).

Under the localization $\lambda_{\R_+}:T_a\rightarrow\R_+$, this evaluates to
\begin{eqnarray}\label{number evaluation}
\overline{N_{p^k}( \x)}
&=&\mathrm{tr}_{\alpha+1}\left[\,(-1)^{\alpha}\gamma(\alpha,s( \x))\,\right]\notag\\
&=&\frac{1}{2\pi i}\int_{\mathcal{C}_{+1}}\frac{\pi\csc(\pi(\alpha+1))}{\Gamma(\alpha+1)}
\left[\,(-1)^{\alpha}\gamma(\alpha,s( \x))\,\right]\;d\alpha\notag\\
&=&\sum_{n=1}^\infty\frac{(-1)^{2n+1}}{n!}\,\gamma(n,s( \x))\notag\\
&=&-\sum_{n=1}^\infty\frac{\Gamma(n)}{\Gamma(n+1)}\,P(n,s( \x))
\end{eqnarray}
where the new contour begins at $\infty$ above the real axis, circles
the point $\{1\}$ counter-clockwise, and returns to $\infty$ below the real
axis. Roughly speaking, this calculation simply
sums the positive integers appropriately adjusted with a
non-homogenous scaling factor and weighted by
$\Gamma(n)/\Gamma(n+1)=1/n$.

The series converges absolutely since
\begin{equation}\label{converges}
\lim_{n\rightarrow\infty} \left|\frac{n!}{(n+1)!}
\frac{\left|\gamma(n+1,s( \x))\right|}
{\left|\gamma(n,s( \x))\right|}\right|
=\lim_{n\rightarrow\infty} \left|\frac{1}{(n+1)}\right|
s( \x)=0\;.
\end{equation}
And observe that
\begin{equation}
\frac{\overline{N_{p^k}( x+\Delta x)}-\overline{N_{p^k}( x)}}{\Delta x}=
-\sum_{n=1}^\infty\frac{1}{n!\Delta x}
\int_{s(x)}^{s( x+\Delta x)}e^{-t}\,t^{n-1}\;dt
\sim\frac{-1}{s(\Delta x)}
\end{equation}
is supposed to represent the average density of prime powers (in this section the symbol $\sim$ means asymptotically). Accordingly, a good and obvious choice for the scaling factor is
$s( \x)=-\log( \x)$ yielding the Poisson functional integral representation
\begin{equation}\label{Poisson}
\overline{N_{p^k}( \x)}_\lambda=- \sum_{n=1}^\infty\frac{(-1)^n}{n!}\,\int_{{\Ta}}\;\mathcal{D}_\lambda\pi_{n,-Id',-\log(\x)}(\tau)\;.
\end{equation}

Crucially,  $\int_1^\x|d\gamma(n,-\log(x))/dx|\,dx=\gamma(n,-\log(\x))$ together with the absolute convergence of $\sum\gamma(n,-\log( \x))/n!$ implies that the sum \emph{(\ref{number evaluation})} can be expressed as
\begin{equation}\label{average N}
    \overline{N_{p^k}(\x)}
    =\mathrm{li}( \x)-\log(\log( \x)).
\end{equation}
It is reasonable, therefore, to view $\overline{N_{p^k}(\x)}$ as an approximation to $J( \x)-w( \x)$ where $J(\x)$ is Riemann's counting function
\begin{equation}
J(\x):=\sum_{p^k\leq \x}\,\frac{1}{k}=\sum_{n\leq \x}
\frac{\Lambda(n)}{\log(n)}
\end{equation}
where $\Lambda(n)$ is the Von Mangoldt function, and $w( \x)$ is the weighted sum of prime power divisors of $ \x$
 \begin{equation}
 w( \x):=\sum_{p^k\mid \x}\,\frac{1}{k}=\sum_{n\,\mid\, \x}\frac{\Lambda(n)}{\log(n)}\;.
\end{equation}
This is justified since  $\overline{N_{p^k}( \x)}=\overline{J( \x)}-\overline{w( \x)}$ where
\begin{equation}
  \overline{J( \x)}:=\mathrm{li}( \x)
  \approx\sum_{n\leq \x}
\frac{\Lambda(n)}{\log(n)},\;\;\;\;\;\;\;
\overline{w( \x)}:=\log(\log( \x))\approx\sum_{n\,\mid\, \x}\frac{\Lambda(n)}{\log(n)}\;.
\end{equation}

What if we had chosen $\beta'=+Id'$ instead? Repeating the calculation for this case yields
\begin{equation}
- \sum_{n=1}^\infty\frac{(-1)^n}{n!}\,\int_{\R_+}\;\mathcal{D}\pi_{n,Id',-\log(\x)}(\tau(\ti))
=-\sum_{n=1}^\infty\frac{(-1)^n\Gamma(n)}{\Gamma(n+1)}\,P(n,-\log( \x))=\mathrm{li}(\x)-\mathrm{li}(\x^2)\;.
\end{equation}
Up to a minus sign this is
\begin{eqnarray}
  \overline{\sigma_{p^k}( \x)}=\mathrm{li}(\x^2)-\mathrm{li}(\x)
&\approx&\sum_{p^k\leq\x}\frac{p^k}{k}-\sum_{p^k|\x}\frac{p^k}{k}\notag\\
&=&\sum_{\begin{array}{c}
                 \scriptstyle{n\leq \x},\;
                 \scriptstyle{ n\,\nmid\, \x}
               \end{array}}\frac{n\Lambda(n)}{\log(n)}\;.
\end{eqnarray}
So, relative to the defined $\alpha$-trace, $\beta'=-Id'$ counts the expected \textbf{number} while $\beta'=Id'$ counts the expected \textbf{sum} of prime powers weighted by $1/k$ that don't divide the cut-off $\x$.

Put $n\rightarrow n+1$ in the integrator. For $\beta'=-Id'$
we get a Poisson functional integral representation of an average Chebyshev function for prime powers up to cut-off $\x$ given by
\begin{eqnarray}
    \overline{Ch_{p^k}( \x)}
    =\sum_{n=1}^{\infty}\,\int_{\R_+}\;\mathcal{D}\pi_{n+1,-Id',-\log(\x)}(\tau(\ti))
    &=&\sum_{n=1}^\infty P(n+1,-\log(\x))\notag\\
    &=&\x-\log(\x)\notag\\
    &\approx&\sum_{p^k\leq\x}\frac{1}{k}\,\log(p^k)
    -\sum_{p^k|\x}\frac{1}{k}\,\log(p^k)\notag\\
    &=&\sum_{\begin{array}{c}
                 \scriptstyle{n\leq \x},\;
                 \scriptstyle{ n\,\nmid\, \x}
               \end{array}}\Lambda(n)\;.
\end{eqnarray}
For $\beta'=Id'$ we get what can be called the `weighted entropy of prime powers':
\begin{eqnarray}
    \overline{H_{p^k}( \x)}
    =\sum_{n=1}^{\infty}  \int_{\R_+}\;\mathcal{D}\pi_{n+1,Id',-\log(\x)}(\tau(\ti))
    &=&\sum_{n=1}^\infty (-1)^nP(n+1,-\log(\x))\notag\\
    &\approx& \sum_{p^k\leq\x}\frac{1}{k}\,p^k\log(p^k)
    -\sum_{p^k|\x}\frac{1}{k}\,p^k\log(p^k)\notag\\
    &=&\sum_{\begin{array}{c}
                 \scriptstyle{n\leq \x},\;
                 \scriptstyle{ n\,\nmid\, \x}
               \end{array}}n\Lambda(n)\;.
\end{eqnarray}

There is a clear pattern emerging, and it starts with the dominant term of $\overline{N_{p^k}( \x)}$ which is a sum of lower incomplete gamma functions that happens to equal $\mathrm{li}(\x)$.  In the spirit of the physical toy model, this average can be interpreted as an expectation value of  a counting operator, say $\langle\x|\widehat{J}({\x})|0\rangle\equiv \overline{J(\x)}=\mathrm{li}(\x)/2\pi i$. When extended to the complex plane as $\overline{J(\x^s)}$ with $s\in\C$ and integrated along a suitable contour with respect to $\log(\zeta(s))$, the average converges to Riemann's exact counting function $J(\x)$. Likewise, the other averages in the pattern converge to exact counting functions.  They can be combined into a compact expression:
\begin{eqnarray}\label{prime counting}
 J(\x;r,i)
 &:=&\int_{\mathcal{C}}\frac{d^i}{ds^i}\overline{J( \x^{s})}
 \,d\log({\zeta}(s-r))\;\,\;\;\;\;\;c>r\notag\\
 &=&\int_{\mathcal{C}}\langle\x|\widehat{J}^{(i)'}({\x}^s)|0\rangle
 \,d\log(\zeta(s-r))\;\,\;\;\;\;\;c>r\notag\\
 &=&\int_{\mathcal{C}}\langle\x|\widehat{J}^{(i)'}({\x}^{s+r})|0\rangle
 \,d\log(\zeta(s))\;\,\;\;\;\;\;c>r\notag\\
 &=&\langle\x|\left[\int_{\mathcal{C}}\widehat{J}^{(i)'}({\x}^{s+r})
 \,d\log(\zeta(s))\right]|0\rangle\;\,\;\;\;\;\;c>r
\end{eqnarray}
where $r,i\in\mathbb{Z}_+$, contour $\mathcal{C}\in\C^\times$ includes the vertical line $s=c+it$ and encloses the singularities of $d\log(\zeta(s))/ds$, and $\log(\zeta(s))$ is the log-zeta function with $s\in\C\backslash\{1\}$. Evidently, exact counting operators $ J(\x;r,i)$ are sums of operators localized on the pole and zeros of zeta. The pole gives the average while the zeros conspire to ``morph'' the smooth average into a step function.

For example, $r,\,i=0$ gives the exact Riemann prime counting function $J(\x)=\sum_{n\leq\x}\frac{\Lambda(n)}{\log(n)}$, and $r=0,\,i=1$ gives the exact second Chebyshev function $\psi(\x)=\sum_{n\leq\x}\Lambda(n)$. Meanwhile, for $r=1,\,i=0$ we get the exact weighted sum of prime powers
\begin{equation}
\sum_{p^k\leq\x}\frac{p^k}{k}= \sum_{n\leq\x}\frac{n\Lambda(n)}{\log(n)}=\mathrm{li}(\x^2)
 -\sum_\rho \mathrm{li}(\x^{1+\rho})-C-\frac{1}{2}-\sum_{k=0}^\infty \mathrm{li}(\x^{-2k+1})\;,
\end{equation}
and for $r=1,\,i=1$ we get the exact weighted entropy of prime powers
\begin{equation}
\sum_{p^k\leq\x}\frac{1}{k}\,p^k\log(p^k)=\sum_{n\leq\x}n\Lambda(n)=\frac{1}{2}\x^2
-\sum_\rho\frac{\x^{1+\rho}}{1+\rho}-C+\arctan(\x^{-1})
\end{equation}
where $\rho$ represents a zeta-zero and $C=12\log(A)-1$ being $A$ the Glaisher-Kinkelin constant.

Defining
\begin{equation}\label{prime zeta}
\log\left(\mathfrak{z}(s)\right):=-\sum_{n=1}^\infty\frac{\mu(n)\Lambda(n)}{\log(n)n^{s}}\;,\;\;\;\;\Re(s)>1\;,
\end{equation}
 we get explicit integrals representing exact prime counting functions:
\begin{equation}\label{3}
 \pi( \x;r,i):=\langle\x|\left[\int_{\mathcal{C}}\widehat{J}^{(i)'}({\x}^{s+r})
 \,d\log(\mathfrak{z}(s))\right]|0\rangle\;\,\;\;\;\;\;c>r\;.
\end{equation}
\end{application}

These explicit integrals for exact counting functions are already known (or can be derived) in number theory (see e.g. \cite{MO, GA}); though in that context they are expressed and interpreted differently than (\ref{prime counting}) and (\ref{3}).

\begin{application}\emph{Counting prime $\kk$-tuples:}

Counting single primes in the previous application provides four clues: average counting follows gamma statistics, the random variables are prime \textbf{powers}, $T_a$ is characterized by logarithmic measures, and expectations do not include ordinals that divide the cut-off $\x$.

These clues suggest the cut-off exerts a strong influence on the counting process and perhaps we were too hasty to construct the toy model on $\R_+$; after all, counting always necessarily includes a finite cut-off as is true of any observation of a quantum system. Hence, replace the space $T_a$ in the definition of the gamma integrator with the space of continuous pointed maps $T_{\x}\ni\tau:(\Ti_+,\ti_a)\rightarrow(\mathfrak{B}^\x,1)$ where $\mathfrak{B}^\x$ is the Banach space with module $(\ZZ/\x\ZZ)^\times$ over the field $\ZZ/p\ZZ$ (for some prime $p$) and localize with $\lambda_\x:T_\x\rightarrow(\ZZ/\x\ZZ)^\times$. Since $\x=\prod_i^{\omega(\x)}p_i^{r_i}$ by prime decomposition, the prime-power $\kk$-tuple quantum states in our model correspond to representations of elements in the direct product group
\begin{equation}
(\ZZ/\x\ZZ)^\times \cong \prod_i^{\omega(\x)}(\ZZ/p_i^{r_i}\ZZ)^\times\
\end{equation}
with $\omega(\x)$ the number of distinct prime factors.

Before getting to prime-power $\mathrm{k}$-tuples, we revisit the previous single prime counting example from this perspective to write (under the map $\lambda_\x:T_\x\rightarrow(\ZZ/\x\ZZ)^\times$)
\begin{equation}
\overline{N_{p^k}( \x)}=-\sum_{n=1}^\infty\frac{(-1)^n}{n!}\, \int_{(\ZZ/\x\ZZ)^\times}\;\mathcal{D}\pi_{n,-Id',-\log(\x)}(\tau(\ti))\;.
\end{equation}
Use the clue that $T_a$ is characterized by logarithmic measures to deduce
\begin{eqnarray}\label{average counting}
\overline{N_{p^k}( \x)}
&=&-\sum_{n=1}^\infty\frac{(-1)^n}{n!}\int_{(\ZZ/\x\ZZ)^\times}e^{\tau(\ti)} \tau(\ti)^{n}\;d(\log(\tau(\ti)))\notag\\
&=&-\sum_{n=1}^\infty\frac{(-1)^n}{n!}\int_{(\ZZ/\x\ZZ)^\times}e^{\log(x)} (\log(x))^{n-1}\;d(\log(x))\notag\\
&=&\int_{(\ZZ/\x\ZZ)^\times}\sum_{n=1}^\infty\frac{(-1)^{n-1}}{n!}(\log(x))^{n-1}\;dx\notag\\
&=&\sum_{2}^\x\frac{x-1}{x\log(x)}\;\;\;\;\;\;\;\;x\in(\ZZ/\x\ZZ)^\times\notag\\
&\approx&C\int_2^\x\frac{r-1}{r\log(r)}\;dr\;\;\;\;\;\;\;\;r\in\R_+\;.
\end{eqnarray}
Interchange of the sum and integral is justified by \emph{(\ref{converges})} and $\overline{N_{p^k}( \x)}<\infty$ for all $\x<\infty$. The normalization constant $C$ comes from the ratio of the counting measure on $(\ZZ/\x\ZZ)^\times$ v.s. the measure on $\R_+$. This is consistent with \emph{(\ref{average N})} (pending determination of constant $C$).

Now, with these preliminaries, we build a model to count \textbf{admissible} prime ${\mathrm{k}}$-tuples. Let $\mathcal{H}_{\mathrm{{\mathrm{k}}}}:=\{0,\ldots,h_{\mathrm{{\mathrm{k}}}}\}$ be a set with $|\mathcal{H}_{\mathrm{{\mathrm{k}}}}|={\mathrm{{\mathrm{k}}}}$ and distinct $h_i\in\ZZ_+$ such that $\mathcal{H}_{\mathrm{{\mathrm{k}}}}$ does not cover the residue classes associated with $(\ZZ/p\ZZ)^\times$ for any prime $p$. Let $(\x,\ldots,\x+h_{\mathrm{{\mathrm{k}}}})\in\mathfrak{N}_+^{\mathrm{{\mathrm{k}}}}\subset\ZZ_+^{\mathrm{{\mathrm{k}}}}$ where $\mathfrak{N}_+^{\mathrm{{\mathrm{k}}}}$ is the pair-wise coprime ${\mathrm{{\mathrm{k}}}}$-lattice and $\x\geq2$. Define the geometric mean of $(\x,\ldots,\x+h_{\mathrm{{\mathrm{k}}}})$ by $\x_{({\mathrm{{\mathrm{k}}}})}:=[\x(\x+h_2)\ldots(\x+h_{\mathrm{{\mathrm{k}}}})]^{1/{\mathrm{{\mathrm{k}}}}}$. An admissible prime $\mathrm{{\mathrm{{\mathrm{k}}}}}$-tuple is a point $(p,\ldots,p+h_{\mathrm{{\mathrm{k}}}})\in\mathfrak{N}_+^{\mathrm{{\mathrm{k}}}}$ such that each $p+h_i$ is prime for any prime $p$. Note that, since we are on the pair-wise coprime lattice, $\ZZ/\x^{\mathrm{{\mathrm{k}}}}_{({\mathrm{k}})}\ZZ\cong\bigoplus_{i=1}^{\mathrm{{\mathrm{k}}}}\ZZ/(\x+h_i)\ZZ$, and counting prime ${\mathrm{{\mathrm{k}}}}$-tuples will require the matrix gamma integrator.

Let $\Bold{T}_{\x_{(\mathrm{k})}}$ be the space of continuous pointed maps $\Bold{\tau}:(\Ti_+,\ti_a)\rightarrow(\mathfrak{B}^{\x_{({\mathrm{{\mathrm{k}}}})}^{\mathrm{{\mathrm{k}}}}},Id)$
where $\mathfrak{B}^{\x_{({\mathrm{{\mathrm{k}}}})}^{\mathrm{{\mathrm{k}}}}}$ is the Banach space with module $(\ZZ/\x^{\mathrm{{\mathrm{k}}}}_{({\mathrm{{\mathrm{k}}}})}\ZZ)^\times$ over the field $\ZZ/p\ZZ$. We follow the single-prime case and choose $\Bold{\beta}'=-\Bold{Id}'$ with cut-off $\mathrm{tr}\,\Bold{\beta}'(\Bold{\tau}_o)=\x_{({\mathrm{{\mathrm{k}}}})}$. The points $(2,2+h_2,\ldots,2+h_{\mathrm{{\mathrm{k}}}})$ and $(\x,\x+h_2,\ldots,\x+h_{\mathrm{k}})$ define a ray $\mathbf{R}_{{\mathrm{k}}}:=(r,r+h_2,\ldots,r+h_\kk)\subset\mathfrak{N}_+^{\mathrm{k}}$ along which we will count prime-power ${\mathrm{k}}$-tuples. Lastly, take $\lambda_{\x_{(\mathrm{k})}}:\Bold{T}_{\x_{(\mathrm{k})}}\rightarrow(\ZZ/\x^{\mathrm{k}}_{({\mathrm{k}})}\ZZ)^\times$.

With the hypothesis that prime powers in $\mathfrak{N}_+^{\mathrm{k}}$ are independent random events, the counting factors into a product  $(\ZZ/\x^{\mathrm{k}}_{({\mathrm{k}})}\ZZ)^\times\cong\prod_{i=1}^{\mathrm{k}}(\ZZ/(\x+h_i)\ZZ)^\times$, and an independent $\alpha$-trace should be taken over each component of $\Bold{\tau}(\ti)$ separately. The expected number of admissible prime $\kk$-tuples along $\mathbf{R}_{{\mathrm{k}}}$ up to cut-off $\x$ is then
\begin{eqnarray}
\overline{N_{p^k_{(\kk)}}( \x)}
&=&\mathrm{tr}_{\alpha+1}\left[\int_{(\ZZ/\x^{\mathrm{{\mathrm{k}}}}_{({\mathrm{k}})}\ZZ)^\times}
\;\widehat{\delta}_{\mathbf{R}_{{\mathrm{k}-1}}}\;
\mathcal{D}\gamma_{\alpha,-\Bold{Id}',-\log(\x_{(\kk)})}(\Bold{\tau}(\ti))\right]\notag\\
&=&-\sum_{(n_1,\ldots,n_k)=1}^\infty\frac{(-1)^{n_i}}{n_i!}
\int_{(\ZZ/\x^{\mathrm{{\mathrm{k}}}}_{({\mathrm{k}})}\ZZ)^\times}
\,\widehat{\delta}_{x_i,x+h_i}\prod_{i=1}^\kk
\;\mathcal{D}\gamma_{n_i,-Id',-\log(\x+h_i)}(x_i)\;.
\end{eqnarray}
The delta term $\widehat{\delta}_{\mathbf{R}_{{\mathrm{k}-1}}}$ only restricts the last $\mathrm{k}-1$ components of $\Bold{\tau}(\ti)$, and the second line uses  $\ZZ/\x^{\mathrm{{\mathrm{k}}}}_{({\mathrm{k}})}\ZZ\cong\bigoplus_{i=1}^{\mathrm{{\mathrm{k}}}}\ZZ/(\x+h_i)\ZZ$. Following the single-prime case we get
\begin{eqnarray}\label{k-tuple conjecture}
\overline{N_{p^k_{(\kk)}}( \x)}
&=&-\sum_{(n_1,\ldots,n_k)=1}^\infty\frac{(-1)^{n_i}}{n_i!}
\int_{(\ZZ/\x^{\mathrm{{\mathrm{k}}}}_{({\mathrm{k}})}\ZZ)^\times}
\,\widehat{\delta}_{x_i,x+h_i}\prod_{i=1}^\kk
e^{\log(x_i)}(\log(x_i))^{n_i-1}\,d(\log(x_i))\notag\\
&=&-\int_{(\ZZ/\x^{\mathrm{{\mathrm{k}}}}_{({\mathrm{k}})}\ZZ)^\times}
\,\widehat{\delta}_{x_i,x+h_i}\sum_{(n_1,\ldots,n_k)=1}^\infty\frac{(-1)^{n_i}}{n_i!}
\prod_{i=1}^\kk\,(\log(x_i))^{n_i-1}\,dx_i\notag\\
&=&(-1)^{\kk-1}\sum_{2}^\x
\frac{Q(x,\mathcal{H}_\kk)}{x_{(\kk)}^\kk\log_{(\kk)}(x)}\;\;\;\;\;\;\;\;x\in\mathbf{R}_{{\mathrm{k}}}\notag\\
&\approx&(-1)^{\kk-1}\, C_{\kk}
\int_2^\x\frac{Q(r,\mathcal{H}_\kk)}{r_{(\kk)}^\kk\log_{(\kk)}(r)}\;dr\;\;\;\;\;\;\;\;r\in\R_+
\end{eqnarray}
where $C_{\kk}$ accounts for the ratio between the counting measure along $\mathbf{R}_{{\mathrm{k}}}$ v.s. the measure on $\R_+$ and $r_{(\kk)}^\kk=(r_{(\kk)})^\kk$. We define $\log_{(\kk)}(x):=\log(x)\log(x+h_2)\cdots\log(x+h_\kk)$, and $Q(x,\mathcal{H}_\kk)$ is a polynomial. For example, $Q(x,\mathcal{H}_1)=(x-1)$ and $Q(x,\mathcal{H}_2)=(x-1)((x+h_2)-1))$ and generally $Q(x,\mathcal{H}_{\mathrm{k}})=\prod_{i=1}^{\mathrm{k}}\left((x+h_i)-1\right)$.

Motivation and formal manipulations aside, we can just take the last two lines of \emph{(\ref{k-tuple conjecture})} as the definition of $\overline{N_{p^k_{(\kk)}}( \x)}$ and $C_{\kk}$. By analogy with the single-prime case, we also define
\begin{equation}
\overline{J_{(\kk)}( \x)}:=(-1)^\kk\sum_{2}^\x\frac{1}{\log_{(\kk)}(x)}
\approx(-1)^\kk \,C_{\kk}\int_2^\x\frac{1}{\log_{(\kk)}(r)}\;dr\;.
\end{equation}

The normalization $C_{\kk}$ allows to compare the measure on $\mathbf{R}_{{\mathrm{k}}}\subset\mathfrak{N}^\kk_+$ with the measure on $\R_+$. We claim the normalization is given by
\begin{equation}
C_{\kk}=\prod_p\left(1-\frac{\nu_p(\mathcal{H}_{\mathrm{k}})}{p}\right)\left(1-\frac{1}{p}\right)^{-{\mathrm{k}}}
\end{equation}
where $\nu_p(\mathcal{H}_{\mathrm{k}})$ is the number of distinct congruence classes mod $p$ covered by the admissible set $\mathcal{H}_{\mathrm{k}}$. To see this, we appeal to a theorem by T\'{o}th\emph{\cite{TO}}:
Let ${\mathrm{k}},m,u\geq1$ and
\begin{equation}
P_{\mathrm{k}}^{(u)}(m)=\!\!\!\!\!\!\!\!\sum_{\begin{array}{c}
            \scriptstyle{1\leq a_1, \ldots, a_{\mathrm{k}}\leq m} \\
            \scriptstyle{(a_i,a_j)=1, i\neq j}\;\;;\;\;
            \scriptstyle{(a_i,u)=1}
          \end{array}}\!\!\!\!\!\!\!\!1
\end{equation}
be the number of ${\mathrm{k}}$-tuples $(a_1,\ldots, a_{\mathrm{k}})$ on the pair-wise coprime lattice with $1\leq a_i\leq m$ and $(a_i,u)=1$ for all $i\in\{1,\ldots, {\mathrm{k}}\}$.
\begin{theorem}\emph{(\cite{T})}
\emph{For a fixed ${\mathrm{k}}\geq1$, we have uniformly for $m,u\geq1$,
\begin{equation}
P_{\mathrm{k}}^{(u)}(m)=A_{\mathrm{k}} f_{\mathrm{k}}(u) m^{\mathrm{k}}+O\left(\theta(u)m^{{\mathrm{k}}-1}\log^{{\mathrm{k}}-1}(m)\right)
\end{equation}
where $\theta(u)$ is the number of squarefree divisors of $u$ and
\begin{eqnarray*}
       A_{\mathrm{k}} &=& \prod_{p'}\left(1-\frac{1}{p'}\right)^{{\mathrm{k}}-1}\left(1+\frac{{\mathrm{k}}-1}{p'}\right) \\
       f_{\mathrm{k}}(u) &=&\prod_{p'|u}\left(1-\frac{{\mathrm{k}}}{p'+{\mathrm{k}}-1}\right)\;.
     \end{eqnarray*}}
\end{theorem}
To apply the theorem along $\mathbf{R}_{{\mathrm{k}}}$, restrict to $u=p$ for some prime $p$ and choose $m>\x+h_{\mathrm{k}}$. Then the density of points in $\mathfrak{N}^\kk_+$ that are coprime to a given prime $p$ (or prime power) is
\begin{equation}
D_{\mathrm{k}}^{(p)}(m):=P_{\mathrm{k}}^{(p)}(m)/m^{\mathrm{k}}=A_{\mathrm{k}} f_{\mathrm{k}}(p)+O\left(\log^{{\mathrm{k}}-1}(m)/m\right)\;.
\end{equation}
Of course $m$ is automatically coprime to all primes $p> m$.

Now, in our case the ${\mathrm{k}}$-tuples $(a_1,\ldots, a_{\mathrm{k}})$ are restricted to the admissible ray $\mathbf{R}_{{\mathrm{k}}}$ so $D_{\mathrm{k}}^{(p)}(m)$ must be divided by the density of points with $\mathrm{GCD}(x+h_i,p)=1$ for each $h_i$ and each $p$. This density is given by $\frac{1}{p}\left(p-\nu_p(\mathcal{H}_{\mathrm{k}})\right)$. Hence, the total density of prime powers along an admissible ray $\mathbf{R}_{{\mathrm{k}}}\subset\mathfrak{N}^\kk_+$ is
\begin{equation}
\lim_{m\rightarrow\infty}\prod_{p\leq m} \frac{D_{\mathrm{k}}^{(p)}(m)}{\left(1-\frac{\nu_p(\mathcal{H}_{\mathrm{k}})}{p}\right)}
=\prod_p\left(1-\frac{1}{p}\right)^{{\mathrm{k}}}
\left(1-\frac{\nu_p(\mathcal{H}_{\mathrm{k}})}{p}\right)^{-1}\;.
\end{equation}
Notice this is precisely the inverse Hardy-Littlewood singular series for prime $\kk$-tuples.

With the prime-power states properly normalized on $\R_+$, we can now count them and assign ordinals. {Appendix C} compares average counts of some prime $2$-tuples predicted by \emph{(\ref{k-tuple conjecture})} v.s. the Hardy-Littlewood $\kk$-tuple conjecture.

We are finally in a position to make use of the interpretation afforded by \emph{(\ref{prime counting})} and \emph{(\ref{3})}. Evidently, explicit counting by means of an integral representation for $\kk>1$ requires a generalized zeta function; to which we now turn.

First, note that an exact counting function for prime $\kk$-tuples given $\mathcal{H}_{\kk}$ and $\x$ is
\begin{equation}
\pi_{(\kk)}(\x)
=(-1)^\kk\sum_{n\leq \x}\mu(n)\cdots\mu(n+h_\kk)\frac{\Lambda(n)}{\log(n)}
\cdots\frac{\Lambda(n+h_\kk)}{\log(n+h_\kk)}=:(-1)^\kk\sum_{n\leq \x}\mu_{(\kk)}(n)\frac{\Lambda_{(\kk)}(n)}{\log_{(\kk)}(n)}\;.
\end{equation}
This suggests to define a generalized zeta function implicitly by
\begin{equation}
\log\left(\zeta_{(\kk)}(s)\right)
:=\sum_{n=1}^\infty\frac{\Lambda_{(\kk)}(n)}{\log_{(\kk)}(n)\,(n_{(\kk)})^s}
=:\sum_{n=1}^\infty\frac{\lambda_{(\kk)}(n)}{n_{(\kk)}^s}
\;\;\;\;\;\;\;\;\Re(s)>1\;.
\end{equation}
Then
\begin{equation}
\log^{(\kk-1)'}\left(\zeta_{(\kk)}(s)\right)
=(-1)^{\kk-1}\sum_{n=1}^\infty\frac{\lambda_{(\kk)}(n)}{n_{(\kk)}^s}\log^{\kk-1}(n_{(\kk)})\;.
\end{equation}
Now define
\begin{equation}
\phi_{(\kk)}(\x):=\sum_{\begin{array}{c}
                 \scriptstyle{n\leq x},\;
                 \scriptstyle{ n\,\nmid\, x}
               \end{array}}\lambda_{(\kk)}(n)\log^{\kk-1}(n_{(\kk)})\;.
\end{equation}
Its average is
\begin{equation}
\overline{\phi_{(\kk)}(\x)}
:=\sum_{2}^\x\frac{Q(x,\mathcal{H}_\kk)}{x_{(\kk)}^\kk\log_{(\kk)}(x)}
\log^{\kk-1}(x_{(\kk)})
\approx C_{\kk}\int_2^\x\frac{Q(r,\mathcal{H}_\kk)}{r_{(\kk)}^\kk\log_{(\kk)}(r)}\log^{\kk-1}(r_{(\kk)})\;dr
\end{equation}
with dominant term
\begin{equation}
\overline{\varphi_{(\kk)}(\x)}
:=\sum_{2}^\x\frac{1}{\log_{(\kk)}(x)}
\log^{\kk-1}(x_{(\kk)})
\approx C_{\kk}\int_2^\x\frac{1}{\log_{(\kk)}(r)}\log^{\kk-1}(r_{(\kk)})\;dr=:C_{\kk}\,\mathrm{Li}_{(\kk)}(\x)\;.
\end{equation}
Asymptotically $\overline{\varphi_{(\kk)}(\x)}\sim C_\kk\,\mathrm{li}(\x)$, so this situation is very much like the single prime case. Together with the interpretation of \emph{(\ref{prime counting})} and \emph{(\ref{3})}, this motivates
\begin{theorem}\label{exact k-tuple}
\emph{Let $\mathcal{R}_{(\kk)}(x):=\log_{(\kk)}(x)/\left(\log(x)\log^{\kk-1}(x_{(\kk)})\right)$ and $\widetilde{ \x}:= \x+\epsilon$ where $ \x\in\N_+$ and $0\leq\epsilon<1$. Let $\sigma_a$ be the abscissa of absolute convergence of $\sum_{n=1}^\infty\frac{\lambda_{(\kk)}(n)\log^{\kk-1}(n_{(\kk)})}{n_{(\kk)}^{s}}$. Then, for $c>\sigma_a$,
\begin{eqnarray}\label{explicit integral}
 \varphi_{(\kk)}(\x)
 &=&\lim_{\epsilon\rightarrow0}\lim_{T\rightarrow\infty}\frac{(-1)^{\kk-1}}{2\pi i}\mathcal{R}_{(\kk)}(1)\int_{c-iT}^{c+iT}
 \mathrm{Li}_{(\kk)}(\widetilde{\x}^s)
 \;d\log^{(\kk-1)'}(\zeta_{(\kk)}(s))\,,
 \;\,\;\;\;\;\;c>\sigma_a\notag\\
 &=&\sum_{n\leq \x}\lambda_{(\kk)}(n)\log^{\kk-1}(n_{(\kk)})\notag\\
 &=&\sum_{n\leq \x}\frac{\Lambda_{(\kk)}(n)}{\log_{(\kk)}(n)}\log^{\kk-1}(n_{(\kk)})\;.
\end{eqnarray}}
The proof is presented in appx. \ref{k-tuple proof}.
\end{theorem}

Define $\langle\x|\widehat{J}_{(\kk)}({\x})|0\rangle:= (-1)^{k-1}\mathcal{R}_{(\kk)}(1)\mathrm{Li}_{(\kk)}(\x)/2\pi i$. Then the $\kk$-tuple explicit integral \emph{(\ref{explicit integral})} together with the heuristic \emph{(\ref{prime counting})} from the single-prime case suggests to define
\begin{equation}
J_{(\kk)}(\x;r,i))
 :=\langle\x|
 \left[\int_{\mathcal{C}}\widehat{J}_{(\kk)}^{(i)'}({\x}^{s+r})
 \,d\log(\zeta_{(\kk)}(s))\right]|0\rangle
 \;\;\;\;\;\;c>k+r\;.
\end{equation}
To count weighted prime-power $\kk$-tuples, use $J_{(\kk)}(\x;0,0)$ which gives
\begin{equation}
J_{(\kk)}(\x)=\sum_{n\leq \x}\frac{\Lambda_{(\kk)}(n)}{\log_{(\kk)}(n)}
\end{equation}
after integration by parts.

Unfortunately, we don't have an explicit expression for $\zeta_{(\kk)}(s)$ so we can't use the integral to do exact counting via residues --- in contrast to the single-prime case. However, we can venture a guess:
\begin{conjecture}
\emph{
\begin{eqnarray}
\zeta_{(\kk)}(s)
\stackrel{?}{=}\sum_{x\in\mathrm{R}_\kk}\frac{1}{x_{(\kk)}^s}
\stackrel{?}{=}\prod_{p\in\mathrm{R}_\kk} \left(1-p_{(\kk)}^{-s}\right)^{-\kk}
\stackrel{??}{=}\prod_{p} \left(\frac{1-\nu_p(\mathcal{H}_\kk)\,p^{-s}}
{1-{p^{-s}}}\right)
\left(1-{p^{-s}}\right)^{-\kk}\;.
\end{eqnarray}
}
\end{conjecture}
Observe that $\zeta_{(\kk)}(1)\stackrel{??}{=}C_\kk\,\zeta(1)$ which supports the further conjecture that $\zeta_{(\kk)}(s)$ has a pole of order $\kk$ at $s=1$ since the pole in the single-prime case rendered average behaviour and $C_\mathrm{k}$ can be viewed as an average spectral density of admissible prime $k$-tuples on $\mathfrak{B}^{\x_{({\mathrm{{\mathrm{k}}}})}^{\mathrm{{\mathrm{k}}}}}$. If $\zeta_{(\kk)}(s)$ can be well defined according to this conjecture or otherwise, then knowledge of its zeros and poles would allow evaluation of the explicit integral \emph{(\ref{explicit integral})} and potentially have important implications in number theory.

\end{application}

 \subsection{Liouville integrators}\label{Liouville}
This subsection further develops the Liouville integrator (which was introduced in subsection \ref{Skew-Gaussian integrators}) beyond the K\"{a}hler case. We give the geometric setting, characterize the integrator, and indicate some of its general features.

Start with a smooth bundle $\mathcal{M}\rightarrow B\stackrel{\pi}{\rightarrow} \ZZZ$. The typical fiber $\mathcal{M}$ is a complex manifold of $\mathrm{dim}_\C(\mathcal{M})=m$ equipped with a Hermitian metric and Levi-Civita connection. The base space is a complex manifold $\ZZZ$ of $\mathrm{dim}_\C(\ZZZ)=n<m$ also equipped with a Hermitian metric. To simplify the exposition, assume $\p\ZZZ=\varnothing$.

With an eye towards field theory, construct the $G^\C$-jet-bundle $\mathcal{J}\rightarrow J^1(B)\stackrel{\tilde{\pi}}{\rightarrow}\ZZZ$. The typical fiber is the $1$-jet space $\mathcal{J}\equiv J^1_\z(\mathcal{Z},\mathcal{M})$ endowed with a symplectic form $\mathbf{\Omega}$. The structure group comprises complex symplectomorphisms $G^\C\equiv Sp(\mathcal{J},\C)$. Given a trivialization of $B$, local coordinates on $J^1(B)$ are $(\z^\mu,\x^I,\x^I_\mu)$ where $\mu\in\{1,\ldots,n\}$ and $I\in\{1,\ldots,m\}$. Associated with $J^1(B)$ is the principal bundle $G^\C\rightarrow \tilde{P}\stackrel{{\Pi}}{\rightarrow} \ZZZ$ assumed to be equipped with a metric compatible, principal Ehresmann connection $\chi$.

In general, we are interested in fields defined to be immersions $\Phi:\ZZZ\rightarrow \mathcal{J}$, and more specifically we will be interested in ``gauge invariant'' fields defined to be smooth emersions $\Phi_\|:\ZZZ\rightarrow \mathcal{J}$ that obey $\mathrm{i}_{\Bold{V}_\mathfrak{h}}d\Phi_\|=0$ (equivalently $\mathcal{L}_{\Bold{V}_\mathfrak{h}}d\Phi_\|=0$) whenever $\Bold{V}_{\mathfrak{h}}$ is an invariant vertical vector field  corresponding to $\mathfrak{h}\in\mathfrak{H}\subset\mathcal{G}\cong T_e G^C$ where $\mathfrak{H}$ is the Lie algebra of a symmetry subgroup $H^\C\subset
G^\C$. We say that $\Phi_\|(\mathcal{Z})$ is a ``gauge invariant'' submanifold. Given a local trivialization $\{{U}_i,\tilde{\varphi}_i\}$ of $\tilde{P}$ and its associated canonical section $\tilde{\varphi}_i\circ\tilde{s}_i:=Id_{\tilde{P}}$, a local gauge invariant field $\Phi_\|$ issues either from an equivariant map $\tilde{\Phi}_i:\tilde{P}\rightarrow \mathcal{J}$ via ${\Phi}_\|(\z)=(\tilde{\Phi}_i\circ\tilde{s}_i(\z)_\|)$ or as a section $\phi_i\in\Gamma(J^1(B))$ via ${\Phi}_\|(\z)=(\tilde{s}_i(\z)^{-1}_\|\circ\phi_i(\z))$ where $\tilde{s}_i(\cdot)_\|$ is \emph{horizontal} with respect to the restricted connection image $\chi(T\mathcal{Z})|_{\mathfrak{H}}$.\footnote{This follows because $p\in\widetilde{\pi}^{-1}(\z)$ is an admissible map: $\mathcal{J}\rightarrow \pi^{-1}(\z)$ so $p\circ\tilde{\Phi}(p)=\phi(\z)\in J^1(B)$.(see e.g. \cite[ch. V BIS.]{CB})}

To specify a classical field theory, we first need a notion of evolution which means we foliate the base manifold $\ZZZ={\Sigma}\times \mathbb{T}$ where $\mathrm{t}\in\mathbb{T}$ is a complex evolution parameter. Next, define a \emph{semi-basic} $1$-form on $\Phi_\|(\mathbb{\ZZZ})\subset \mathcal{J}$ by
\begin{equation}
\Bold{\omega}=\omega_I(\s^I,\s^I_\mu)d\x^I
:=\left[\int_{\Phi_\|({\Sigma})}\varpi_I^\mu(\x^I,\x_\mu^I)\; \Bold{\tau}(\Phi_\|({\Sigma}))_\mu\right]d\x^I
=\left[\int_{\Phi_\|({\Sigma})}\varpi_I(\x^I,\x_\mu^I)\cdot\; \Bold{\tau}(\Phi_\|({\Sigma}))\right]d\x^I
\end{equation}
where $\varpi^\mu_I(\x^I,\x_\mu^I)d\x^I\otimes\Bold{e}_\mu
=:\varpi^\mu_I(\x)d\x^I\otimes\Bold{e}_\mu$ is a vector-valued $1$-form with $\{\Bold{e}_\mu\}$ a basis in $T_\z\ZZZ$, and $\Bold{\tau}(\Phi_\|({\Sigma}))$ is the induced infinitesimal surface element on $\Phi_\|({\Sigma})$.

Use this $1$-form to construct an action functional $\mathrm{S}:\Gamma\Phi_\|\rightarrow\C_+$ where $\Gamma\Phi_\|$ denotes the space of  compactly supported, gauge invariant emersions $\Phi_\|:\mathcal{Z}\rightarrow\mathcal{J}$:
\begin{equation}\label{action}
\mathrm{S}(\Phi_\|)= \int_{\Phi_\|(\mathbb{T})}\Bold{\omega}
=\int_{\Phi_\|(\mathbb{T})}\int_{\Phi_\|({\Sigma})}\varpi_I(\x)\cdot\; \Bold{\tau}(\Phi_\|({\Sigma}))\wedge d\x^I
=:\int_{\ZZZ}\Phi_\|^\ast\Bold{\varpi}\,.
\end{equation}
with $\Bold{\varpi}:=\varpi_I\cdot\; \Bold{\tau}(\Phi_\|({\Sigma}))\wedge d\x^I$. Since $(\Phi_\|)'(\Bold{v}_\mathfrak{h})=0$, (here prime indicates a derivative map)
\begin{equation}
(\Phi_\|^\ast\varpi)^\mu(\Bold{\mathrm{t}})
=\varpi^\mu((\tilde{\Phi}\circ\tilde{s}_i(\z))_{\|}'(\Bold{\mathrm{t}}))
=\varpi^\mu(\tilde{\Phi}'(\Bold{v}_{\mathfrak{h}}))
=\varpi^\mu(D_{{H}}\tilde{\Phi}_{\|}(\Bold{v}_{\mathfrak{h}}))
\end{equation}
where $\Bold{\mathrm{t}}\in T_{\mathrm{t}}\mathbb{T}$ and $\Bold{v}_{\mathfrak{h}}:=\tilde{s}'_i(\z)_\|(\Bold{\mathrm{t}})\in T_\z\tilde{{P}}$ and $D_{{H}}$ is the covariant derivative induced by the symmetry group ${H}^\C$. Therefore, in local coordinates the action reads
\begin{equation}
\mathrm{S}(\Phi_\|)=\int_{\ZZZ}\varpi_I^\mu(\x(\z))
\left[\frac{\p\x^I}{\p\z^\mu}-(\Bold{V}_{(\mu)}\x(\z))^I\right]\,\Bold{\tau}_{\ZZZ}
=:\int_{\ZZZ}\left\{\varpi_I^\mu(\x(\z))
v_\mu^I-\mathrm{H}_{\Bold{V}}(\x(\z))\right\}\,\Bold{\tau}_{\ZZZ}
\end{equation}
with $\Bold{V}:=\{\Bold{V}_{(\mu)}\}$ the set of fundamental vector fields on $\mathcal{J}$ associated with the group action coming from the Lie subalgebra $\chi(\frac{\p}{\p\z^\mu})\in\mathfrak{H}$, the volume form is $\Bold{\tau}_{\ZZZ}$, and we have defined $\mathrm{H}_{\Bold{V}}(\x(\z)):=\varpi_I^\mu(\x(\z))(\Bold{V}_{(\mu)}\x(\z))^I$. There are no derivative terms for $\x_\mu^I$ because $\Bold{\omega}$ is semi-basic by definition.

Extend this action functional to non-gauge invariant emersions $\Gamma\Phi$, and instead impose gauge invariance according to
\begin{definition}
Let $\Bold{\omega}$ be an absolute integral invariant under a subgroup $H^\C\subset G^\C$. The action functional on $\Gamma\Phi$ is defined to be
\begin{equation}\label{horizontal action}
\mathrm{S}(\Phi)
:=\int_{\ZZZ}\left\{\varpi_I^{\Bold{\mu}}(\x(\z))
v_{\mu}^I-\mathrm{H}_{\Bold{V}}(\x(\z))\right\}\,\Bold{\tau}_{\ZZZ}
\end{equation}
where $\mathrm{H}_{\Bold{V}}(\x(\z)):=\varpi_I^\mu(\x(\z))(\Bold{V}_{(\mu)}\x(\z))^I$ can be viewed as an effective Hamiltonian density.
\end{definition}
Observe that $\mathrm{S}(\Phi)$ is manifestly covariant as it does not depend on the foliation, and it is gauge invariant by construction iff $\mathcal{L}_{\Bold{V}_{(\mu)}}\Bold{\omega}=0$. (Since $\Bold{\omega}$ is an absolute integral invariant by definition, then $\int_{\mathbb{T}}\left(\varphi_{\Bold{V}}\circ\Phi\right)^\ast\Bold{\omega}
=\int_{\mathbb{T}}\Phi^\ast\Bold{\omega}$ where $\varphi_{\Bold{V}}$ is the flow generated by some $\Bold{V}_{(\mu)}$ and therefore $\mathcal{L}_{\Bold{V}_{(\mu)}}\Bold{\omega}=0$.)

\begin{application}\emph{Covariant field equations}\label{covariant Field Equations}

View $\mathcal{M}$ as a $\mathrm{dim}(\mathcal{M})_\R=2m$ real Riemannian manifold and give $\mathcal{J}$ canonical/Darboux coordinates $(\phi^I,\pi^I,\phi^I_\mu,\pi^I_\mu)$. Suppose the action takes the special form
\begin{equation}\label{simple action}
\mathrm{S}(\Phi)
=\int_{\ZZZ}\left\{\pi_I^{\Bold{\mu}}(\z)
\phi^I_{,\,\mu}(\z)
-\mathrm{H}_{\Bold{V}}(\phi^I(\z),
\pi_\mu^I(\z))\right\}\Bold{\tau}_{\ZZZ}\;.
\end{equation}
Covariant variation of the action
with respect to the Levi-Civita connection on $\mathcal{M}$ pulled back to $\ZZZ$ via $\Phi$ yields covariant Hamilton's field equations
\begin{equation}
\phi^I_{,\,\mu}=\dfrac{\p\mathrm{H}_{\Bold{V}}}{\p\pi_I^{\mu}}
\;\;\;\;\;\;\;\;\;\;\;\;
{\pi_I^{\mu}}_{;\,\mu}= -\dfrac{\p\mathrm{H}_{\Bold{V}}}{\p\phi^I}
\end{equation}
where ${\pi_I^{\mu}}=g^{\mu\nu}\pi_\nu^J g_{JI}$.

The second variation gives the Jacobi quadratic form coming from the Jacobi operator applied to a variational vector field $\Bold\zeta\in T_{\Phi}\Gamma\Phi\cong\Gamma(\Phi^\ast T\mathcal{J})$ with $\Bold\zeta(\z)=:(\Bold{\xi}(\z),\Bold{\eta}(\z))\in T_{\Phi(\z)}\mathcal{J}$;
\begin{equation}\label{Jacobi}
\left({\xi}^I,\eta_I^\nu\right)
\left(
   \begin{array}{cc}
    \pi_L^{\mu} \tensor{R}{^L_{KIJ}}\phi^K_{\,,\,\mu} -\dfrac{\p^2\mathrm{H}_{\Bold{v}}}{\p\phi^I \p\phi^J}
    \;\;\;\; &J'^J_I D_\mu-\dfrac{\p^2\mathrm{H}_{\Bold{v}}}{\p\phi^I \p\pi_J^\mu} \\ \\
     -J_J^I D_{\,\nu}-\dfrac{\p^2\mathrm{H}_{\Bold{v}}}{\p\pi_I^\nu \p\phi^J}
    \;\;\;\; & -\dfrac{\p^2\mathrm{H}_{\Bold{v}}}{\p\pi_I^\nu \p\pi_J^\mu} \\
   \end{array}
 \right)
\left(
    \begin{array}{c}
        {\xi}^J \\
\eta^\mu_J\\
    \end{array}
\right)
 \end{equation}
where the pair $(\xi^J,\eta^J_\mu)$ is the variation of the conjugate pair $(\phi^J,\pi^J_\mu)$, $\tensor{R}{^L_{KIJ}}$ are components of the curvature tensor on $\mathcal{M}$, and $D_\mu$ is the covariant derivative pulled back to $\ZZZ$.\emph{\cite{LA6}}

1) For example, consider a scalar field of mass $m$ on a real Riemannian manifold $\M^4\subset\ZZZ$ such that $B=\M^4\times\C$ and  $\mathrm{H}_{\Bold{V}}=1/2(g_{\mu\nu}\pi^\nu\pi^\mu+m^2\phi^2)$. The field equations are $\phi_{,\,\mu}=g_{\mu\nu}\pi_\nu$ and $\pi^\mu_{\,;\,\mu}=-m^2\phi$. Together they yield the expected scalar wave equation $(\Delta+m^2)\phi=0$ where $\Delta=g^{\mu\nu} D_\mu D_\nu$. Meanwhile, the Jacobi equation for the variational fields $(\xi,\eta^\mu)$ gives $\xi_{\,,\,\mu}=g_{\mu\nu}\eta^\nu$ and $\eta^\mu_{\;;\,\mu}=-m^2\xi$ which also happens to yield a wave equation $(\Delta+m^2)\xi=0$. In fact, one can show that Hamilton's equations for fields are the same as the Jacobi equation for the variational vector fields when $\mathrm{H}_{\Bold{V}}$ is quadratic in the fields.

2) As a second example, now on $\R^{3,1}\subset\ZZZ$, suppose the index $I=(i,\mu)$ comprises a vector index $i$ and a one form index $\mu$ such that $\phi^I\sim A^i_\mu$ represents a massless vector-valued one form in the adjoint representation of $H^\C$ and $\pi_I^\nu\sim F^{\mu\nu}_i$ represents a two form in the adjoint representation. In other words, as a manifold $[\mathcal{M}]\cong T_\z^\ast\R^{3,1}\times T_e H^\C$. Since $\mathcal{L}_{\Bold{V}_{(\mu)}}\Bold{\omega}=0$, the conjugate pair $(A,F)$ can be used to construct a gauge-invariant action, and we expect the group algebra $\mathfrak{H}$ is the direct sum of a gauge Lie algebra and the generator of time evolution.

Gauge flow is generated by $[A_\mu,A_\nu]$, and the vector field generating the flow associated with time evolution derives from the energy density $\mathrm{tr}\left(\star F\wedge F\right)$ of momentum-type fields; that is from $F_i^{\mu\nu}$. Then the group action induces vector fields $\Bold{V}_{(\mu)}= ([A_\mu,A_\nu]^i+F_{\mu\nu}^i)\Bold{e}_{A^i_\nu}+\mathrm{more}$ where the index $i\in\{1,2,\ldots,\mathrm{dim}(H^\C)\}$ and the ``more'' term (which is in the $\Bold{e}_{F_i^{\mu\nu}}$ direction) won't contribute because $\Bold{\varpi}$ is semi-basic. The Hamiltonian is $\mathrm{H}_{\Bold{V}}=F_i^{\mu\nu}([A_\mu,A_\nu]^i+ F_{\mu\nu}^i)$ and the action is
\begin{equation}\label{Yang-Mills}
\mathrm{S}(A,F)
=\int_{\M^4}\left\{F_i^{\mu\nu}\p_{[\,\mu}A^i_{\nu]}
-\tfrac{1}{2}F_i^{\mu\nu}\left(c^i_{jk}A_{[\,\mu}^jA_{\nu]}^k+ F^i_{\mu\nu}\right)\right\}\,\Bold{\tau}_{\M^4}
\end{equation}
where $c^i_{jk}$ are the structure constants of the gauge group. The field equations are\footnote{We get both the Yang-Mills equation \emph{and} the curvature relation $F=DA$ (and hence $DF=0$) but only on the subspace of classical solutions in $\Gamma\Phi_\|\,$; just as $p\propto v=dq/dt$ holds only classically since $p,\dot{q}$ do not generally commute quantum mechanically.}
\begin{equation}
A^i_{[\,\mu,\,\nu]}=c^i_{jk} A^j_\mu A^k_\nu+{F}^i_{\mu\nu}
\;\;\;\;\;\;\;\;\tensor{F}{^{i\mu\nu}_{,\,\mu}}=c^i_{jk}A_\mu^j{F}^{k\mu\nu}\;.
\end{equation}

\end{application}

Motivation aside, we posit the action functional of (\ref{horizontal action}) describes a classical field theory. Compare its Jacobi operator with (\ref{supersymmetric form}): they describe similar dynamical systems. Moreover, the classical fields and classical variational fields obey the same equations of motion for quadratic $\mathrm{H}_{\Bold{V}}$.  So the idea is to define a Liouville integrator using the linearized version of (\ref{horizontal action}); that is, from its Jacobi operator applied to fields in $T_\Phi\Gamma\Phi$.

Of course, in deriving the Jacobi operator in the above example we took a real cotangent bundle, but the goal is to formulate everything for complex fields. Accordingly, rerunning the variations for a complex $\mathcal{M}$ and general $\varpi^{\Bold{\mu}}_I(\x(\z))$ will give both even-grade and odd-grade variational fields $w^J,w_J^\mu\in T_\Phi\Gamma\Phi$ along with a sesquilinear Jacobi operator that decomposes into a Hermitian $\mathrm{Q}$ and a skew-Hermitian $\Omega$. As the fields in $T_\Phi\Gamma\Phi$ are now variational, the skew-Hermitian form $\Omega$ issues from the symplectic form $\mathbf{\Omega}$.

We are ready to define the Liouville integrator. For general manifolds $\ZZZ$ and $\mathcal{M}$, the space of smooth emersions $\Gamma\Phi$ will not be a topological group. However, at any $\Phi\in\Gamma\Phi$, the tangent space $T_\Phi\Gamma\Phi\cong\Gamma(\Phi^\ast T\mathcal{J})$ is a point-wise abelian group that can be used to define a functional integral and subsequently to parametrize $\Gamma\Phi$ (see Application \ref{CDM for fields}). We can summarize the relevant geometric structures by the pull-back diagram
\[\begin{tikzcd}
\Phi^\ast T\mathcal{J} \arrow{r}{\mathrm{pr}} \arrow[swap]{d}{\pi^\ast} & T\mathcal{J} \arrow{d}{\pi} \\
 \mathcal{Z}\arrow{r}{\Phi} & \mathcal{J}
\end{tikzcd}
\]
 where both bundles have structure group $G^\C=Sp(T_{\Phi(\z)}\mathcal{J},\C)$ and typical fiber $T_{\Phi(\z)}\mathcal{J}$ which furnishes a representation of $G^\C$. Recall $\mathcal{J}$ is endowed with a gauge invariant $1$-form $\Bold{\omega}$ and a symplectic form $\mathbf{\Omega}$. There is an associated principal bundle $G^\C\rightarrow \tilde{P}^\ast\stackrel{\Pi^\ast}{\rightarrow}\ZZZ$ equipped with a metric compatible connection.

Endowed with a suitable topology, $\Gamma(\Phi^\ast T\mathcal{J})$ is an abelian topological group.  As in the Gaussian/skew-Gaussian case, $\Gamma(\Phi^\ast T\mathcal{J})=:W$ inherits a complex structure from $\mathcal{M}$. The induced complex structure then leads to a $\mathbb{Z}_2$-grading $W=:W^{\mathbf{e}}\oplus W^{\mathbf{o}}$. It is useful to think of $W$ as an extended quantum phase space containing both even-grade and odd-grade fields and their conjugate momenta. This will be the domain of the Liouville integrator family.

\begin{definition}
Given the geometric constructions above, the Liouville family of integrators $\mathcal{D}_\Lambda l_{\bar{w}^{\mathbf{e}},\mathrm{Q},\,\Omega}(w^{\mathbf{e}})$ with domain $W^{\mathbf{e}}$ is characterized by
\begin{eqnarray}
&&\Theta_{\bar{w}^{\mathbf{e}},\mathrm{Q},\,\Omega}(w^{\mathbf{e}})
=\mathrm{Pf}(\Omega/\hat{\s})e^{-(\pi/\check{\s})\mathrm{Q}(w^{\mathbf{e}}-\bar{w}^{\mathbf{e}})}\notag\\
&&\mathrm{Z}_{\bar{w}^{\mathbf{e}},\mathrm{Q},\,\Omega}=\mathrm{Vol}^{1/2}_\lambda\left(W^{\mathbf{o}}\,/\,W^{\mathbf{e}}\right)
_{\mathrm{Q},\,\Omega}
\end{eqnarray}
where $w^{\mathbf{e}}\in W^{\mathbf{e}}$ is an even-grade field on $\ZZZ$, and $\mathrm{Vol}(\cdot)$ is interpreted (formally) as a normalized volume functional  that localizes (via $\lambda$) to a well-defined Pfaffian/determinant ratio determined by quadratic forms $\Omega,\,\mathrm{Q}$ whose ranks are not necessarily equal. As usual, for the scaling parameters,  $\hat{\s},\check{\s}\in\C_+$. There is no boundary term since we assume $\p\mathcal{Z}=\varnothing$.
\end{definition}

The Liouville family is defined in terms of the primitive integrators $\mathcal{D}_\lambda w^{\mathbf{e}}$ and $\mathcal{D}_\lambda w^{\mathbf{o}}$ by
\begin{equation}
\mathcal{D}_\lambda l_{\bar{w}^{\mathbf{e}},\mathrm{Q},\,\Omega}(w^{\mathbf{e}})
:=\mathrm{Pf}(\Omega/\hat{\s})
\,e^{-(\pi/\check{\s})\mathrm{Q}(w^{\mathbf{e}}-\bar{w}^{\mathbf{e}})}
\mathcal{D}_\lambda w^{\mathbf{e}}
\end{equation}
with
\begin{equation}
 \mathrm{Pf}_\lambda(\Omega/\hat{\s}):=\int_{W^{\mathbf{o}}}
 e^{-\pi\hat{\s}\,\Omega(w^{\mathbf{o}}-\bar{w}^{\mathbf{o}})}\;\mathcal{D}_\lambda w^{\mathbf{o}}
\end{equation}
where $w^{\mathbf{o}}\in W^{\mathbf{o}}$ is an odd-grade field. Restoring the Pfaffian to integral form gives
\begin{equation}\label{restored integral}
\int_{W}
 \!\!e^{-(\pi/\check{\s})\mathrm{Q}(w^{\mathbf{e}}-\bar{w}^{\mathbf{e}})
 -\pi\hat{\s}\,\Omega(w^{\mathbf{o}}-\bar{w}^{\mathbf{o}})}
 \;\mathcal{D}_\lambda (w^{\mathbf{e}},w^{\mathbf{o}})
 =\mathrm{Vol}^{1/2}_\lambda\left(W^{\mathbf{o}}\,/\,W^{\mathbf{e}}\right)_{\mathrm{Q},\Omega}\,.
\end{equation}
The integral representation over the full space $W$ helps to illuminate the volume tag: since $\mathrm{Vol}^{1/2}(W^{\mathbf{o}}\,/\,W^{\mathbf{e}})
\sim\mathrm{Pf}(\Omega/\hat{\s})\,/\,
\sqrt{\mathrm{Det}(\mathrm{Q}/\check{\s})}\,$, then for the K\"{a}hler case  $\Omega\circ J=\mathrm{Q}$ if we identify the scaling parameters $\check{\s}\equiv\hat{\s}$, we get $\left|\mathrm{Vol}^{1/2}_\lambda(W^{\mathbf{o}}\,/\,W^{\mathbf{e}})\right|\sim \left|\mathrm{pf}(\mathrm{Q})/\sqrt{\det(\mathrm{Q})}\right|$ --- indicating even-odd balance.

\begin{remark}
The definition of Liouville still makes sense (with suitable modifications) if $W^{\mathbf{e},\mathbf{o}}$ are non-abelian topological groups. In that case, Liouville suggests a hand-waiving characterization of non-abelian Fourier transform: Since $W^{\mathbf{e}}$ represents an even-grade quantum phase space, quantum fields will inhabit a Lagrangian subspace $L\subset W^{\mathbf{e}}$ relative to the symplectic form on $W^{\mathbf{e}}$ induced from $\mathbf{\Omega}$. Evidently then, Fourier duals can be regarded as orthogonal Lagrangian subspaces $(L,L')$ in $W^{\mathbf{e}}$, and \textbf{non-abelian Fourier transform} would correspond to the function $\int_{W^{\mathbf{e}}_{\hat{\bar{w}}^{\mathbf{e}}}}\delta(L)\;\mathcal{D}_\lambda l_{\bar{w}^{\mathbf{e}},\mathrm{Q},\,\Omega}(w^{\mathbf{e}})
\mapsto \int_{W^{\mathbf{e}}_{\hat{\bar{w}}^{\mathbf{e}}}}\delta(L')\;\mathcal{D}_\lambda l_{\bar{w}^{\mathbf{e}},\mathrm{Q},\,\Omega}(w^{\mathbf{e}})$.
 \end{remark}

\begin{application}\emph{Berline-Vergne and Duistermaat-Heckman localization:}

By taking different limits of $\,\check{\s}$, the Liouville family will be shown to be consistent with the Berline-Vergne theorem\emph{\cite{BV}} and, with suitable specialization, the Duistermaat-Heckman theorem\emph{\cite{DH}}.

To get Berline-Vergne, construct the space $W^{\mathbf{e}}:=\bigsqcup_{\bar{w}^{\mathbf{e}}(\z)}W^{\mathbf{e}}_{\hat{\bar{w}}^{\mathbf{e}}}$ where $W^{\mathbf{e}}_{\hat{\bar{w}}^{\mathbf{e}}}:=W^{\mathbf{e}}\backslash\mathrm{Ker}(\Gamma')$ and $\bar{w}^{\mathbf{e}}$ is a critical point of the effective action\footnote{Since $\mathrm{Q}$ is quadratic, the effective action $\Gamma$ coincides with $\mathrm{Q}$ and so $\bar{w}^{\mathbf{e}}=w^{\mathbf{e}}_{cr}$ where $Dw^{\mathbf{e}}_{cr}=0$.} associated with $\mathrm{Q}$. Consider the integral
\begin{equation}\label{Pfaffian integral}
\int_{W^{\mathbf{e}}_{\hat{\bar{w}}^{\mathbf{e}}}}
 \mathrm{Det}_\lambda(\mathrm{Q}/\check{\s})^{1/2}
 \;\mathcal{D}_\lambda l_{\bar{w}^{\mathbf{e}},\mathrm{Q},\,\Omega}(w^{\mathbf{e}})
\end{equation}
where $\mathcal{D}_\lambda l_{\bar{w}^{\mathbf{e}},\mathrm{Q},\,\Omega}(w^{\mathbf{e}})
:=\mathrm{Pf}(\Omega/\hat{\s})e^{-(\pi/\check{\s})\mathrm{Q}(w^{\mathbf{e}}-\bar{{w}}^{\mathbf{e}})}
\mathcal{D}_\lambda w^{\mathbf{e}}$. Taking the limit $|\check{\s}|\rightarrow0$ with $\hat{\s}=1$ gives
\begin{eqnarray}
\lim_{|\check{\s}|\rightarrow0}\int_{W^{\mathbf{e}}_{\hat{\bar{w}}^{\mathbf{e}}}}
 \mathrm{Det}_\lambda(\mathrm{Q}/\check{\s})^{1/2}\;\mathcal{D}_\lambda l_{\bar{w}^{\mathbf{e}},\mathrm{Q},\,\Omega}(w^{\mathbf{e}})
 &=:&\int_{W^{\mathbf{e}}_{\hat{\bar{w}}^{\mathbf{e}}}}\mathcal{D}_\lambda l_{\bar{w}^{\mathbf{e}},\mathrm{Q},\,\Omega}(\delta_{\bar{w}^{\mathbf{e}}}(w^{\mathbf{e}}))\;.
\end{eqnarray}
We can account for the non-trivial zero modes of operator $D$ by translation on ${W^{\mathbf{e}}_{\hat{\bar{w}}^{\mathbf{e}}}}$ to get
\begin{equation}
\int_{W^{\mathbf{e}}_{0}}\mathcal{D}_\lambda l_{0,\mathrm{Q},\,\Omega}(\delta_0(w^{\mathbf{e}}))
=\mathrm{Pf}_\lambda({\Omega})\;.
\end{equation}
Then, since $\bar{w}^{\mathbf{e}}$ is a critical point it represents a horizontal lift with initial point $(\z,0)$; which means $\bar{w}^{\mathbf{e}}$ is the constant function $\bar{w}^{\mathbf{e}}(\z)=\z$. Therefore, with $\mathrm{Pf}_\lambda= \mathrm{pf}$,
\begin{equation}
 \int_{W^{\mathbf{e}}}\mathcal{D}_\lambda l_{\bar{w}^{\mathbf{e}},\mathrm{Q},\,\Omega}(\delta_{\bar{w}^{\mathbf{e}}}(w^{\mathbf{e}}))
 =\int\!\!\!\!\!\!\!\!\sum_{\{\bar{w}^{\mathbf{e}}\}}\int_{W^{\mathbf{e}}_{0}}\mathcal{D}_\lambda l_{0,\mathrm{Q},\,\Omega}(\delta_0(w^{\mathbf{e}}))
=\int_{\ZZZ}\mathrm{pf}({\Omega})
\end{equation}
which expresses the localization $W^{\mathbf{e}}|_{\{\bar{w}^{\mathbf{e}}\}}\cong \ZZZ$ for horizontal fields.

More generally,  one can localize $W^{\mathbf{e}}$ onto the zero locus of a suitable endomorphism $\mathrm{L}:W^{\mathbf{e}}\rightarrow W^{\mathbf{e}}$ in which case \emph{(\ref{delta integrator})} yields (assuming isolated zeros of $\left(\mathrm{L}w^{\mathbf{e}}\right)(\z):=\mathrm{L}_\lambda(w^{\mathbf{e}}(\z))$ such that $\mathrm{L}_\lambda$ is an isometry with non-degenerate derivative map $\mathrm{L}'_\lambda$)
\begin{equation}\label{localization}
\int\!\!\!\!\!\!\!\!
\sum_{\{\bar{w}^{\mathbf{e}}\}}\int_{W^{\mathbf{e}}_{0}}F(w^{\mathbf{e}})\mathcal{D}_\lambda l_{0,\widetilde{\mathrm{S}},\,\Omega}(\delta(\mathrm{L}w^{\mathbf{e}}))
=\sum_{\{w^{\mathbf{e}}_o(\z)\}}F(w^{\mathbf{e}}_o(\z))\frac{
\mathrm{Pf}_\lambda({\Omega})}{\mathrm{Det}_\lambda \mathrm{L}'_{w^{\mathbf{e}}_o}}
\end{equation}
where $\widetilde{\mathrm{S}}:=\mathrm{Q}\circ \mathrm{L}$, and $w^{\mathbf{e}}_o(\z)\in\mathrm{Ker}(\mathrm{L}_\lambda)$. In particular, restricting to an equivalence class of isometries such that $\mathrm{L}'_\lambda\sim{{\Omega}}\circ J$ with equivalence relation relative to symplectic transformations, \emph{(\ref{localization})} reduces to
\begin{equation}
\int\!\!\!\!\!\!\!\!
\sum_{\{\bar{w}^{\mathbf{e}}\}}\int_{W^{\mathbf{e}}_{0}}F(w^{\mathbf{e}})\mathcal{D}_\lambda l_{0,\widetilde{\mathrm{S}}}(\delta(\mathrm{L}w^{\mathbf{e}}))
=\sum_{\{w^{\mathbf{e}}_o(\z)\}}\frac{F(w^{\mathbf{e}}_o(\z))}
{\mathrm{pf} (\mathrm{L}'_{w^{\mathbf{e}}_o(\z)})}\;.
\end{equation}
This can be viewed as the functional integral analog of Berline-Vergne localization. From here,  one could use the Berline-Vergne theorem to show that, for a suitable choice of $\lambda$, the left-hand side is a localized Liouville functional integral given by $\int_{\ZZZ}\Bold{F}_{\Bold{\tau}}$ where $\Bold{F}_{\Bold{\tau}}$ is an equivariantly closed top-form on $\ZZZ$.

Alternatively, return to \emph{(\ref{Pfaffian integral})} and instead take the limit $|\check{\s}|\rightarrow\infty$ to get (still with $\hat{\s}=1$)
\begin{equation}
\int_{W^{\mathbf{o}}_0}
 e^{-\pi\,\Omega(w^{\mathbf{o}}-\bar{w}^{\mathbf{o}})}
 \;\mathcal{D}_\lambda w^{\mathbf{o}}
 =\mathrm{Pf}_\lambda(\Omega)\;.
\end{equation}
To make use of this, include some integrable $F(w^{\mathbf{e}})$ in \emph{(\ref{Pfaffian integral})}, localize by\footnote{Technically, as a convolution algebra, $\mathbf{F}(G)$ doesn't contain an identity, but it does possess an approximate identity. In other words, here $\lambda$ represents a net in $\mathbf{F}(G)$.} $\lambda:w^{\mathbf{e}}\rightarrow id$ so that $w^{\mathbf{e}}(\z)=\z$,  and take the limit $|\check{\s}|\rightarrow\infty$ to get
\begin{equation}
\int\!\!\!\!\!\!\!\!\sum_{\{\bar{w}^{\mathbf{e}}\}}\int_{W_{0}}F(w^{\mathbf{e}})
 e^{-\pi\,{\Omega}(w^{\mathbf{o}})}
 \;\mathcal{D}_\lambda (w^{\mathbf{e}},w^{\mathbf{o}})
 =\int_{\ZZZ} F(\z)\mathrm{pf}({\Omega})
 =:\int_{\ZZZ}\Bold{F}(w^{\mathbf{e}})(\z)\;.
\end{equation}

Duistermaat-Heckman is just a few short steps away. Assume that $ G^\C$ furnishes a Hamiltonian action on $\mathcal{J}$, and  let $\Bold{V}_{\!\!\!g}\in T\mathcal{J}$ represent the vector field induced by the action of some $g\in G^\C$. Choose a fixed element $\mu\in T^\ast\mathcal{J}$ that generates a moment(um) map by
\begin{equation}
H_{\Bold{V}_{\!\!\!g}}(\z):=\langle\mu(\z),\Bold{V}_{\!\!\!g}(\z)\rangle\;.
\end{equation}
Given the $2$-form $\mathbf{\Omega}$ and its associated equivalence class of isometries $\mathrm{L}_\lambda$, choose (if it exists) a suitable $\mu$ satisfying
\begin{equation}\label{momentum map}
d\langle\mu(\z),(\mathrm{L}w^{\mathbf{e}})'(\z)\rangle=\mathrm{i}_{(\mathrm{L}w^{\mathbf{e}})'(\z)}{\Omega}
\end{equation}
where the right-hand side denotes the interior product. Restore the scaling $\hat{\s}$ and take
\begin{equation}
 \Bold{F}(w^{\mathbf{e}})(\z)=e^{-\pi\hat{\s}\, \langle\mu,(\mathrm{L}w^{\mathbf{e}})(\z)'\rangle}=e^{-\pi \langle\mu,\,\hat{\s}\,(\mathrm{L}w^{\mathbf{e}})'(\z)\rangle}\;.
\end{equation}
It turns out that $\Bold{F}(w^{\mathbf{e}})(\z)$ is an equivariantly closed top-form.\emph{\cite{DH}} Since $\mathrm{dim}_\R(\ZZZ)=2n$, we verify Duistermaat-Heckman (assuming $\check{\s},\hat{\s}$-independence, isolated zeros of $\mathrm{L}w^{\mathbf{e}}$, and using shorthand notation for the Pfaffian ratio)
 \begin{equation}
 \int_{\ZZZ} \Bold{F}(w^{\mathbf{e}})(\z)=\sum_{\{w^{\mathbf{e}}_o(\z)\}}\frac{e^{-\pi \langle\mu,\,\hat{\s}\,(\mathrm{L}w^{\mathbf{e}}_o)'(\z)\rangle}}{\mathrm{pf} (\hat{\s}\,\mathrm{L}'_{w^{\mathbf{e}}_o(\z)})}
 =\hat{\s}^{-n}\sum_{\{w^{\mathbf{e}}_o(\z)\}}\frac{e^{-\pi\hat{\s}\, \langle\mu,\,(\mathrm{L}w^{\mathbf{e}}_o)'(\z)\rangle}}{\mathrm{pf} (\mathrm{L}'_{w^{\mathbf{e}}_o(\z)})}\;.
 \end{equation}

The localization displayed by the Duistermaat-Heckman theorem hinges on three conditions: i) the integrals do not depend on $\check{\s}$ and $\hat{\s}$; ii) up to symplectomorphisms, $\mathrm{L}'_\lambda\sim{\Omega}\circ J$; and iii) there exists a momentum map $\mu$ such that all three objects $\mathrm{L}'_\lambda$, ${\Omega}$, and $\mu$ are related through \emph{(\ref{momentum map})} for all $\z\in\ZZZ$.
\end{application}

Although functional Liouville was defined over $W^{\mathbf{e}}$, reflecting a bias towards correlation phase space variables over dynamical phase space variables, the underlying integral over the full space $W$ is more flexible and probably more fundamental. The point is, not surprisingly, some applications are better understood from the $W^{\mathbf{e}}$ perspective and some from $W^{\mathbf{o}}$.  We should stress that Liouville is defined for $T_\Phi\Gamma\Phi$ an abelian topological group and quadratic $\mathrm{Q},\Omega$. Of course, if the action functional is non-quadratic, one must resort to perturbation of the Hamiltonian.

With perturbation theory in mind, define a non-quadratic Liouville functional integral in terms of a non-quadratic action functional $\mathrm{S}$ on a \emph{non-linear} space of fields $\mathcal{N}\Phi$ that has been parametrized by a suitable topological group $\mathcal{G}$ (see e.g. \cite{LA3} for the abelian topological group case);
\begin{eqnarray}\label{non-linear Liouville}
\int_{\mathcal{N}\Phi^+\times\mathcal{N}\Phi^-}
e^{-\pi\mathrm{S}(\Phi-\bar{\Phi})}
 \;\mathcal{D}_\lambda \Phi
 :=\int_{\mathcal{G}^+\times\mathcal{G}^-}
 e^{-\pi\tilde{\mathrm{S}}(\Theta-\bar{\Theta})}
 \;\mathcal{D}_\lambda \Theta
 &:=&\mathrm{Vol}^{1/2}_\lambda(\mathcal{G}^-/\mathcal{G}^+)
 _{\tilde{\mathrm{S}}}\notag\\
 &=:&\mathrm{Vol}^{1/2}_\lambda(\mathcal{N}\Phi^-/\mathcal{N}\Phi^+)
 _{{\mathrm{S}}}\notag\\
\end{eqnarray}
where the parametrization is a map $P: \mathcal{G}\rightarrow\mathcal{N}\Phi$ and $\tilde{\mathrm{S}}=\mathrm{S}\circ P$. Actually, this characterization is cheating a bit because $\mathrm{Vol}^{1/2}_\lambda\left(\mathcal{G}^-/\mathcal{G}^+\right)_{\tilde{\mathrm{S}}}$ is not really defined. According to current technology, the right-hand side must be expressed as a perturbation about $\bar{\Theta}$ in terms of the Liouville integrator: it must be understood in a linear approximation as the ratio of two determinants of quadratic operators on an abelian topological group updated at each order of perturbation.

But even in the free-field case where $\tilde{\mathrm{S}}$ is quadratic, the functional integrals over $\mathcal{N}\Phi$ and $\mathcal{G}$ don't agree in general, because the parametrization won't capture the zero modes in $\mathcal{N}\Phi$ (unless $\mathcal{N}\Phi$ is a non-abelian group). This clearly demonstrates the limitation of perturbation theory: we are trying to represent a QFT over a non-linear space of fields and a  non-linear Hamiltonian by a QFT with fields on a group manifold and a quadratic Hamiltonian that is updated at each order of perturbation.

\section{Conclusion}
A common property seemingly possessed by \emph{useful} functional integrals is that they eventually reduce to a set of bona fide integrals due to some form of localization in the domain of integration. In other words, an induced localization in the infinite-dimensional functional-integral domain leads to a quantifiable/measurable object or objects that can be associated with the original functional integral. Promeasures in particular exemplify this observation.

Such a view suggests defining functional integrals on topological groups in terms of families of well-defined Haar measures on locally-compact topological groups. Further, the rigorous integration theory on locally compact groups allows to consider Banach-valued integrable functions. So, we construct a space of Banach-valued functionals on a topological group that inherits any algebraic structure possessed by the target Banach space; and vice versa. With this approach, functional integrals serve a dual purpose: i) they are a tool to formally manipulate a Banach algebra of integrable functionals on a topological group; and ii) given a `topological localization', they become bona fide  measure-theoretic objects. The second point in particular offers an interpretation of the measurement process in quantum theory (see Remark \ref{measurement}).

Of course the nature of the Banach algebra of functionals $\mathbf{F}(G)$ and its image under integral operators depend on the underlying topological group and its associated family of Haar measures. This immediately leads one to contemplate functional integrals beyond the familiar Gaussian type. Being obvious infinite-dimensional counterparts to finite-dimensional integrals, one can develop functional integrators from useful finite-dimensional integrals. In this paper we detailed skew-Gaussian, Liouville, gamma, and Poisson functional integrals as specific examples and gave some particular applications of each relevant to physics and mathematics. Significantly, the skew-Gaussian and Liouville integrators can  generate exterior algebras of functionals. Their construction exposes inherent SUSY in the context of sesquilinear forms on complex topological vector spaces, but it suggests an unorthodox physical interpretation of SUSY based on a correlation/dynamics duality that is reminiscent of (and perhaps related to) wave/particle duality. In appendix D, we presented evidence of BRST symmetry and strong-weak duality as a simple consequence of the extended quantum phase space domain of Liouville. As for non-quadratic integrators, gamma and Poisson-type integrators yield an interesting physical model for assigning ordinals to primes. Although we focused on prime \emph{numbers} here, the approach seems applicable to physical models that count prime \emph{cycles} and their $\kk$-fold correlations in the context of spectral determinants and dynamical zeta functions.

One aspect of these integrators that we ignored (with the exception of translation invariance of the Gaussian integrator and the rather sketchy discussions in section \ref{Liouville is a tool}) is their invariance under relevant symmetries of the domains of integration. This is of particular importance and will be explored in future work. Also, for the most part we merely offered functional integral representations of some pertinent objects: The more difficult task of solving non-trivial integrals remains, but hopefully the framework proposed here, augmented with a study of symmetry properties of integrators, will enable reliable manipulations and perhaps suggest new solution strategies.

Although our pragmatic definition of functional integrals could be fairly characterized as mere bookkeeping, it does provide an uncluttered landscape by which to envision and construct new functional integral tools that should be of value in mathematical physics. More importantly, it provides a bridge between the concrete and abstract Banach $\ast$-algebras $\mathfrak{B}$ and $\mathbf{F}(G)$ respectively. As such, it combines operator and functional integral methods under one roof. In particular, one can characterize a $C^\ast$-algebra representing a quantum system through a family of integrators taking values in $L_B(\mathcal{H})$. In this regard, beyond what has been presented here, we have in mind generalizing gamma integrators in the context of functional Mellin transforms. Functional Mellin transforms are useful for both constructing and representing (quantum) $C^\ast$-algebras. Hence, one can quantize from the top down: That is, instead of quantizing a commutative algebra of functions on some phase space, one identifies a pertinent topological group and constructs a non-commutative algebra directly using the machinery of functional Mellin transforms. We develop functional Mellin transforms in a companion paper.

\appendix
\section{Preliminary mathematics}\label{topological groups}
The key mathematical components of the proposed definition are topological groups and Banach-valued integration. In this appendix, we present relevant definitions and theorems. (Proofs of theorems in this appendix will be omitted since they are not particularly germane to our development. The proofs, and much more, can be found in \cite{HM}--\cite{BNEEB}.)
\begin{definition}

A Hausdorff topological group $G$ is a group endowed with a topology
such that; (i) multiplication $G\times G\rightarrow G$ by
$(g,h)\mapsto gh$ and inversion $G\rightarrow G$ by $g\mapsto
g^{-1}$ are continuous maps, and (ii) $\{e\}$ is closed.
\end{definition}
Remark that the closure hypothesis on $\{e\}$ together with the
topology and group structure allow the closure property to be
`transported' to every element in $G$. That is, $G$ is Hausdorff iff
$\{e\}$ is closed. Moreover, since $G$ is Hausdorff, it is locally
compact iff every $g\in G$ possesses a compact neighborhood.

Clearly, there is not a lot of structure imposed on the group to work with at this point.  The most basic but perhaps most useful tool to probe topological groups at this level of generality is the exponential map. The motivation for the following definition comes from analogy with the exponential map for finite-dimensional Lie groups.
\begin{definition}\emph{(\cite[defs. 5.7, 5.39]{HM})} A
one-parameter subgroup ${\phi}:\R\rightarrow G$ of a topological group is the unique extension of a continuous
homomorphism $f\in
\mathrm{Hom}_C(I\subseteq\R,G)$ such that $f(t+s)=f(t)f(s)$ and $f(0)=e\in G$. Let $\mathfrak{L}(G)$ denote the set
of all one-parameter subgroups $\mathrm{Hom}_C(\R,G)$ endowed with the uniform convergence topology on compact sets in
$\R$. The exponential map induced by $G$ is defined by
\begin{eqnarray}
\mathrm{exp}_G:\mathfrak{L}(G)&\rightarrow& G \notag\\
\phi&\mapsto&\mathrm{exp}_G\,\phi=:\phi(1)\;.
\end{eqnarray}
\end{definition}
In particular and an important starting point, if $G$ is an abelian topological group then $\mathfrak{L}(G)\equiv\mathrm{Hom}_C(\R,G)$ is a topological vector space with the uniform convergence topology on compact sets.

There are  additional useful structures that can be introduced on $G$. For instance,  a projective system of finite-dimensional Lie groups allows to define a `pro-Lie group' as the projective limit.\cite{HM2} This provides a large class of $G$ with interesting structures. For a number theory example, the absolute Galois group
\begin{equation}
Gal(\bar{F}/F)=\lim_{\stackrel{\longleftarrow}{i}}Gal(E_i/F)
\end{equation}
is a `profinite' topological group, i.e. the inverse limit of finite Galois groups $Gal(E_i/F)$ where $E_i/F$ is a finite Galois extension of the field $F$.

Examples particularly relevant to physics (and hence part of our focus), draw from the group of units $A$ of some topological algebra $\mathfrak{A}\,$; in which case $A$ comes equipped with certain structures inherited from $\mathfrak{A}$. We highlight two examples: complete continuous inverse algebras and Banach algebras.

\subsubsection{$G$ from a CCIA}
\begin{definition}\emph{(\cite[\S 1.3]{GLO})}
A complete continuous inverse algebra(\emph{CCIA}) is a Mackey complete, unital topological algebra $\mathfrak{A}$ whose group of units(invertible elements) $A$ is open and group inversion is continuous.

A locally convex space $S_\diamond$ is Mackey complete iff the integral $\int_a^b\gamma(\ti)\;d\ti$ exists for any smooth curve $\gamma:(\alpha,\beta)\rightarrow S_\diamond$ with $\alpha<a<b<\beta$.
\end{definition}

A \emph{complex} CCIA is enough to construct a functional calculus on $\mathfrak{A}$ associated to holomorphic functions.\cite{GLO} In particular, one can construct complex-analytic exponential and logarithm maps;
\begin{definition}\emph{(\cite[def. 5.1]{GLO})}
Let $B_1(\textbf{\emph{1}})$ be the unit ball about the identity
element $\textbf{\emph{1}}\in\mathfrak{A}$. The exponential
$\mathrm{exp}_{\mathfrak{A}}:\mathfrak{A}\rightarrow \mathfrak{A}$ and
logarithm
$\log_{\mathfrak{A}}:B_1(\textbf{\emph{1}})\rightarrow\mathfrak{A}$
are defined by
\begin{eqnarray}
&&\mathrm{exp}_{\mathfrak{A}}(\mathfrak{a})
:=\frac{1}{2\pi i}\int_\Gamma \mathrm{exp}(z)(z\mathbf{1}-\mathfrak{a})^{-1}\;dz
\;\;\;\;\;\;\;\;\forall\,\mathfrak{a}\in\mathfrak{A},
\;\notag\\
&&\log_{\mathfrak{A}}(\textbf{\emph{1}}+\mathfrak{a})
:=\frac{1}{2\pi i}\int_\Gamma \log(z)(z\mathbf{1}-\mathfrak{a})^{-1}\;dz\;\;\; \;\;\;\;\;
\forall\,\mathfrak{a}\in B_1(\textbf{\emph{1}})
\end{eqnarray}
for $\Gamma$ a partially smooth contour in $\C$ enclosing the spectrum $\sigma(\mathfrak{a})$.
\end{definition}
Since $\mathrm{exp}_{\mathfrak{A}}$ and $\log_{\mathfrak{A}}$ are complex-analytic, these contour integrals yield the usual series
\begin{eqnarray}
&&\mathrm{exp}_{\mathfrak{A}}(\mathfrak{a})
=\sum_0^\infty\frac{1}{n!}\mathfrak{a}^n
\;\;\;\;\;\;\;\;\forall\,\mathfrak{a}\in\mathfrak{A},
\;\notag\\
&&\log_{\mathfrak{A}}(\textbf{{1}}+\mathfrak{a})
=\sum_1^\infty\frac{(-1)^{n+1}}{n}\mathfrak{a}^n \;\;\;\;\;\;\;\;
\forall\,\mathfrak{a}\in B_1(\textbf{{1}})\;.
\end{eqnarray}

 Being an open subset of a locally convex $\mathfrak{A}$, the group of units $A$ inherits a manifold structure (c.f. \cite[ex. III.1.3]{NEEB}). Further, the complex-analytic structure of a complex CCIA is enough to endow the units of a CCIA with a Lie group structure.
\begin{definition}
A topological group $A$ is a Lie group if there exists a neighborhood $U$ of $\{e\}$
such that, for every subgroup $H$, if $H\subseteq U$ then $H=\{e\}$.
\end{definition}
\begin{theorem}\emph{(\cite[prop. 3.2, prop. 3.4]{GLO})}
Let $A$ be the set of units of a \textbf{complex} \emph{CCIA} $\mathfrak{A}$. Then group inversion $\mathrm{inv}:A\rightarrow A$ is complex-analytic and, hence, $A$ is a complex-analytic Lie group  with exponential map $\mathrm{exp}_A\equiv\mathrm{exp}_{\mathfrak{A}}|_A:T_eA\rightarrow A$ by $\mathfrak{a}\mapsto \mathrm{exp}_{\mathfrak{A}}(t\mathfrak{a})=:\phi_{\mathfrak{a}}(t)$ such that $d\phi_{\mathfrak{a}}(d/dt)=\mathfrak{a}\in T_eA$ and $t\in \R$. If $\mathfrak{A}$ is \textbf{real} \emph{CCIA}, then inversion is real-analytic  and $A$ is a real-analytic Lie group.
\end{theorem}
Observe that $\phi_{\mathfrak{a}}\in\mathfrak{L}(G)$ is the unique one-parameter subgroup generated by $\mathfrak{a}$ given that $d\phi_{\mathfrak{a}}(d/dt)=\mathfrak{a}$ and $\phi_{\mathfrak{a}}(1)=\mathrm{exp}_A\mathfrak{a}$.
In addition, since a CCIA is Mackey complete, $A$ is a \textrm{BCH}-Lie group:
\begin{definition}\emph{(\cite[def. 5.5]{GLO})}
Define a \emph{\textrm{BCH}}-Lie group $A_{[\,,\,]}$ to be a complex-analytic Lie group such that: i) there exists an open $0$-neighborhood $U\subset T_eA_{[\,,\,]}$  with $V:=\mathrm{exp}_{A_{[\,,\,]}}(U)$ open in $A_{[\,,\,]}$ such that $\varphi:=\mathrm{exp}_{A_{[\,,\,]}}|_U^V:U\rightarrow V$ is a diffeomorphism; and ii) there exists a $(0,0)$-neighborhood $W\subseteq U\times U$ such that $\mathrm{exp}_{A_{[\,,\,]}}(\mathfrak{a})\mathrm{exp}_{A_{[\,,\,]}}(\mathfrak{b})\subseteq V$ and $\varphi^{-1}(\varphi(\mathfrak{a})\varphi(\mathfrak{b}))$ is the \emph{BCH} series for all $\mathfrak{a},\mathfrak{b}\in W$.
\end{definition}
\begin{proposition}\emph{(\cite[th. 5.6]{GLO})}
If $\mathfrak{A}$ is complex \emph{CCIA}, then the group of units $A$ is a \emph{BCH}-Lie group.
\end{proposition}
Suppose $G\leq A$ is open. A continuous group inverse renders a topological $G$, and the additional structure possessed by $A$ implies $G$ issues as units of $\mathfrak{A}$ such that $G\subset\mathfrak{A}$ is open. It follows that such a $G$ coming from a complex CCIA is a BCH-Lie group.

\subsubsection{$G$ from a Banach algebra}
For the second example, let $\mathfrak{A}$ be a unital Banach algebra and $A$ its set of units. There exists a homeomorphism $\eta:\mathfrak{A}\rightarrow \mathfrak{L}(A)$ such that
$ \mathrm{exp}_A(\eta(\mathfrak{a}))=:\phi_{\mathfrak{a}}(1)$.
Moreover, for any subgroup $G^A\leq A$, there is an induced homeomorphism $\eta_{G^A}:\mathfrak{A}\rightarrow
\mathfrak{L}(G^A)$. Consequently, the exponential function extends to the algebra level
$\mathrm{exp}_{\mathfrak{A}}:\mathfrak{A}\rightarrow A$  by
$\mathfrak{a}\mapsto\phi_{\mathfrak{a}}(t)=\mathrm{exp}_{\mathfrak{A}}(t\mathfrak{a})$, and it enjoys the standard
properties if $\mathfrak{A}$ is endowed with a Lie bracket.\cite[ch. 5]{HM}

\begin{definition}\emph{(\cite[def. 5.1]{HM})}\label{exponential}
Let $B_1(\textbf{\emph{1}})$ be the unit ball about the identity
element $\textbf{\emph{1}}\in\mathfrak{A}$. The exponential,
$\mathrm{exp}_{\mathfrak{A}}:\mathfrak{A}\rightarrow A$, and
logarithm,
$\log_{\mathfrak{A}}:B_1(\textbf{\emph{1}})\rightarrow\mathfrak{A}$,
are defined by
\begin{eqnarray}
&&\mathrm{exp}_{\mathfrak{A}}(\mathfrak{a})
:=\sum_0^\infty\frac{1}{n!}\mathfrak{a}^n
\;\;\;\;\;\forall\,\mathfrak{a}\in\mathfrak{A},
\;\notag\\
&&\log_{\mathfrak{A}}(\textbf{\emph{1}}+\mathfrak{a})
:=\sum_1^\infty\frac{(-1)^{n+1}}{n}\mathfrak{a}^n \;\;\;\;\;
\mathrm{for}\,\|\mathfrak{a}\|<1\;.
\end{eqnarray}
\end{definition}
The two functions are absolutely convergent and analytic for the indicated
$\mathfrak{a}\in\mathfrak{A}$.

\begin{proposition}\emph{(\cite[prop. 5.3]{HM})}\label{log/exp}\\
Let $N_0$ be the connected component of the $0$-neighborhood of
$\mathrm{exp}_{\mathfrak{A}}^{-1}B_1(\textbf{\emph{1}})$. Then
\begin{eqnarray}
&&i)\,\log_{\mathfrak{A}}(\mathrm{exp}_{\mathfrak{A}}\mathfrak{a})
=\mathfrak{a}\,\;\;\;\;\;\forall\,\mathfrak{a}\in N_0\;.\notag\\
&&ii)\,\mathrm{exp}_{\mathfrak{A}}(\log_{\mathfrak{A}}\mathfrak{a})=\mathfrak{a}
\,\;\;\;\;\;\forall\,\mathfrak{a}\in B_1(\textbf{\emph{1}})\;.\notag\\
&&iii)\,\left.\mathrm{exp}_{\mathfrak{A}}\right|_{N_0}:N_0\rightarrow
B_1(\textbf{\emph{1}})\,\text{is an analytic homeomorphism with}\notag\\
&&\hspace{.3in}\text{analytic inverse }\log_{\mathfrak{A}}:B_1(\textbf{\emph{1}})\rightarrow
N_0\;.
\end{eqnarray}
\end{proposition}

\begin{definition}\emph{(\cite[def. 5.32]{HM})}\label{linear Lie}
Let $\mathfrak{A}_L\subseteq(\mathfrak{A},[\cdot,\cdot])$ be a
closed Lie subalgebra of a Banach algebra with identity equipped with a Lie bracket. Let
$G^{A_l}$ be a subgroup of the units of $\mathfrak{A}_L$
such that $\mathrm{exp}_{\mathfrak{A}_L}$ is a homeomorphism mapping
a neighborhood of $\{0\}\in\mathfrak{A}_L$ into a neighborhood of
$\{e\}\in G^{A_l}$. A topological group is a \textbf{linear Lie group} if it is
isomorphic to $G^{A_l}$.
\end{definition}

\begin{proposition}\emph{(\cite[cor. 5.37, th. 5.41]{HM})}
If $G$ is a linear Lie group, then $G$ is an analytic group, the set $\mathfrak{L}(G)$ is a completely
normable topological real Lie algebra, and $\mathrm{exp}_G$ is a
homeomorphism from a $0$-neighborhood of $\mathfrak{L}(G)$ to an
$e$-neighborhood in $G$. The image $\mathrm{exp}_G\mathfrak{L}(G)$ generates the connected component $G_0$ in $G$.
\end{proposition}
Remark that the matrix algebra $M_n(\mathfrak{A})$ with $n\in\mathbb{N}$ of a CCIA is again a CCIA if given the product topology through identification with $\mathfrak{A}^{n\times n}$. In consequence, $GL(n,\mathfrak{A})$ is a Lie group that contains the linear Lie groups.\cite{NEEB}

\subsubsection{Locally compact $G$}
We continue to add structure to $G$:
\begin{definition}
A neighborhood basis at $g\in G$ is a family
$\mathcal{N}$ of neighborhoods such that, given any neighborhood $U$
of $g$, there exists an $N\in\mathcal{N}$ with $N\subset U$. $G$ is \textbf{locally compact} if every $g\in G$ has a neighborhood
basis comprised of compact sets.
\end{definition}
\begin{proposition}\emph{(\cite[sec. 1]{W})}
Let $G$ be a locally compact topological group. A subgroup $H$ is locally compact iff it is locally closed in $G$, i.e. $H$ is open in $\overline{H}$. A homogenous space $G/H$ is locally compact if $H$ is a closed, normal subgroup. A group extension $e\rightarrow N\stackrel{i}{\rightarrow}E\stackrel{j}{\rightarrow}G\rightarrow e$ is locally compact iff both $N$ and $G$ are locally compact.
\end{proposition}

The most important consequence of local compactness for our purpose is the
well-known result:
\begin{theorem}
If $G$ is locally compact, then there exists a unique (up to
positive scalar multiplication) Haar measure.
\end{theorem}
Evidently, locally compact topological groups can be used as a
footing on which to ground functional integration, because they supply
measure spaces on which to model functional integrals and their associated integrators. With this in mind, recall Banach-valued integration on locally compact topological groups.
\begin{proposition}\emph{(\cite[prop. B.34]{W})}\label{banach integration}
Let $G_\lambda$ be a locally compact topological group, $\mu$ its associated
Haar measure, and $\mathfrak{B}$ a Banach space possibly with an
algebraic structure. Then the set of integrable
functions $L^1(G_\lambda,\mathfrak{B})\ni f$, consisting of equivalence classes of measurable functions equal almost everywhere
with norm $\|f\|_1:=\int_{G_\lambda}\|f(g_\lambda)\|d\mu(g_\lambda)\leq\|f\|_\infty\,\mu(\mathrm{supp}\,f)<\infty$, is a
Banach space. Moreover, $f\mapsto\int_{G_\lambda}f(g_\lambda)d\mu(g_\lambda)$ is a linear map
such that
\begin{equation}
\|\int_{G_\lambda}f(g_\lambda)\;d\mu(g_\lambda)\|\leq\|f\|_\infty\,\mu(\mathrm{supp}\,f)
\end{equation}
for all $f\in L^1(G_\lambda,\mathfrak{B})$,
\begin{equation}
\varphi\left(\int_{G_\lambda}f(g_\lambda)\;d\mu(g_\lambda)\right)
=\int_{G_\lambda}\varphi\left(f(g_\lambda)\right)\;d\mu(g_\lambda)
\end{equation}
for all $\varphi\in\mathfrak{B}'$, and
\begin{equation}
L_B\left(\int_{G_\lambda}f(g_\lambda)\;d\mu(g_\lambda)\right) =\int_{G_\lambda}L_B\left(f(g_\lambda)\right)\;d\mu(g_\lambda)
\end{equation}
for bounded linear maps
$L_B:\mathfrak{B}\rightarrow\mathfrak{B}_2$. Moreover, Fubini's
theorem holds for all equivalence classes $f\in L^1(G_1\times G_2,\mathfrak{B})$.
\end{proposition}

\begin{corollary}\emph{(\cite[lemma. 1.92]{W})}\label{H-representation}
Let $\mathfrak{B}^\ast$ be a $C^\ast$-algebra and
$\pi:\mathfrak{B}^\ast\rightarrow L_B(\mathcal{H})$ a
representation with
$L_B(\mathcal{H})$ the algebra of bounded linear operators on Hilbert space
$\mathcal{H}$. Then
\begin{equation}
\left\langle \pi\left(\int_{G_\lambda}f(g_\lambda)\;d\mu(g_\lambda)\right)v|w\right\rangle
=\int_{G_\lambda}\left\langle \pi\left(f(g_\lambda)\right)v|w\right\rangle \;d\mu(g_\lambda)\;,
\end{equation}
\begin{equation}
\left(\int_{G_\lambda}f(g_\lambda)\;d\mu(g_\lambda)\right)^\ast =\int_{G_\lambda}f(g_\lambda)^\ast \;d\mu(g_\lambda)\;,
\end{equation}
and
\begin{equation}
a\int_{G_\lambda}f(g_\lambda)\;d\mu(g_\lambda)b=\int_{G_\lambda}af(g_\lambda)b\;d\mu(g_\lambda)
\end{equation}
where $v,w\in\mathcal{H}$ and $a,b\in M(\mathfrak{B}^\ast)$ with
$M(\mathfrak{B}^\ast)$ the multiplier algebra of
$\mathfrak{B}^\ast$.
\end{corollary}

It can be shown (\cite[appx. B]{W}) that  $L^1(G_\lambda,\mathfrak{B}^\ast)$ is a Banach $\ast$-algebra when equipped with: i) the $\|\cdot\|_1$ norm, ii)
the convolution
\begin{equation}
f_1\ast f_2 (g_\lambda):=\int_{G_\lambda}f_1(h_\lambda)f_2(h_\lambda^{-1}g_\lambda)\;d\mu(h_\lambda)\;,
\end{equation}
and iii) the involution
\begin{equation}
f^\ast (g_\lambda):={f(g_\lambda^{-1})}^\ast\Delta(g_\lambda^{-1})
\end{equation}
where $\Delta$ is the modular function on $G_\lambda$.

\section{Subsumed approaches}\label{approaches}
The proposed definition of functional integrals is only useful to the extent that it includes known successful approaches. So it is important to check that this is the case.

\begin{application}\emph{The Wiener and Feynman path integral via time slicing:}

Consider $\mathcal{P}_0\C^m$, the infinite dimensional vector space of piece-wise continuous pointed paths $x:[\ti_a,\ti_b]\subset\R\rightarrow\C^m$ with $x(\ti_a)=0$, and take $G$ to be its underlying abelian  group equipped with a suitable topology. Choose $\mathfrak{B}\equiv\C$, and let $\Lambda=\{\lambda_n:G\rightarrow\mathcal{H}_\lambda^n\;\forall\;n\in\mathbb{N}_+\}$ by $x\mapsto(x(\ti_1),x(\ti_2),\ldots,x(\ti_n))$ where $\ti_1<\ti_2\,\ldots\,<\ti_n$ and the Hilbert space $\mathcal{H}_\lambda^n$ comprises states characterized by a mean and covariance.\footnote{Technically, $\lambda_n$ maps $G$ to the abelian group underlying $\mathcal{H}_\lambda^n$, but this distinction in not necessary here and it is better to use familiar notation.}

Evidently the pair $(\mathcal{H}_\lambda^n,P_{n n'})$ is a projective system for $G$ with maps $P_{n n'}$ determined by $P_{n n'}\circ\lambda_{n'}=\lambda_{n}$. It achieves the reduction $\mathbf{F}(G)|_{G_\Lambda}=\bigoplus_{\lambda_n} L^1(\mathcal{H}_\lambda^n,\C)$ for finite time slices. Under the restrictions $\mathfrak{B}\equiv\R$ and $x:\mathbb{T}\rightarrow\R^m$  with $\mathbb{T}:=[\ti_a,\ti_b]$, the projective system can be used to define a promeasure with Gaussian weight, i.e. the Wiener measure.

But in the generic case, with $\mathfrak{B}\equiv\C$ and $x:\mathbb{T}\rightarrow\C^m$, to get the Feynman path integral one must work harder and either: i) analytically continue the restricted case; ii) use the projective system to construct time-sliced integrals defined using the Trotter product formula; or iii) use the topological dual space, its associated projective system, and Fourier duality to define projective distributions according to \emph{\cite{D-W1,HPS}}. Of course one must still determine the class of integrable functions $f\in L^1(\mathcal{H}_\lambda^n,\C)$  allowed by each approach through functional analysis.

Remark that one can contemplate more general projective systems that are not based on time slicing. This was rigorously achieved in \emph{\cite{AL/BR}} in the case where $\mathcal{P}_0\C^m$ carries the structure of an infinite-dimensional real, separable Hilbert space. The projections are ordered according to their finite dimension. Coupled with the theory of oscillatory integrals, the projective system for $G$ gives rigorous access to Feynman-type path integrals and their localization by stationary phase. Further, in the context of gauge field theory, projective systems have been constructed \emph{\cite{AS/LE,SA/BA}} based on the holonomy of a connection. Once again, in these more general projective systems the most difficult work is quantifying integrable functions.

\end{application}

Unfortunately, projective systems derived from time slicing run into issues if non-cartesian coordinates are used on $\C^m$ (see e.g. \cite[ch. 8]{KL}). Complications arise because there are consistency conditions that must be obeyed by the projective system, and it may be difficult to find a suitable projective system and/or integrable functions. Even more troublesome; in the case the target manifold is more general than $\C^m$, the space of pointed paths is generically no longer a Banach space and the projective method cannot be applied directly. In either case, one must be careful to pay close attention to delicate mathematical issues --- undermining the intuitive and formal appeal of path integrals.

For Feynman path integrals at least, the shortcomings of the projective method can be sidestepped by utilizing dual abelian groups in the framework of Fourier/Pontryagin duality as exemplified in \cite{AL/KR,CA/D-W3}. In this approach, one no longer attempts to define a rigorous measure on the integration domain. Instead the path integral is related through Fourier duality to a bona fide integral. There is, however, no (direct) topological localization in this context since the dual space is assumed measurable from the outset. The next example is a brief outline of the Cartier/DeWitt-Morette (CDM) approach which illustrates this idea.

\begin{application}\label{CDM}$\mathrm{Cartier/DeWitt}$-$\mathrm{Morette\,
\,\,functional\,integration\,scheme}$\emph{:\cite{CA/D-W3}}

The CDM scheme for functional integration corresponds to the
particular case of $\mathfrak{B}\equiv\C$ and (as above) $G$ an abelian topological group underlying an infinite dimensional Banach space.
More precisely, $G$ is the abelian (additive under point-wise addition) group
of continuous pointed maps $x:(\mathbb{T},\ti_a)\rightarrow (\C^m,0)$ equipped with a suitable topology, and $X_0:=\mathcal{P}_0\C^m$ is its associated Banach space over $\C$. To be consistent with the notation of CDM, we will abuse notation and write $G\equiv X_0$ keeping in mind that scalar multiplication is strictly not allowed. Consequently, any question regarding scale must ultimately be referred to scalar multiplication in $\mathbf{F}(X_0)$ through the definition of $\mathrm{int}_\Lambda$.\footnote{CDM uses $X_0$ as their domain of integration. But the field structure is not relevant to the integration in the sense that their integrators are not invariant under scalar multiplication. Of course \emph{scale} is an important issue, but it is better handled within the algebraic structure of $\mathbf{F}(X_0)$. We accomplish this by including a scale factor $\s\in\C_+$  in the definition of $\mathcal{D}_\lambda x$ (where $\C_+$ is the right-hand complex plane). Otherwise said; the scale $\s$ is part of the specification of $\Lambda$.}

Since $X_0$ is abelian, the space of one-parameter subgroups $\mathfrak{L}(X_0)$ is a topological vector space. The abelian group $X'_0$ underlying the topological dual $(\mathcal{P}_0\C^m)'$ is assumed to be a locally compact Polish space when equipped with a suitable topology. Hence, $X'_0$ can be equipped with a positive measure $\mu$, $\mathfrak{L}(X_0)'$ is a locally compact Banach
space, and $L^1(X'_0,\C)$ is a Banach algebra under convolution.

The space of integrable functionals $\mathbf{F}(X_0)$ is the set of functionals defined by
\begin{equation}\label{integrable functional}
\mathrm{F}_\mu(x):=\int_{X'_0}\Theta(x,x')\;d\mu(x')
\end{equation}
where $\Theta(x,x'):X_0\times X'_0\rightarrow\C$ is continuous,
bounded and integrable with respect to $\mu$. Then $\mathbf{F}(X_0)$ is a Banach space
with an induced norm defined as the total variation of $\mu$.
Bounded linear integral operators $\int_X\mathcal{D}_\lambda x$ with
$\|\int_X\mathrm{F}_\mu\mathcal{D}_\lambda x\|\leq\|\mathrm{F}_\mu\|$ on
$\mathbf{F}(X_0)$ are defined by
\begin{equation}\label{CDM def}
\int_{X_0} \mathrm{F}_\mu(x)\mathcal{D}_\lambda
x:=\int_{X'_0}\widehat{\mathrm{F}}_\lambda(x')\;d{\mu}(x')
\end{equation}
where
\begin{equation}
\int_{X_0}\Theta(x,x')\mathcal{D}_\lambda x:=\widehat{\mathrm{F}}_\lambda(x')
\end{equation}
defines the integrator family $\mathcal{D}_\Lambda x$ (with $\widehat{\mathrm{F}}_\lambda\in L^1(X_0',\C)$). In particular, given the standard choice $\Theta(x,x')=e^{-2\pi i\langle x',x\rangle}$, then \emph{(\ref{integrable functional})} can be interpreted as the Fourier transform of the measure $\mu$. Note that an affine transformation $x\mapsto x+x_a$ along with the translation invariance
$\mathcal{D}_\lambda (x+x_a)=\mathcal{D}_\lambda x$ yields
integration on $X_a$, which is the space of pointed maps $x:(\mathbb{T},\ti_a)\rightarrow (\C^m,x_a)$.

It is evident that a choice of $\lambda$ corresponds to a class of functions $\widehat{\mathrm{F}}_\lambda(x')$ integrable with respect to $\mu$. Conversely, a choice of $\mu$ corresponds to a class of functions $\mathrm{F}_\mu(x)$ integrable with respect to $\lambda$. In this sense, $\lambda$ and $\mu$ are Fourier dual. So here we have an explicit determination of $\Lambda$: it is the Fourier dual to the set $\{\mu\}$ of measures on $X_0'$. For the archetypical Gaussian case with $\Theta(x,x')=e^{-2\pi i\langle x',x\rangle}$, parameter $\lambda$ characterizes the mean and covariance of the Gaussian paths of interest. The functional integral on the left-hand side of \emph{(\ref{CDM def})} is exact (in the sense it is also specified by the same $\lambda$), because there is a one-to-one correspondence between the two sides for Gaussian paths by duality. That is, the Fourier transform of a Gaussian is still a Gaussian.

To handle spaces of pointed maps
$\mathcal{P}_{\m_a}\mathbb{M}$  (which are not topological groups) where now
$m:\mathbb{T}\rightarrow\mathbb{U}\subseteq\mathbb{M}$ with
$m(\ti_a)=\m_a$ and $\mathbb{U}\subseteq\mathbb{M}$ an open
neighborhood of a smooth dimension $\mathrm{dim}(\mathbb{M})=m$ Riemannian manifold, CDM uses the
left-invariant vector field Lie algebra $\mathfrak{G}_a$ at a point $\m_a$ to
identify the non-abelian linear Lie group $\widetilde{G}$ underlying
$\mathcal{P}_{\m_a}\mathbb{M}$. In this case, a one-parameter subgroup morphism
$p:\mathfrak{L}(X_0)\rightarrow\mathfrak{L}(\widetilde{G})$
induces a morphism
\begin{eqnarray}
\mathrm{Exp}:\mathfrak{L}(X_0)&\rightarrow& \mathcal{P}_{\m_a}\mathbb{M}\notag\\
\mathfrak{x}&\mapsto&\mathrm{Exp}(\mathfrak{x})=\left(\mathrm{exp}_{\widetilde{G}}\circ p
\right)(\mathfrak{x})\;.
\end{eqnarray}
Given $\mathrm{Exp}$ and the fact $\mathcal{P}_{\m_a}\mathbb{M}$
is contractible since it is a pointed space, the parametrization
\begin{eqnarray}
P:X_0&\rightarrow& \mathcal{P}_{\m_a}\mathbb{M}\notag\\
x&\mapsto&\mathrm{Exp}(\mathrm{log}_{X_0}(x))
\end{eqnarray}
allows the integral on
$\mathcal{P}_{\m_a}\mathbb{M}$ to be defined by
\begin{equation}\label{manifold integral}
\int_{\mathcal{P}_{\m_a}\mathbb{M}}\mathrm{F}(m)\mathcal{D}_\lambda
m:=\int_{X_0}\mathrm{F}_\mu(P(x))\mathcal{D}_\lambda (P(x))
:=\int_{X_0}\mathrm{F}_\mu(P(x))\,|\mathrm{Det}_\lambda
P'_{(x)}|\,\mathcal{D}_\lambda x\;.
\end{equation}
The left-hand side furnishes the path integral route to QM. Note that it has limited applicability if $\M$ is not
geodesically complete.

Meanwhile, if it happens that $\mathbb{M}=\mathbb{G}$ is a Lie
group manifold, then \emph{(\ref{manifold integral})} can be readily used
since the Lie algebra of left-invariant vector fields is automatically available at each point of the group manifold;
\begin{equation}
\int_{\mathcal{P}_{\mathrm{g}_a}\mathbb{G}}\mathrm{F}(m)\mathcal{D}_\lambda
m:=\int_{X_0}\mathrm{F}_\mu(P(x))\mathcal{D}_\lambda (P(x))
=\int_{\mathbb{G}}f(g_\lambda)\,d\nu(g_\lambda)\;.
\end{equation}
In particular, (as is well-known) this means that the free point-to-point propagator on
a group manifold is `exact' in the sense that it can be expressed as
a sum over relevant $\lambda$ of finite dimensional integrals. Again, the left-hand side is exact due to Pontryagin duality.

Alternatively, CDM can use the soldering form $\Bold{\theta}$ on the
frame bundle $F\mathbb{M}$ equipped with a connection to construct
the development map parametrization. The explicit
construction of the development map uses the identification
$\Bold{\theta}(\mathrm{hor}(v_p))=\dot{z}$ where $z:\mathbb{T}\rightarrow\C^n$
and vector $\mathrm{hor}(v_p)\in T_pF\mathbb{M}$ is tangent to the
horizontal lift $\widetilde{m}(\mathbb{T})$ at point $p\in F\mathbb{M}$ of a curve $m(\mathbb{T})\in\mathbb{M}$. The tangent to the horizontal lift $\widetilde{m}(\mathbb{T})$ induces the morphism
\begin{eqnarray}
\mathrm{Dev}:\dot{\mathcal{P}}_{\m_a}\mathbb{M}&\rightarrow& \mathcal{P}_{\m_a}\mathbb{M}\notag\\
\dot{m}&\mapsto&\mathrm{Dev}(\dot{m})=(\pi\circ\widetilde{m})
\end{eqnarray}
where $\dot{\mathcal{P}}_{\m_a}\mathbb{M}$ is the abelian topological group of pointed maps
$\dot{m}:\mathbb{T}=[\ti_a,\ti_b]\rightarrow\mathbb{U}\subseteq
T_{\m_a}\mathbb{M}$. Here $\dot{m}(t):=\rho_{m(t)}(\dot{z}(t))$ such that $\dot{m}(\ti_a)=0$ where the frame $\rho_{m(t)}:\C^n\rightarrow T_{m(t)}\mathbb{M}$ and $\pi$ is the
projection on the frame bundle. Then
\begin{equation}
\int_{\mathcal{P}_{\m_a}\mathbb{M}}\mathrm{F}(m)\mathcal{D}_\lambda
m:=\int_{\dot{\mathcal{P}}_{\m_a}\mathbb{M}}\mathrm{F}_\mu(\mathrm{Dev}(\dot{m}))\mathcal{D}_\lambda
(\mathrm{Dev}(\dot{m}))\;.
\end{equation}
When $\mathbb{M}=\mathbb{G}$ and the connection is Riemannian,
$\mathrm{Dev}$ and $\mathrm{Exp}$ amount to the same thing.
\end{application}
It should be noted that \cite[\S19.1]{CA/D-W3} suggested generalizing the space of pointed paths in their scheme to include locally compact \emph{abelian} groups, and the localization/projective system in the CDM framework is effected indirectly through Pontryagin duality.

\begin{application}\emph{Loop groups:}

The previous two examples can be readily applied to continuous loops $x:\mathrm{S}^1\rightarrow\mathbb{G}$ yielding functional integrals whose domains are the free loop group $L\mathbb{G}=\mathrm{Hom}_C(\mathrm{S}^1,\mathbb{G})$ or based loop group $L_{g_a}\mathbb{G}=\mathrm{Hom}_C((\mathrm{S}^1,s_a),(\mathbb{G},g_a))$ of some Lie group $\mathbb{G}$. Utilizing a suitable parametrization \`{a} la the CDM scheme, these can be extended to loop spaces $L\mathbb{M}$ and $L_{\mathrm{m}_a}\mathbb{M}$ of a Riemannian manifold. We are, of course, glossing over symmetry issues regarding invariance of the initial point for paths in $L\mathbb{M}$. (We explicate the CDM parametrization for $L_{\mathrm{m}_a}\mathbb{M}$ in the non-trivial example presented in \emph{\S\ref{Skew-Gaussian integrators}}.)
\end{application}

The next obvious generalization is to promote paths to fields with suitable analytic properties; $x:\mathbb{D}\rightarrow\M$ where $\D$ is a smooth Riemannian manifold with $\mathrm{dim}(\D)=d\leq m$. The CDM scheme can be used to construct functional integrals for fields $x:\mathbb{D}\rightarrow\M$  according to the definition, but we will not verify that specifying $\Lambda$ and $\mathrm{F}|_{G_\Lambda}$ captures the intricacies of renormalization. Consequently, we can only claim that the proposed definition of functional integrals includes \emph{free} fields since in this case they are characterized as Gaussian.

\begin{application}\emph{CDM for fields:\cite{LA3}}\label{CDM for fields}

Let $\mathcal{F}\C^m$ be the Sobolev space
$W^{k,p}(\U)$ of $L^p$ maps $x:\U\subseteq\D\rightarrow\C^m$ with $\U$
open and $\D$ a compact (with or without boundary) or open Riemannian manifold. If $\D$ has
boundary $\p\D$, let $\mathcal{F}_0\C^m=W_0^{k,p}(\U)$ be the closure in $W^{k,p}(\U)$ of  the vector
space of $C^{\infty}$ maps with compact support in $\U$. Recall that
$W^{k,p}(\U)$ and $W_0^{k,p}(\U)$ are Banach. Continue to take $\mathfrak{B}\equiv\C$, and again abuse notation by writing $\mathcal{F}\C^m\equiv  X$ or $\mathcal{F}_0\C^m\equiv X$. Of course, particular applications require consideration of some type of boundary conditions or functional constraints implemented through $\Lambda$.

Let $\mathcal{F}\mathbb{M}$ denote the space of fields $x:\mathbb{D}\rightarrow\M$ for both open and compact $\D$ for notational simplicity. Introduce the
exterior differential system
\begin{equation}
\left\{\Bold{\omega}_{I}=0\right\},\;\;\Bold{\omega}_{I}\in\mathit{\Lambda}(\mathcal{F}\M)
\end{equation}
with $I\in\{1,\ldots,N\}$ and $N\leq m$.
This system defines a parametrization $P:X\rightarrow \mathcal{F}\mathbb{M}$ by
\begin{equation}\label{field parametrization}
P^*\Bold{\omega}_{I}=0\;\;\forall\,I\;.
\end{equation}
As with the previous case of paths, two particularly prevalent parametrizations arise from Pfaff exterior
differential systems associated with the exponential map $\mathrm{Exp}:T_{x}\mathcal{F}\M\rightarrow\mathcal{F}\M$ and the
development map
$\mathrm{Dev}:\mathcal{F}T_{\m_a}\M\rightarrow\mathcal{F}\M$.

Finally, define
\begin{equation}\label{field integral def.}
  \int_{\mathcal{F}\mathbb{M}}\mathrm{F}(m)\mathcal{D}_\lambda
m:=\int_{X}\mathrm{F}_\mu(P(x))\mathcal{D}_\lambda (P(x))
:=\int_{X}\mathrm{F}_\mu(P(x))\,|\mathrm{Det}_\lambda
P'_{(x)}|\,\mathcal{D}_\lambda x
\end{equation}
where $\mathrm{F}_\mu\in\mathbf{F}(X)$ is defined by
\begin{equation}
\mathrm{F}_\mu(x):=\int_{X'}\Theta(x',x)\;d\mu(x')
\end{equation}
with $\mu$ the Haar measure on the dual group $X'$  underlying the topological dual $(\mathcal{F}\C^m)'$. (As in CDM for paths, $X'$ is assumed locally compact Polish.)

Structurally, functional integrals for fields are quite similar to path integrals. But there are complications lurking in $\mathcal{F}\mathbb{M}$ concerning the localization associated with $\Lambda$.  Borrowing from the path integral case, one approach is to specify $\Lambda$ by means of finite projections. In the context of QFT, the convention is to construct a projective system based on causal ordering in $\R^{3,1}$ and account for the spatial dependence through the formal device $\lim_{m\rightarrow\infty}\C^m$: One considers a field on space-time to be a path with an infinite number of components indexed by some space-like surface (tensored with any non-trivial representation carried by the field).

More generally, in the context of FQFT one decomposes $\mathbb{D}=\Sigma\times \mathbb{T}$ and constructs the projective system $\Sigma\times\{\ti\}$ with $\ti\in\mathbb{T}$. The sewing axiom is then a consequence of the composition of projections. As in the case of paths, Fourier duality and dual projective systems allow for the definition of projective Sobolev distributions. And the CDM scheme allows  to effectively transfer $\Lambda$ from $\mathcal{F}\mathbb{M}$ to the set of measures on the space of integrable functionals on the dual $X'$.

The projective approach for fields is a direct generalization of the original $d=0+1$ path version, and it gains legitimacy through comparison with canonical quantization and operator methods. In favorable circumstances, one is able to find a fairly simple description of $\Lambda$ and integrable functionals $\mathrm{F}|_{G_\Lambda}$ leading to exactly solvable models. Examples include generating functions and local $n$-point functions in free-field QFT, rational CFT, and TQFT.

Conversely, specifying $\Lambda$ and $\mathrm{F}|_{G_\Lambda}$ in the context of perturbative QFT is far more involved. For one thing, one must find integrable functionals $\mathrm{F}|_{G_\Lambda}$ at each order of perturbation. In addition, one typically requires invariance of various objects under some kind of symmetry, and this redundancy must be accounted for consistently at each level of perturbation.

 Resolution of these two issues requires the programs of renormalization and gauge theory. For QFT applications, it is important to implement these programs within this framework. We take some preliminary steps towards gauge theory in appendix \emph{\ref{Liouville is a tool}}, but a detailed analysis of renormalization and gauge theory lies far outside our present scope. We do want to make two brief remarks regarding renormalization, however. First, one cannot interpret renormalization as
  a rescaling of fields in this framework because scalar multiplication in $X$ is not supported --- strictly (despite notation) $X$ is an abelian group. So interactions that necessitate rescaling through renormalization require modified (i.e. rescaled) integrators $\mathcal{D}_\lambda(m)$ at each order of perturbation.  It is
   tempting to speculate that physical considerations might lead (through appropriate choices of $\lambda$) to some topology on the dual space $X'$ that would effectively
    act as a cut-off for continuous fields and hence regulate unruly integrals. In this sense, the renormalization group appears to fit into the framework via $\Lambda$ and $\mathrm{F}|_{G_\Lambda}$. Second, the idea of effective field theories seems to be captured by the notion of
      topologically induced localization if the topological cover somehow corresponds to energy scale. (Since the cover refinement characterizes continuity in $\mathbf{F}(X)$, it implies a minimum time-interval resolution for realizing continuous dynamical evolution.)
\end{application}

\section{Comparing gamma and Hardy-Littlewood}
In this appendix we compare the counts of prime doubles between (\ref{k-tuple conjecture}) and Hardy-Littlewood (c.f. \cite{TO2}). Observe
\begin{equation}
\frac{1}{\log(r)\log(r+h_{2i})}\sim\frac{1}{\log(r)^2}-\frac{h_{2i}}{r\log(r)^3}
+O\left(\frac{h_{2i}^2}{r^2\log(r)^3}\right)\;.
\end{equation}
So, for given cut-off $\x$ and off-set $h_{2i}$,  we don't expect much difference when $\x\gg h_{2i}$. But for the other way around $h_{2i}> \x$, there may be. The table below contains the exact number of prime doubles $(p,p+h_{2i})$ for several values of $h_{2i}$ and $\x$ (numerics by Mathematica 9.0).
\begin{table}[H]
\centering
\begin{tabular}{l||c|c|c|c|c|c|c|c|}
  $\x\,\diagdown \;h_{2i}$& $10^1$ & $10^2$ & $10^3$ & $10^4$ & $10^5$ & $10^6$ & $10^7$& $10^8$ \\\hline\hline
  $10^2$ & 11 & 9 & 5 & 5 & 3 & 2 & 2&3\\\hline
  $10^3$ & 51 & 49 & 37 & 34 & 23 & 20 & 17&16\\\hline
  $10^4$ & 270 & 260 & 253 & 224 & 186 & 163 & 142&112\\\hline
  $10^5$ & 1624 & 1615 & 1631 & 1556 & 1431 & 1219 & 1050&918\\\hline
 $10^6$ & 10934 & 10906 & 10993 & 10798 & 10629 & 9766 & 8592&7539\\\hline
 $10^7$ & 78211 & 78248 & 78265 & 77850 & 77680 & 76212 & 71247&63352\\\hline
 $10^8$ & 586811 & 586908 & 586516 & 586587 & 585883 & 583976 & 573938&540323\\\hline
\end{tabular}
\caption{Exact number of prime doubles for the indicated cut-off $\x$ and off-set $h_{2i}$.}
\end{table}

For the Hardy-Littlewood estimate, we have for $h_{2i}=10^1$
\begin{equation}
C_{2}=\frac{1-\frac{1}{2}}{\left(1-\frac{1}{2}\right)^2}\cdot\frac{1-\frac{2}{3}}{\left(1-\frac{1}{3}\right)^2}
\cdot\frac{1-\frac{1}{5}}{\left(1-\frac{1}{5}\right)^2}
\cdot\frac{1-\frac{2}{7}}{\left(1-\frac{1}{7}\right)^2}
\cdot\prod_{7<p}\frac{1-\frac{2}{7}}{\left(1-\frac{1}{7}\right)^2}\approx 1.76\;.
\end{equation}
All subsequent values of $h_{2i}$ in the table produce the same leading terms in the product since they are all powers of $10$. So $C_{2}\approx 1.76$ for all $h_{2i}$.

Using
\begin{equation}
C_{2}\int_2^\x\frac{1}{\log(x)^2}\;dx\;,
\end{equation}
the percentage deviation of the Hardy-Littlewood estimate relative to the exact count
is tabulated below:
\begin{table}[H]
\centering
\begin{tabular}{l||c|c|c|c|c|c|c|c|}
  $\x\,\diagdown \;h_{2i}$& $10^1$ & $10^2$ & $10^3$ & $10^4$ & $10^5$ & $10^6$ & $10^7$& $10^8$ \\\hline\hline
  $10^2$ & 64 & 101 & 261 & 261 & 502 & 802 & 802&802\\\hline
  $10^3$ & 19.7 & 24.6 & 65 & 79.6 & 165 & 205 & 259&282\\\hline
  $10^4$ & 5.8 & 9.8 & 12.9 & 27.5 & 53.5 & 75.2 & 101&155\\\hline
  $10^5$ & 2.5 & 3.1 & 2.1 & 7 & 16.3 & 36.5 & 58.5&81.3\\\hline
$10^6$ & .6 & .8 & .2 & 1.8 & 3.4 & 12.6 & 28&45.8\\\hline
 $10^7$ & .1 & .1 & .07 & .6 & .8 & 2.8 & 10&23.6\\\hline
 $10^8$ & .03 & .02 & .08 & .07 & .2 & .5 & 2.3&8.6\\\hline
\end{tabular}
\caption{Percentage deviation between exact and Hardy-Littlewood estimates of prime doubles for the indicated cut-off $\x$ and off-set $h_{2i}$.}
\end{table}
\noindent Since the Hardy-Littlewood estimate is asymptotic, it is not surprising that percentages are fairly high for smaller cut-offs. But notice the general trend of increasing deviation across rows as the ratio $h_{2i}/x$ increases.

Now to compare; the estimate of prime doubles (instead of weighted prime-power doubles) coming from (\ref{k-tuple conjecture}) requires the Moebius inverse
\begin{equation}
C_{2}\sum_{m=1}^\infty\frac{\mu(m)}{m}{\int_2}^{\,\x^{1/m}}
\frac{(r-1)((r+h_{2i})-1)}{r(r+h_{2i})\log(r)\log(r+h_{2i})}\;dr\;.
\end{equation}
The percentage deviation of the gamma conjecture estimate from the exact count is:
\begin{table}[H]
\centering
\begin{tabular}{l||c|c|c|c|c|c|c|c|}
  $\x\,\diagdown \;h_{2i}$& $10^1$ & $10^2$ & $10^3$ & $10^4$ & $10^5$ & $10^6$ & $10^7$ & $10^8$ \\\hline\hline
  $10^2$ & .2 & 3.7 & 24.8 & 5.8 & 25.7 & 57.2 & 34.7 &21.4\\\hline
  $10^3$ & .5 & 1.9 & 9.5 & 6.4 & 11.2 & 6.6 & 7.5 &.05\\\hline
  $10^4$ & .5 & 3.2 & 1.2 & .6 & .4 & 4.2 & 5.7 &4.6\\\hline
  $10^5$ & 1.1 & 1.5 & .3 & .9 & .7 & .1 & .3 &.2\\\hline
$10^6$ & .2 & .4 & .5 & .7 & .5 & .2 & .5 &.5\\\hline
 $10^7$ & .04 & .007 & .04 & .4 & .1 & .2 & .4 &.01\\\hline
 $10^8$ & .007 & .01 & .06 & .03 & .07 & .01 & .2 &.1\\\hline
\end{tabular}
\caption{Percentage deviation between exact and gamma conjecture estimates of prime doubles for the indicated cut-off $\x$ and off-set $h_{2i}$. (The sum over $m$ converges rapidly so only the first $50$ terms were used.)}
\end{table}
\noindent At least in these parameter ranges the gamma estimates are superior, and there is no reason not to expect similar comparisons throughout parameter space and for all prime $\kk$-tuples.

\section{Proof of Theorem \ref{exact k-tuple}}\label{k-tuple proof}
The proof modifies mostly standard arguments.\cite{MO} To begin, integrate (\ref{explicit integral}) by parts. The boundary terms won't contribute because:

1) A comparison test between the series representations of $\log^{(\kk-1)'}(\zeta_{(\kk)}(s))$ and $\log(\zeta(s))$ yields a finite $\sigma_a$. So, for $s=c+it$ with $c\in\R$, we have $\lim_{t\rightarrow\infty}|\log^{(\kk-1)'}(\zeta_{(\kk)}(c+it))|<\infty$ provided $c>\kk$. To see this, observe that
\begin{eqnarray}
 &&|\log^{(\kk-1)'}(\zeta_{(\kk)}(c+it))|
 \leq\sum_{p^m}\left|\frac{\log^{\kk-1}(p_{(\kk)}^m)}{m^\kk\,p_{(\kk)}^{ms}}\right|
=\sum_{p^m}\frac{\log^{\kk-1}(p_{(\kk)})}{m\,p_{(\kk)}^{mc}}\notag\\
&&\hspace{1.6in}=\sum_{p^m}\frac{1}{m\,p_{(\kk)}^{m(c-(\kk-1))}}
\frac{\log^{\kk-1}(p_{(\kk)})}{p^{m(\kk-1)}_{(\kk)}}\notag\\
&&\hspace{1.6in}<\sum_{p^m}\frac{1}{m\,{(p_{(\kk)}^m)}^{(c-(\kk-1))}}\notag\\
&&\hspace{1.6in}<\sum_{p^m}\frac{1}{m\,{(p^m)}^{(c-(\kk-1))}}\notag\\
&&\hspace{1.6in}=|\log(\zeta((c-(\kk-1))+it))|\;.
 \end{eqnarray}

2) The inequality $\log^{\kk-1}(x_{(\kk)})/\log_{(\kk)}(x)\leq1/\log(x)$ implies $\mathrm{Li}_{(\kk)}(x)\leq \mathrm{li}(x)$. Hence, $\lim_{t\rightarrow\infty}|\mathrm{Li}_{(\kk)}(x^{s})|=0$ because
\begin{eqnarray}
\lim_{t\rightarrow\infty}\left|\mathrm{Li}_{(\kk)}(x^{s})\right|
\leq\lim_{t\rightarrow\infty}\left|\mathrm{li}(x^{(c+it)})\right|
&=&\lim_{t\rightarrow\infty}\left|\frac{x^{(c+it)}}{(c+it)\log(x)}
\left(1+O\left(\frac{1}{(c+it)\log(x)}\right)\right) \right|\notag\\
&\leq&\frac{x^c}{\log(x)}\lim_{t\rightarrow\infty}
\left|\frac{1}{(c+it)}
\left(1+O\left(\frac{1}{(c+it)}\right)\right) \right|\notag\\
&=&0\;.
\end{eqnarray}

Next, we will need the truncating integral
\begin{lemma}For $n\in \mathbb{N}_+$,
\begin{equation}
\frac{1}{2\pi i}\int_{c-i T}^{c+i T}\left(\frac{x}{n_{(\kk)}}\right)^s\log(x)\frac{\log^{\kk-1}(x^s_{(\kk)})}{\log_{(\kk)}(x^s)}ds
=\left\{\begin{array}{l}
\mathcal{R}_{(\kk)}(1)^{-1}
+O\left(\frac{\left(\frac{x}{n}\right)^c}{T\log(x/n)}\right)\;\;\;\;\frac{x}{n}>1 \\
O\left(\frac{\left(\frac{x}{n}\right)^c}{T\log(x/n)}\right)\;\;\;\;0<\frac{x}{n}<1
\end{array}\right.\;.
\end{equation}
\end{lemma}
\emph{proof}:
For $\frac{x}{n}>1$ integrate over a rectangle with left edge $(L-i T,L+i T)$ and right edge $(c-i T,c+i T)$. We have
\begin{eqnarray}
\lim_{L\rightarrow-\infty}\left|\int_{L-i T}^{L+i T}\left(\frac{x}{n_{(\kk)}}\right)^s\log(x)\frac{\log^{\kk-1}(x^s_{(\kk)})}{\log_{(\kk)}(x^s)}\;ds\right|
&\leq&\lim_{L\rightarrow-\infty}\left|\int_{L-i T}^{L+i T}\left(\frac{x}{n_{(\kk)}}\right)^s\log(x)\frac{1}{\log(x^s)}ds\right|\notag\\
&&\hspace{-2.5in}\leq\lim_{L\rightarrow-\infty}\int_{-T}^{T}\frac{\left(\frac{x}{n_{(\kk)}}\right)^L}{|L+ i t|}dt
<\lim_{L\rightarrow-\infty}\frac{T \left(\frac{x}{n_{(\kk)}}\right)^L}{L}
<\lim_{L\rightarrow-\infty}\frac{T \left(\frac{x}{n}\right)^L}{L}=0\;.
\end{eqnarray}
The top and bottom contribute
\begin{eqnarray}
\lim_{L\rightarrow-\infty}\left|\int_{L\pm i T}^{c\pm i T}\left(\frac{x}{n_{(\kk)}}\right)^s\log(x)\frac{\log^{\kk-1}(x^s_{(\kk)})}{\log_{(\kk)}(x^s)}\;ds\right|
\leq\left|\int_{-\infty\pm i T}^{c\pm i T}\frac{\left(\frac{x}{n_{(\kk)}}\right)^s}{s}ds\right|
&\leq&\int_{0}^{\infty}\frac{-\left(\frac{x}{n_{(\kk)}}\right)^{c-r}}{|(c-r)\pm i T|}dr\notag\\
&<&\int_{0}^{\infty}\frac{-\left(\frac{x}{n}\right)^{c-r}}{T}dr\notag\\
&=&\frac{-\left(\frac{x}{n}\right)^{c}}{T\log(x/n)}\;.
\end{eqnarray}
Lastly, the pole at $s=0$ contributes residue $\mathrm{Res}=\mathcal{R}_{(\kk)}(1)^{-1}$.

For $\left(\frac{x}{n}\right)<1$ integrate over a rectangle with left edge $(c-i T,c+i T)$ and right edge $(R-i T,R+i T)$. Then
\begin{equation}
\lim_{R\rightarrow\infty}\left|\int_{R-i T}^{R+i T}\frac{\left(\frac{x}{n_{(\kk)}}\right)^s}{s}ds\right|
\leq\lim_{R\rightarrow\infty}\int_{-T}^{T}\frac{e^{-R|\log(x/n_{(\kk)})|}}{|R+ i t|}dt<\lim_{R\rightarrow\infty}\frac{T e^{-R|\log(x/n)|}}{R}=0\;.
\end{equation}
The top and bottom for $\frac{x}{n}<1$ contribute the same order as for $\frac{x}{n}>1$ by the analogous argument, and the lemma is established. $\qed$

Finally,  making use of the truncating integral with $x\rightarrow\widetilde{\x}$ (for $c>\sigma_a$) in the integration by parts of (\ref{explicit integral}) yields
\begin{eqnarray}
\lim_{\epsilon\rightarrow0}\lim_{T\rightarrow\infty}\frac{(-1)^{\kk-1}}{2\pi i}\mathcal{R}_{(\kk)}(1)\int_{c-iT}^{c+iT}\log^{(\kk-1)'}(\zeta_{(\kk)}(s))
\widetilde{\x}^{s}\log(\widetilde{\x})\frac{\log^{\kk-1}(\widetilde{\x}_{(\kk)}^s)}{\log_{(\kk)}
(\widetilde{\x}^s)}\,ds\notag\\
&&\hspace{-4in}
=\lim_{\epsilon\rightarrow0}\lim_{T\rightarrow\infty}\frac{\mathcal{R}_{(\kk)}(1)}{2\pi i}\int_{c-iT}^{c+iT}\sum_{n=1}^\infty\frac{\lambda_{(\kk)}(n)\log^{\kk-1}(n_{(\kk)})}{n_{(\kk)}^{s}}
 \widetilde{\x}^{s}\log(\widetilde{\x})\frac{\log^{\kk-1}(\widetilde{\x}_{(\kk)}^s)}
 {\log_{(\kk)}(\widetilde{\x}^s)}\,ds\notag\\
 &&\hspace{-4in}=\lim_{\epsilon\rightarrow0}\lim_{T\rightarrow\infty}
 \sum_{n=1}^\infty\lambda_{(\kk)}(n)\log^{\kk-1}(n_{(\kk)})
 \frac{\mathcal{R}_{(\kk)}(1)}{2\pi i}\int_{c-iT}^{c+iT}
 \frac{\widetilde{\x}^{s}}{n_{(\kk)}^s}
 \log(\widetilde{\x})\frac{\log^{\kk-1}(\widetilde{\x}_{(\kk)}^s)}
 {\log_{(\kk)}(\widetilde{\x}^s)}\,ds\notag\\
 &&\hspace{-4in}=\lim_{\epsilon\rightarrow0}\sum_{n\leq\lfloor\widetilde{\x}\rfloor}
 \lambda_{(\kk)}(n)\log^{\kk-1}(n_{(\kk)})\notag\\
 &&\hspace{-4in}=\sum_{n\leq \x}\lambda_{(\kk)}(n)\log^{\kk-1}(n_{(\kk)})\;.
 \end{eqnarray}
The third equality follows from the lemma, and the final line follows from $\widetilde{\x}:=\x+\epsilon$. Justifying the interchange of the sum and integral is straightforward, and interchange of the $T$-limit and sum is allowed because the summand contains $O(n^{-c})$ with $c>1$.
$\QED$

\section{Liouville is such a tool}\label{Liouville is a tool}
This appendix presents evidence that Liouville, being a functional integral over the dual-extended quantum phase space $W\sim Z\times Z'$ with symplectic symmetry, possesses structure that displays BRST symmetry and $S$-duality. Our treatment is preliminary and meant to encourage further investigation.

\subsection{BRST symmetry}
Section \ref{Quadratic-type integrators} displayed a hint of supersymmetry by treating $\eta/\psi'$ as Fourier duals relative to integrators characterized by a K\"{a}hler form, and in \S\ref{Liouville} supersymmetry appeared more clearly. But up until now we have avoided addressing symmetry aspects, because a thorough treatment deserves its own paper. Notwithstanding a thorough treatment, this subsection will briefly examine symmetry invariance of the Liouville integrator to argue that functional Liouville can accommodate the BRST formalism.

Recall the geometrical setup for the Liouville integrator in \S\ref{Liouville} and the well-known story of symplectic reduction\cite{MW}. For simplicity, specialize to the case where $\mathcal{J}$ is a vector space with symplectic form $\mathbf{\Omega}=d\Bold{\omega}$ and suppose the action on $\mathcal{J}$ of the symmetry subgroup $H^\C\subset
G^\C$ is Hamiltonian. Hence its Lie algebra $\mathfrak{H}$ induces Hamiltonian vector fields $\Bold{V}_{\!\!\mathfrak{h}}$ on $\mathcal{J}$  such that $\mathcal{L}_{\Bold{V}_{\!\!\mathfrak{h}}}\Bold{\omega}=0$ for all ${\mathfrak{h}}\in\mathfrak{H}$. This implies $\mathrm{i}_{\Bold{V}_{\!\!\mathfrak{h}}}\mathbf{\Omega}=-dh_{\Bold{V}_{\!\!\mathfrak{h}}}$, and so to each $\Bold{V}_{\!\!\mathfrak{h}}$ one can associate a generalized Hamiltonian\footnote{If the map from $\mathfrak{H}$ to the algebra of smooth functions with respect to the Poisson bracket defined by $\mathbf{\Omega}$ is a Lie algebra homomorphism and if $\mathbf{\Omega}=d\Bold{\omega}$, then the generalized Hamiltonian is given by $h_{\Bold{V}_{\!\!\mathfrak{h}}}=\mathrm{i}_{\Bold{V}_{\!\!\mathfrak{h}}}\Bold{\omega}$.} $h_{\Bold{V}_{\!\!\mathfrak{h}}}:\mathcal{J}\rightarrow\R$ and construct a moment(um) map $P:\mathcal{J}\rightarrow\mathfrak H^\ast$ defined by $P(\x)({\mathfrak{h}}):=h_{\Bold{V}_{\!\!\mathfrak{h}}}(\x)$. If the derivative map $P'$ is surjective, then $\mathrm{Ker}(P)$ is a coisotropic submanifold and $\mathrm{Ker}(P')$, which is spanned by $\{\Bold{V}_{\!\!\mathfrak{h}}:\mathfrak{h}\in\mathfrak{H}\}$, is a null distribution of $\mathbf{\Omega}$ pulled-back to $\mathrm{Ker}(P)$. Under a mild assumption $\mathrm{Ker}(P')$ is integrable, and hence $\mathrm{Ker}(P)$ projects onto its space of leaves where the induced $2$-form is now non-degenerate.

The crux of BRST is that the symplectic reduction story can be implemented on the space of fields using algebraic tools. Specifically, one introduces a double complex based on the Koszul resolution of $\mathrm{Ker}(P)$ and the Chevalley-Eilenberg complex of $\mathfrak{H}$ along with a differential operator $Q$ such that the gauge invariant fields on $\mathcal{J}$ are represented by the $0$-degree cohomology of the complex. The cohomology algebra can then be represented by a bi-graded Poisson algebra on the supermanifold $\Pi T\mathcal{J}$. The method is useful because it is much easier to implement a classical gauge theory on $\Pi T\mathcal{J}$ via the global BRST generator $Q$ than on $T\mathcal{J}$ via the local gauge symmetry generators. Indeed, BRST symmetry remains intact even if the gauge symmetry is absent/broken.

From the geometric picture underlying the Liouville functional integral, it is clearly possible to implement the restriction to $\Phi_\|$ by employing BRST on the space of fields instead of imposing $\mathcal{L}_{\Bold{V}_{\!\!\mathfrak{h}}}\Bold{\omega}=0$. So Liouville can naturally incorporate BRST symmetry: or rather, the geometry of Liouville is amenable to BRST methods. The algebraic and geometric interpretations of BRST symmetry have been well-understood for decades and there is no use in repeating it here for Liouville  (for a contemporary Lie algebroid formulation see \cite{CL}).

Instead, we want to use Liouville to give a perhaps more physical interpretation of BRST symmetry. The first task is to characterize the various degrees of freedom, say for $\mathrm{dim}_\C(\ZZZ)=4$. To make contact with gauge theory we take $\mathcal{J}$ to be a vector space furnishing a representation of a gauge symmetry group. Consequently, we don't have to linearize the space of fields and the domain of Liouville is simply $\Gamma\Phi$. We have already seen the pair $(\phi^I,\pi_I^\mu)$ corresponds to vector-valued canonical configuration and momentum fields. Moving on to the pair $(\pi_I,\phi^I_\mu)$, we can view $\phi_\mu^I$ as the derivative map associated with $\phi^I$. To interpret $\pi_I$, return to the example of (\ref{simple action}) to gain some intuition.  Since the simple form displayed there corresponds to canonical classical phase space, we can, by analogy with $\Bold{\omega}$, interpret $\pi_I$ as a momentum current associated with a momentum flux $\pi_I^\mu$.

To see that these interpretations are reasonable, take the Hamiltonian associated with evolution-time flow to be given by the energy density of the momentum flux $\mathrm{tr}(\star\pi\wedge \pi)$. Including the conjugate pair $(\pi_I,\phi^I_\mu)$, the action functional in canonical coordinates gives
\begin{equation}\label{full action}
\mathrm{S}(\Phi)=\int_{\ZZZ}\left\{\pi_I^\mu(\z)\phi^I_{\,,\,\mu}(\z)
+\phi_I^\mu(\z)\pi^I_{\,,\,\mu}(\z)
-\mathrm{H}_{\Bold{V}}(\phi^I(\z),\pi^\mu_I(\z)),\pi^I(\z),\phi^\mu_I(\z))\right\}
\Bold{\tau}_{\ZZZ}\;.
\end{equation}
Hamilton's equations are
\begin{eqnarray}
\phi^I_{,\,\mu}&=&{\pi}^I_{\mu}
\;\;\;\;\;\;\;\;{\pi_I^\mu}_{;\,\mu}=0\notag\\
\notag\\
\pi^I_{,\,\mu}&=&0
\;\;\;\;\;\;\;\;\;\;{\phi_I^\mu}_{;\,\mu}=0\;.
\end{eqnarray}
For this case, the gauge invariant $1$-form flux $\Bold{\varpi}$ on $\mathcal{J}$ will give rise to gauge invariant fields with divergence-free derivative maps,  conserved momentum currents, and divergence-free momentum fluxes. In particular, let $\mathcal{Z}=\R^{3,1}$ and choose  $\mathcal{M}\cong T^\ast\R^{3,1}\times T_e H^\C$ so that index $I=(i,\mu)$. The field equations are
\begin{eqnarray}
\phi^i_{[\,\mu,\,\nu]}&=&c^i_{jk} \phi^j_\mu \phi^k_\nu+{\pi}^i_{\mu\nu}
\;\;\;\;\;\;\;\;\tensor{\pi}{_i^{\mu\nu}_{,\,\mu}}=c^j_{ik}{\pi}_j^{\mu\nu}\phi_\mu^k\notag\\
\notag\\
\pi^i_{[\,\mu,\,\nu]}&=&c^i_{jk} \pi^j_\mu \phi^k_\nu
\hspace{.8in}
\tensor{\phi}{_i^{\mu\nu}_{,\,\mu}}=c^j_{ik}{\phi}_j^{\mu\nu}\phi_\mu^k\;.
\end{eqnarray}
Observe that $\mathrm{S}(\Phi)$ is invariant under the on-shell nilpotent variations
\begin{equation}
 \delta_\varepsilon\phi^I=\varepsilon\pi^I\,,\;\;\;\;
 \delta_\varepsilon\pi^I_{\mu}=\varepsilon \p_\mu\pi^I\,,\;\;\;\;
 \delta_\varepsilon\pi^I=0\,,\;\;\;\;
 \delta_\varepsilon\phi^I_\mu=\varepsilon \p_\mu\phi^I
\end{equation}
where $\varepsilon$ is a complex variational parameter. Evidently, we should regard $(\phi^I,\pi_I)$ as configuration and momentum-current fields on $\mathcal{M}$ and $(\pi^\mu_I,\phi_\mu^I)$ as their dynamically-conjugate fields. In other words, if $Z$ represents a quantum phase space and $Z'$ its topological dual, then $z:=(\phi^I,\pi^\mu_I)\in Z$ and $z':=(\pi^I,\phi_\mu^I)\in Z'$. The duality is defined point-wise by the relation $\langle z'(\z),z(\z)\rangle\in\C$ where $z:\mathcal{Z}\rightarrow T_{\Phi(\z)}\mathcal{J}$. Then the dual Lie group transformations associated with $z'$ are defined by
\begin{equation}\label{dual Lie group}
\langle\sigma'_g z'(\z),z(\z)\rangle
:=\langle z'(\z),\sigma_{g^{-1}}z(\z)\rangle
\end{equation}
where $\{\sigma_g\}$ is the Lie group of transformations on $\Phi^\ast T\mathcal{J}$ induced by all $g\in Sp(T_{\Phi(\z)}\mathcal{J},\C)$. This implies $\mathcal{D}_\lambda(w^{\mathbf{e}},w^{\mathbf{o}})$ is invariant under $Sp(T_{\Phi(\z)}\mathcal{J},\C)$.

These fields, being linear combinations of $w^{\mathbf{e},\mathbf{o}}$, represent both Gaussian and skew-Gaussian complex degrees of freedom.  By now it is fairly clear Liouville possesses the requisite degrees of freedom to display \emph{classical} BRST symmetry.\footnote{It is clear the BRST symmetry is a manifestation of the underlying symplectic symmetry on $\mathcal{J}$. In fact, both the BRST and SUSY-like symmetry seen in \S\ref{Skew-Gaussian integrators} are just particular symplectic transformations on the tangent space $T_{\Phi}\Gamma\Phi$. Notice that extending the quantum phase space to $Z\times Z'$ converts the  on-shell non-nilpotent SUSY-like variation in \S\ref{Skew-Gaussian integrators} into the on-shell nilpotent BRST variation seen here.} Keep in mind these characterizations are predicated on assuming that time evolution is derived from the momentum flux energy density only --- which seems like a reasonable physical expectation.

We would now like to use an action functional like (\ref{full action}) in the Liouville integral of \S\ref{Liouville}. But with a gauge symmetry present, the Jacobi matrix will be degenerate. The degeneracy can be lifted in the usual way by adding a gauge fixing functional, say $\mathrm{S}_{\mathrm{gf}}(w)$, to the original action so that $\mathcal{L}_{\Bold{V}_{(\mu)}}\Bold{\omega}\neq0$. Not surprisingly, the classical BRST symmetry remains since, after all, it is a symplectic transformation.

To see this, consider $\int_We^{-\mathrm{S}(w^{\mathbf{e}},w^{\mathbf{o}})-\mathrm{S}_{\mathrm{gf}}(w^{\mathbf{e}},w^{\mathbf{o}})}\;\mathcal{D}_\lambda(w^{\mathbf{e}},w^{\mathbf{o}})$
with the symmetry breaking action functional $\mathrm{S}_{\mathrm{gf}}(\Phi)
=\mathrm{Q}_{\mathrm{gf}}(w^{\mathbf{e}})+\Omega_{\mathrm{gf}}(w^{\mathbf{o}})$. To insure the gauge-fixed integral represents gauge-fixed degrees of freedom, choose $\Omega_{\mathrm{gf}}= \mathrm{Q}_{\mathrm{gf}}\circ J$. Then, from the definition of Liouville get\footnote{This is strictly true only for quadratic gauge fixing. Non-quadratic gauge fixing can be used while maintaining the BRST symmetry, but the functional integral relations will hold only perturbatively.}
\begin{eqnarray}
\int_W\!\!\!e^{-\mathrm{S}(w)-\mathrm{S}_{\mathrm{gf}}(w-\bar{w})}\;\mathcal{D}_\lambda (w^{\mathbf{e}},w^{\mathbf{o}})
&=&\int_{W^{\mathbf{e}}}\!\!\!\mathrm{Pf}(\Omega)e^{-\mathrm{Q}(w^{\mathbf{e}}-\bar{w}^{\mathbf{e}})}
\;\mathcal{D}_\lambda l_{\mathrm{Q}_{\mathrm{gf}},\Omega_{\mathrm{gf}}}(w^{\mathbf{e}})\notag\\
&=&\int_{W^{\mathbf{e}}_{\mathrm{gf}}}\!\!\!\!\mathrm{Pf}(\Omega)
e^{-\mathrm{Q}(w^{\mathbf{e}}-\bar{w}^{\mathbf{e}})}\;\mathcal{D}_\lambda w^{\mathbf{e}}\;.
\end{eqnarray}
The second equality follows from the identities\footnote{These follow from the affine linear change of variable $\tilde{Q}=Q\circ L$ applied to the definition of the Gaussian and skew-Gaussian integrators. (see e.g. \cite[ch. 4]{CA/D-W3})} $\int_{Z_0}e^{-\tilde{Q}(z)}\mathcal{D}_\lambda\omega_{\mathrm{Q}}(z)
=\mathrm{Det}(\mathrm{Q})^{1/2}/\mathrm{Det}(\tilde{\mathrm{Q}})^{1/2}$ and $\int_{Z_0}e^{-\tilde{\Omega}(\eta)}\mathcal{D}_\lambda\omega_{\Omega}(\eta)
=\mathrm{Pf}(\tilde{\Omega})/\mathrm{Pf}(\Omega)$ which yield
\begin{equation}
\int_{W^{\mathbf{e}}}\mathrm{Pf}(\Omega)e^{-\mathrm{Q}(w^{\mathbf{e}}-\bar{w}^{\mathbf{e}})}
\;\mathcal{D}_\lambda l_{\mathrm{Q}_{\mathrm{gf}},\Omega_{\mathrm{gf}}}(w^{\mathbf{e}})
=\frac{\mathrm{Det}_\lambda(\mathrm{Q}_{\mathrm{gf}})^{1/2}}
{\mathrm{Det}_\lambda(\mathrm{Q})^{1/2}}
\frac{\mathrm{Pf}_\lambda(\Omega)}
{\mathrm{Pf}_\lambda(\Omega_{\mathrm{gf}})}
=\left.\frac{\mathrm{pf}(\Omega)}
{\sqrt{\mathrm{det}(\mathrm{Q})}}\right|_{W_{\mathrm{gf}}}
\end{equation}
since $\Omega_{\mathrm{gf}}= \mathrm{Q}_{\mathrm{gf}}\circ J$. In words, gauge-fixed Liouville integrates the gauge-invariant action functional with a Liouville integrator characterized by gauge-fixed quadratic and skew-quadratic forms. Alternatively, it can be viewed as integration of a gauge invariant action functional over gauge-fixed fields with respect to the primitive Liouville integrator (characterized by trivial quadratic forms).

We have assumed there is a symmetry group $H^\C$ acting by translations on $\mathcal{J}$ such that $\mathcal{L}_{\Bold{V}_{(\mu)}}\Bold{\omega}=0$. It follows that $\mathcal{L}_{\Bold{V}_{(\mu)}}\mathbf{\Omega}=0$ since we specialized to $\mathbf{\Omega}=d\Bold{\omega}$; implying $\{\Bold{V}_{(\mu)}\}$ are symplectomorphisms. But the structure group is $Sp(T_{\Phi(\z)}\mathcal{J},\C)$, so the symmetry group is isomorphic to a subgroup of the structure group. Although there may be subgroups remaining in $Sp(T_{\Phi(\z)}\mathcal{J},\C)/H^\C$, they will not induce gauge transformations unless they leave $\Bold{\omega}$ invariant. Furthermore, the Liouville integrator is automatically gauge invariant if the action functional is gauge invariant. We conclude that
\begin{equation}
\int_{W^{\mathbf{e}}_{\mathrm{gf}}}\mathrm{Pf}(\Omega)e^{-\mathrm{Q}(w^{\mathbf{e}}-\bar{w}^{\mathbf{e}})}\;\mathcal{D}_\lambda w^{\mathbf{e}}
=\mathrm{Vol}_\lambda^{1/2}(W^{\mathbf{o}}/W^{\mathbf{e}})
_{\mathrm{Q}+\mathrm{Q}_{\mathrm{gf}},\Omega+\Omega_{\mathrm{gf}}}\;.
\end{equation}

These integrals can be applied to Yang-Mills QFT when $\mathcal{M}$ is the abelian group of differential $1$-forms ${\Lambda}^1
(\mathcal{Z},\mathfrak{H}^\C)$ where $\mathfrak{H}^\C$ is the Lie algebra of the symmetry group. First, use the complex structure $J$ to project onto real fields $w^{\mathbf{e},\mathbf{o}}_R$ and take $\bar{w}=0$. Then, with the identifications ${w_R^{\mathbf{e}}}^I\sim A^i_\mu$, ${w_R^{\mathbf{e}}}^I_\nu\sim F^i_{\mu\nu}$, ${w_R^{\mathbf{o}}}^I\sim c^I$, and ${w_R^{\mathbf{o}}}^I_{\nu}\sim \p_\nu\bar{c}^{\,I}$, take the first-order Lagrangian density in (\ref{Yang-Mills}) and add the usual ghost and (quadratic) gauge-fixing terms --- with the understanding that the $(c^I,\,\p_\nu\bar{c}^I)$ ghosts stem from the Maurer-Cartan $1$-form\cite{BR}.\footnote{More precisely, $c^I$ is the \emph{vertical} Maurer-Cartan $1$-form extended to the principal bundle $G^\C\rightarrow \tilde{P}\stackrel{{\Pi}}{\rightarrow} \ZZZ$ via the Ehresmann connection $\chi$,  and $\p_\mu\bar{c}^i$ is its conjugate ``momentum'' (see \S\ref{Liouville}). Being vertical, $c^I$ only has $c^i$ non-zero components.} The integral over $F$ can be performed if one wishes to see the standard second-order formulation.

\subsection{Shades of duality}
The Mathai-Quillen Thom class representative displayed in (\ref{duality}) exhibits $S$-duality for holomorphic maps if we take the linear operator $\mathrm{D}$ to be the Dolbeault operator $\bar{\p}$ and integrate over $\psi'$.

Start with
\begin{equation}
\int_{Z_0\times Z_0'}e^{2\pi i \langle
{\bar{\p}'}\psi',\eta\rangle-\pi\s\Omega_\mathrm{B}(\eta)}
\,\mathcal{D}_\lambda(\eta,\psi') =\mathrm{Pf}_\lambda(\Omega_\mathrm{B}/\s)
\int_{Z_0'}e^{-(\pi /\s)|\psi'|^2}\,\mathcal{D}_\lambda\psi'\;.
\end{equation}
Doing the integration over $\psi'$ yields (using $\langle \bar{\p}'\psi',\eta\rangle=\langle \psi',\bar{\p}\eta\rangle$)
\begin{equation}
\int_{Z_0\times Z_0'}e^{2\pi i \langle
\psi',{\bar{\p}}\eta\rangle-\pi\s\Omega_\mathrm{B}(\eta)}
\,\mathcal{D}_\lambda(\eta,\psi')
=\int_{Z_0}\delta(\bar{\p}\eta)e^{-\pi\s\Omega_\mathrm{B}(\eta)}
\,\mathcal{D}_\lambda\eta
=\int_{Z^{(\mathrm{hol})}_0}
e^{-\pi\s\Omega_\mathrm{B}(\eta)}\,\mathcal{D}_\lambda\eta
\end{equation}
where $Z^{(\mathrm{hol})}_0\subset Z_0$ is the subspace of holomorphic pointed maps. Conclude that
\begin{equation}
\int_{Z^{(\mathrm{hol})}_0}e^{-\pi\s\Omega_\mathrm{B}(\eta)}
\,\mathcal{D}_\lambda\eta
=\mathrm{Pf}_\lambda(\Omega_\mathrm{B}/\s)
\int_{Z_0'}e^{-(\pi /\s)|\psi'|^2}\,\mathcal{D}_\lambda\psi'
\end{equation}
which expresses a Dolbeault strong-weak duality for the scaling factor $\s$ of the dual fields $\eta$ and $\psi'=\mathrm{C}\eta'$.\footnote{Zeta regularization of $\mathrm{Pf}_\lambda(\Omega_\mathrm{B}/\s)$ will induce an exponent of $\s$ that depends on the zeta function associated with $\Omega$ and, hence, will depend on the Euler characteristic of the target manifold of $\eta$.} In this simple example, the $S$-duality is seen to be a direct consequence of the topological duality between $Z_0$ and $Z_0'$.

To see the duality in terms of $W^{\mathbf{e}}\oplus W^{\mathbf{o}}$, recall the full integral representation (\ref{restored integral}) of Liouville. When $\bar{w}=0$ and $\hat{\s}=\check{\s}\equiv\s$, the action takes the form
\begin{equation}\label{s-dual action}
\mathrm{S}(w^{\mathbf{e}},w^{\mathbf{o}})=\pi\int_{\ZZZ}\left(
  \begin{array}{cc}
    w^{\mathbf{e}} ,& w^{\mathbf{o}} \\
  \end{array}
\right)
\left(
  \begin{array}{ll}
    1/\s\,\mathrm{Q} & 0 \\
   0 & \s\,\Omega \\
  \end{array}
\right)
\left(
  \begin{array}{c}
    w^{\mathbf{e}} \\
    w^{\mathbf{o}} \\
  \end{array}
\right)\Bold{\tau_{\ZZZ}}
\end{equation}
where $\s\in\C_+$. By definition of Liouville, $\mathrm{Q}$ and $\Omega$ are allowed to have unequal ranks, say $\mathrm{rank}(\mathrm{Q})=\alpha$ and $\mathrm{rank}(\Omega)=\beta$. Then the action is invariant under global $\mathfrak{osp}(\alpha,\beta)$. And, since $\mathcal{D}_\lambda(w^{\mathbf{e}},w^{\mathbf{o}})$ is invariant under $Sp(T_{\Phi(\z)}\mathcal{J},\C)$, it follows that the functional integral $\mathrm{Vol}^{1/2}_\lambda\left(W^{\mathbf{o}}\,/\,W^{\mathbf{e}}\right)_{\mathrm{Q},\,\Omega}$ is invariant under global $\mathfrak{osp}(\alpha,\beta)$. This superalgebra has been classified\cite{KAC}
\begin{eqnarray}
B(a,b)&:=&\mathfrak{osp}(2a+1,2b)\;,\,\,\,\,\,\,a\geq0,\;b>0\notag\\
C(b)&:=&\mathfrak{osp}(2,2b-2)\;,\,\,\,\,\,\,b\geq2\notag\\
D(a,b)&:=&\mathfrak{osp}(2a,2b)\;,\,\,\,\,\,\,a\geq2,\;b>0
\end{eqnarray}
whose even subalgebras are products of the classical Lie algebras $B_n,C_n,D_n$
\begin{equation}
\begin{array}{lll}
B^+(a,b)&\cong B_a\times C_b &=\mathfrak{so}(2a+1)\times\mathfrak{sp}(2b)\\
C^+(b)&\cong C_{b-1}\times \C &=\C\times\mathfrak{sp}(2(b-1))\\
D^+(a,b)&\cong D_a\times C_b &=\mathfrak{so}(2a)\times\mathfrak{sp}(2b)\;.
\end{array}
\end{equation}

If  $\mathrm{rank}(\mathrm{Q})=\mathrm{rank}(\Omega)$, then integration over $W^{\mathbf{e}}$  yields a manifestly $\mathfrak{sp}(2a)$ invariant $\int_{W^{\mathbf{o}}}\mathrm{Det}(\mathrm{Q}/\s)^{-1/2}e^{-\pi\s\,\Omega(w^{\mathbf{o}})}
\;\mathcal{D}_\lambda w^{\mathbf{o}}$ with a ``hidden'' $\mathfrak{so}(2a)$ symmetry. On the other hand, integration over $W^{\mathbf{o}}$ instead gives a manifestly $\mathfrak{so}(2a)$ invariant $\int_{W^{\mathbf{e}}}\mathrm{Pf}(\Omega/\s)e^{-(\pi/s)\mathrm{Q}(w^{\mathbf{e}})}
\;\mathcal{D}_\lambda w^{\mathbf{e}}$ with a ``hidden'' $\mathfrak{sp}(2a)$ symmetry. Meanwhile, if $\mathrm{rank}(\mathrm{Q})=1$ we get $\C\leftrightarrow\mathfrak{sp}(2a)$, and the Langlands dual pair appears for  $\mathrm{rank}(\mathrm{Q})=\mathrm{rank}(\Omega)+1$. Of course the integrated dual symmetry is not really hidden since functional $\mathrm{Pf}(\Omega)$ and $\mathrm{Det}(\mathrm{Q})$ are suitably invariant.

There are two caveats to add here: we're working on $T_\Phi\Gamma\Phi$ and not $\Gamma\Phi$, and the action functional will not be quadratic in general. Therefore, it seems unlikely the global symmetry will hold in general. This suggests the global symmetry is not generically realized quantum mechanically on the abelian topological group $T_\Phi\Gamma\Phi$ at every point $\Phi\in\Gamma\Phi$, but perhaps it may be the expression of a local symmetry on the non-abelian topological group $\Gamma\Phi$ --- at least semi-classically. If so, we should investigate gauged $OSp(\alpha,\beta)$ QFT.

In particular, consider Liouville for super Yang-Mills with a non-quadratic action functional $\mathrm{S}$ in the case $\mathrm{rank}(\mathrm{Q})=\mathrm{rank}(\Omega)+1$. Referring to the discussion around (\ref{non-linear Liouville}), we formally have
\begin{equation}
\int_{W}
 e^{-\pi{\mathrm{S}}(\Phi-\bar{\Phi})}
 \;\mathcal{D}_\lambda \Phi
 :=\mathrm{Vol}^{1/2}_\lambda(W^{\mathbf{o}}/W^{\mathbf{e}})
 _{{\mathrm{S}}}
\end{equation}
where $W^{\mathbf{e}}$ is the space of $1$-forms and $2$-forms with values in $\mathfrak{so}(2a+1)$ and $W^{\mathbf{o}}$ is the space of 1-forms and 2-forms with values in $\mathfrak{sp}(2a)$. Restricting to real fields, using the Yang-Mills action functional (\ref{Yang-Mills}), and integrating over the $1$-forms $A\in W^{\mathbf{e},\mathbf{o}}$
 gives
\begin{equation}
 \int_{W^{\mathbf{e}}\oplus W^{\mathbf{o}}}
 e^{-(\pi/\s)\mathrm{S}(F_{A_{(\mathfrak{so})}})-(\pi\s)\mathrm{S}(F_{A_{(\mathfrak{sp})}})}
 \;\mathcal{D}_\lambda F_A
 =\mathrm{Vol}^{1/2}_\lambda(W^{\mathbf{o}}/W^{\mathbf{e}})
 _{{\mathrm{S}}}
\end{equation}
where $\mathrm{S}(F_A)=\int_{\mathbb{M}^4}\mathrm{tr}(\star F_A\wedge F_A)$. Since $\mathrm{S}(F_A)$ is quadratic this can be expressed as
\begin{eqnarray}
 \int_{W^{\mathbf{e}}}\mathrm{Pf}(\mathrm{S}_{(\mathfrak{sp})}/\s)
 e^{-(\pi/\s)\mathrm{S}(F_{A_{(\mathfrak{so})}})}
 \;\mathcal{D}_\lambda F_{A_{(\mathfrak{so})}}
 &=& \int_{W^{\mathbf{o}}}\mathrm{Det}(\mathrm{S}_{(\mathfrak{so})}/\s)^{-1/2}
 e^{-(\pi\s)\mathrm{S}(F_{A_{(\mathfrak{sp})}})}
 \;\mathcal{D}_\lambda F_{A_{(\mathfrak{sp})}}\notag\\
 &=&\mathrm{Vol}^{1/2}_\lambda(W^{\mathbf{o}}/W^{\mathbf{e}})
 _{{\mathrm{S}}}\;.
\end{eqnarray}
According to the previous subsection, the action functionals in these integrals must be augmented by gauge fixing terms; in which case the $\mathrm{Pf}$ and $\mathrm{Det}$ terms can be interpreted as ghost functional integrals for even and odd fields respectively. In the end we have
\begin{eqnarray}
 \int_{W^{\mathbf{e}}_{\mathrm{gf}}}\mathrm{Pf}(\mathrm{S}_{(\mathfrak{sp})}/\s)
 e^{-(\pi/\s)\mathrm{S}(F_{A_{(\mathfrak{so})}})}
 \;\mathcal{D}_\lambda F_{A_{(\mathfrak{so})}}
 &=& \int_{W^{\mathbf{o}}_{\mathrm{gf}}}\mathrm{Det}(\mathrm{S}_{(\mathfrak{so})}/\s)^{-1/2}
 e^{-(\pi\s)\mathrm{S}(F_{A_{(\mathfrak{sp})}})}
 \;\mathcal{D}_\lambda F_{A_{(\mathfrak{sp})}}\notag\\
 &=&\mathrm{Vol}^{1/2}_\lambda(W^{\mathbf{o}}/W^{\mathbf{e}})
 _{\mathrm{S}+\mathrm{S}_{\mathrm{gf}}}\;.
\end{eqnarray}

This shows strong-weak duality between Langlands-dual Yang-Mills QFTs in the context of the Liouville integrator. Recall $\s\in\C_+$ so $\s\rightarrow 1/\s$ is equivalent to the usual $\tau\rightarrow -1/\tau$ for $\tau\in\mathbb{H}$ under the exchange $x\leftrightarrow iv$ and $iy\leftrightarrow u$ for $\s=x+iy$ with $x\geq0$ and $\tau=u+iv$ with $v\geq0$. We put $\hat{\s}=\check{\s}\equiv\s$ in (\ref{s-dual action}) by hand so the duality misses the expected lacing number. A more complete study would have to explain if and how the lacing number enters the picture if we don't assume $\hat{\s}=\check{\s}\equiv\s$. Note that if we re-scale $\check{\s}\rightarrow 2\check{\s}$ then duality implies $\check{\s}\leftrightarrow 1/\hat{\s}$ and the lacing number is not present as explained in \cite[pg. 13]{KW}.

\end{document}